\documentclass[conference]{IEEEtran}
\IEEEoverridecommandlockouts
\AtBeginDocument{%
  }

\usepackage[switch]{lineno}

\usepackage{tikz}

\usepackage{amsmath}

\usepackage{amssymb}

\usepackage{amsthm}

\usepackage{array}

\usepackage{multirow}

\usepackage{stmaryrd}

\usepackage[inline]{enumitem}

\usepackage{mathtools}

\usepackage{centernot}

\usepackage{verbatim}

\usepackage{prftree}

\usepackage{xspace}

\usepackage{thmtools}

\usepackage{thm-restate}

\usepackage{marginnote}

\usepackage{cleveref}

\Crefname{example}{Ex.}{Exs.}
\Crefname{equation}{Eq.}{Eqs.}
\Crefname{figure}{Fig.}{Figs.}
\Crefname{tabular}{Tab.}{Tabs.}
\Crefname{section}{Sec.}{Secs.}
\Crefname{definition}{Def.}{Defs.}
\Crefname{theorem}{Thm.}{Thms.}
\Crefname{proposition}{Prop.}{Props.}
\Crefname{lemma}{Lem.}{Lems.}
\Crefname{corollary}{Cor.}{Cors.}
\Crefname{appendix}{App.}{Apps.}

\usetikzlibrary{fit,positioning,arrows,shapes,automata,shapes.callouts}

\newenvironment{sizeddisplay}[1]
  {\par\nopagebreak#1\noindent\ignorespaces}
 {\nopagebreak\ignorespacesafterend}

\newcommand{\cL}{\ensuremath{\mathcal{L}}\xspace}

\makeatletter
\newcommand{\actiont}[1]{%
	\@tempswafalse
	\@for\next:=#1\do
	{\if@tempswa.\else\@tempswatrue\fi\texttt{\next}}%
}

\makeatother

\newcommand{\kP}[1][]{\if\relax\detokenize{#1}\relax\ensuremath{P}\else\ensuremath{P_{#1}}\fi\xspace}
\newcommand{\kQ}[1][]{\if\relax\detokenize{#1}\relax\ensuremath{Q}\else\ensuremath{Q_{#1}}\fi\xspace}

\newcommand*{\eg}{\textit{e.g.}\xspace}
\newcommand*{\Eg}{\textit{E.g.}\xspace}
\newcommand*{\ie}{\textit{i.e.,}\xspace}

\newcommand*{\etal}{\textit{et~al.}\xspace}
\newcommand*{\wrt}{w.r.t.\xspace}

\newcommand*{\resp}{resp.\xspace}

\newcommand*{\sth}{\text{s.t.}\xspace}

\newcommand*{\Wlog}{\textit{W.l.o.g.}\xspace}

\newcommand*{\sHML}{\textsc{sHML}\xspace}
\newcommand*{\sHMLw}{\ensuremath{\textsc{sHML}^\vee}\xspace}
\newcommand*{\sHMLwDet}{\ensuremath{\sHMLw_{\scriptscriptstyle \Det}}\xspace}
\newcommand*{\sHMLnf}{\ensuremath{\sHMLw_\textsc{nf}}\xspace}
\newcommand*{\recHML}{\textsc{recHML}\xspace}

\usepackage{preamble/shorthand}
\usepackage{preamble/pic}
\usepackage{preamble/monitor}
\usepackage{preamble/mathpartir}
\usepackage{preamble/rechml}
\usepackage{preamble/actor}

\usepackage{graphicx}
\usepackage{mathptmx}
\usepackage{amsmath}
\usepackage{amssymb}
\usepackage{amsthm}

\newtheorem{theorem}{Theorem}[section]
\newtheorem{lemma}[theorem]{Lemma}
\newtheorem{proposition}[theorem]{Proposition}
\newtheorem{corollary}{Corollary}
\theoremstyle{definition}
\newtheorem{definition}{Definition}[section]
\newtheorem{example}{Example}[section]
\newtheorem{remark}{Remark}

\begin{document}

\title{If At First You Don't Succeed: Extended Monitorability through Multiple Executions
\thanks{Supported by the doctoral student grant of the Reykjavik University Reseach Fund and 
``Mode(l)s of Verification and Monitorability'' (MoVeMent) (grant no~217987) of the Icelandic Research Fund.
}
}

\author{\IEEEauthorblockN{Antonis Achilleos}
\IEEEauthorblockA{\textit{Reykjavik University} \\
Reykjavik, Iceland \\
antonios@ru.is}
\and
\IEEEauthorblockN{Adrian Francalanza}
\IEEEauthorblockA{\textit{University of Malta} \\
Msida, Malta \\
adrian.francalanza@um.edu.mt}
\and
\IEEEauthorblockN{Jasmine Xuereb}
\IEEEauthorblockA{\textit{Reykjavik University and University of Malta} \\
Reykjavik, Iceland, and Msida, Malta \\
jasmine.xuereb.15@um.edu.mt}
}

\maketitle


\begin{abstract}
  This paper studies the extent to which branching-time properties can be adequately verified using runtime monitors.
  We depart from the classical setup where monitoring is limited to a single system execution and investigate the enhanced observational capabilities when monitoring a system over multiple runs.
  To ensure generality, 
  we focus on branching-time properties expressed in the modal $\mu$-calculus, a well-studied foundational logic.
  %
  Our results show that the proposed setup can systematically extend 
  established monitorability limits for branching-time properties.
  We 
  validate our results by instantiating them to verify actor-based systems. 
  We also prove bounds that capture the correspondence between the syntactic structure of a property and the number of required system runs.
  %
\end{abstract}



\section{Introduction}
\label{sec:intro}

Branching-time properties have long been considered the preserve of static analyses, verified using established techniques such as model checking~\cite{Clarke1999Book,Baier2008Book}.
Unfortunately, these verification techniques cannot be used 
when the system model is either too expensive to build and analyse (\eg state-explosion problems), poorly understood (\eg system logic governed by machine-learning procedures) or downright unavailable (\eg restrictions due to intellectual property rights).
Recent work has shown that runtime monitoring can be used  effectively (in isolation or in conjunction with other verification techniques) to verify 
certain branching-time properties~\cite{
KestenP05:CTL-RV,
PnueliZaks06:RV-MC,
DaSilvaMelo09:MC-RV,
HinrichsSZ14:MC-RV,
AhrendtCPS15FM:Static-RV,
FrancalanzaAI17:monitorability-branching,
DesaiDS17:MC-RV,
KejstovaRB17:MC-RV,
AcetoAFIL19:monitorability-linear,
DBLP:conf/forte/AttardAAFIL21,
StuckiSSBFMSD21:Gray-box-RV,
AudritoDSTV22:CTL-RV,
FerrandoM22:MC-RV,
AcetoCFI23}.
Specifically, (execution) \emph{monitors} (or sequence recognisers)~\cite{Schneider:2000,Ligatti05,BielM11:EditAut,DBLP:journals/iandc/Francalanza21} passively observe the \emph{execution} of a system-under-scrutiny (SUS), possibly aided by auxiliary information, 
to compare the observed behaviour (instead of its state space) against a correctness property of interest. 

The 
use of monitors for verification purposes is called runtime verification (RV)~\cite{BartocciFFR18,DBLP:journals/jlp/LeuckerS09}.
It is weaker
than static techniques for verifying both linear-time and 
branching-time properties: 
%
monitor observations are 
constrained to  
the current (single) computation path of the SUS
limiting 
the \emph{range} of verifiable properties.
%
For instance, the linear-time property $G\, \psi$ (always $\psi$) can only be 
monitored for violations but not satisfactions, whereas infinite renewal properties such as $G\, F\, \psi$ \emph{cannot} be monitored for at all.
Monitorability limits are more acute for branching-time properties: 
the \emph{maximal} monitorable subset for the modal $\mu$-calculus was shown to be semantically equivalent to the syntactic fragment  $\sHML {\cup} \textsc{cHML}$~\cite{FrancalanzaAI17:monitorability-branching,AcetoAFIL19:monitorability-linear}.  

\begin{example}\label{ex:intro}
  Consider 
  a server SUS exhibiting four events: 
  receive queries ($r$), 
  service queries ($s$),
  allocate memory ($a$)
  and close connection ($c$).  
  Modal $\mu$-calculus 
  properties\footnote{
  Formula 
  $\Um{\alpha}{\F}$ 
  describes 
  states that \emph{cannot} perform $\alpha$ transitions whereas 
  its dual, 
  $\Em{\alpha}{\T}$ 
  describes states that \emph{can} perform $\alpha$ transitions.}
  such as \textsl{``all interactions can only start with a receive query''}, 
  \ie $\varphi_0 \deftxtS \Um{s}{\F} \wedge \Um{a}{\F} \wedge \Um{c}{\F} \in \sHML$ 
  can be runtime verified since any SUS execution observed 
  that starts 
  with
   event $s$, $a$ or $c$ confirms that the running SUS violates the property (irrespective of any execution events that may follow). 
  %
  However, the branching-time property  \textsl{``systems that can perform a receive action, \Em{r}{\T}, \emph{cannot} also 
  close, \Um{c}{\F}''}, \ie 
  %
  \begin{center}
    $\varphi_1 \deftxtS\ \Em{r}{\T} \Rightarrow \Um{c}{\F} \ \ \equiv\ \ \Um{r}{\F} \vee  \Um{c}{\F} \ \not\in \ \sHML \cup \textsc{cHML}$
  \end{center}
  is \emph{not} monitorable for either satisfactions or violations.
  No (single) trace prefix provides enough evidence to conclude that a system satisfies this property,
  whereas an observed trace starting with 
  $r$ (dually $c$) is not enough to conclude that the emitting SUS (state) violates the property: 
  %
  one also needs 
  evidence that the same 
  state can also emit 
  $c$ (dually $r$).
  \qed
\end{example}

There are various approaches for extending the set of monitorable properties.
One method is to weaken the detection requirements expected of the 
monitors~\cite{AcetoAFIL21:sosym,DBLP:conf/csl/AcetoAFIL21}
 (\eg 
 allowing certain 
violations to go undetected). 
This, in turn, 
impinges on 
what it means for a property to be monitorable. 
Another 
approach 
is to increase the monitors' observational capabilities.
Aceto \etal~\cite{AceAFI:18:FOSSACS} investigate the increased 
observational power 
after augmenting the information recorded in the trace: apart from reporting computational steps \emph{that happened},
they consider trace events that can also record branching information such as the computation steps that \emph{could have} happened \emph{at a particular state}, 
or the computation steps that could \emph{not} have happened.
This approach 
treats the SUS as a \emph{grey-box}~\cite{StuckiSSBFMSD21:Gray-box-RV,ZhangLD12:Gray-box-RV} since the augmented traces reveal information about the SUS states reached.
This paper builds 
on Aceto \etal's work while sticking, as much as possible, to a black-box treatment of the SUS. 
%
We study the increase in 
observational power obtained from considering multiple execution traces for the same SUS \emph{without} relying directly on information about the specific intermediary states reached during monitoring. 

\begin{example}
  \label{ex:intro2}
  Property $\varphi_1$ from \Cref{ex:intro} can be monitored for violations over \emph{two} executions of the same system: 
  a first trace starting with event, $r$, and a second trace starting with event, $c$, is sufficient evidence to conclude that the SUS violates $\varphi_1$. 
  \qed 
\end{example}

Analysing multiple traces is \emph{not} always sufficient 
to conclude that a system violates a property with 
disjunctions since the same prefix could, in principle, reach different states.

\begin{example}\label{ex:non-det}
  Consider the property \textsl{``after any receive query, \Um{r}{\ldots}, if a SUS can service it, \Em{s}{\T}, then (it 
  takes precedence and) it should not 
  allocate more memory, \Um{a}{\F}''}, expressed as 
  %
  \begin{center}
    $\varphi_2 \deftxtS \Um{r}{(\Em{s}{\T} \Rightarrow \Um{a}{\F})} \ \equiv\ \Um{r}{(\Disj{\Um{s}{\F}}{\Um{a}{\F}})} $
  \end{center}
  Intuitively, $\varphi_2$ 
  is violated 
  when the state reached after 
  event $r$ can perform \emph{both} events $s$ and $a$. 
  %
  Observing traces 
  $rs\cdots$ and $ra\cdots$ along two 
  executions is not enough
   to conclude that the SUS violates $\varphi_2$: 
  although both executions start from the same state, say $p$, distinct states could 
  be reached after 
  event $r$, \ie $p \traS{r} p_1 \traS{s} p_2$ and $p \traS{r} p'_1 \traS{a} p'_2$ where $p_1 \neq p'_1$.   
  \qed
\end{example}

Although non-deterministic SUS behaviour cannot be ruled out in general, 
%
many systems are deterministic \wrt a \emph{subset} of actions, such as asynchronous LTSs and output actions~\cite{Selinger97Concur,HondaTokoro91} (\eg if $r$ was an asynchronous output in \Cref{ex:non-det} then $p_1 = p'_1$.) 
Moreover, deterministic behaviour is \emph{not} necessarily required to runtime-verify all the behaviours specified.

\begin{example}\label{ex:non-det-det} 
  Consider the property that, in addition to the behaviour described by $\varphi_2$, 
  it requires that 
  \textsl{``\ldots  the SUS does \emph{not} exhibit any action after a close event''}, formalised as  $\varphi_3$. 
  \begin{center}
    $\varphi_3 \deftxtS  \Conj{\bigl(\Um{r}{(\Disj{\Um{s}{\F}}{\Um{a}{\F}})}\bigr)\ }{\ \bigl(\Um{c}{(\Um{r}{\F} \wedge \Um{s}{\F} \wedge \Um{a}{\F} \wedge \Um{c}{\F} )}\bigr)}$
  \end{center} 
  It might be reasonable to assume that a SUS behaves deterministically for receive actions (\eg when a single thread is in charge of receiving). 
  %
  Moreover, \emph{no} determinism assumption is required for close actions  to runtime verify the subformula \Um{c}{(\Um{r}{\F} \wedge \Um{s}{\F} \wedge \Um{a}{\F} \wedge \Um{c}{\F})};  
  %
  any trace from either $cr\cdots, cs\cdots$, $ca\cdots$ or $cc\cdots$ suffices to 
  infer the violation of 
  $\varphi_3$. \exqed   
\end{example}

%
The properties discussed in this paper
are formalised in terms of a variant of the modal $\mu$-calculus~\cite{DBLP:journals/tcs/Kozen83} called Hennessy-Milner Logic with Recursion~\cite{LARSEN1990265}, \recHML.
%
This logic is a natural choice for describing branching-time properties and is employed by state-of-the-art model checkers, including \textsf{mCRL2}~\cite{DBLP:conf/tacas/CranenGKSVWW13} and \textsf{UPPAAL}~\cite{Behrmann2004}, as well as \textsf{detectEr} ~\cite{DBLP:conf/forte/AttardAAFIL21,DBLP:journals/scp/AcetoAAEFI24}, a stable RV tool.
%
It 
has been shown to embed 
standard logics such as LTL, CTL and CTL*~\cite{Baier2008Book,Clarke1999Book,AcetoAFIL21:sosym}. 
%
%
%
Moreover, existing maximality results for branching-time logics~\cite{FrancalanzaAI17:monitorability-branching,AceAFI:18:FOSSACS,DBLP:conf/csl/AcetoAFIL21}
have only been established for \recHML.
%
Our exposition focusses on ``safety'' properties that can be monitored for violations; monitoring for satisfactions of branching-time properties is symmetric~\cite{FrancalanzaAI17:monitorability-branching}.
%
%
This paper presents an augmented monitoring setup that repeatedly analyses a (potentially non-deterministic) SUS across multiple executions, so as to study how the monitorability limits established in~\cite{AcetoAFIL19:monitorability-linear,FrancalanzaAI17:monitorability-branching} are affected.  
Our contributions are: 


%
\begin{enumerate}
  \item A formalisation of a monitoring setup that gathers information over multiple 
  system runs
   (\Cref{sec:mon-setup}).
  \item An analysis, formalised as a proof system, that uses sets of partial traces to runtime verify the system against a branching-time property (\Cref{sec:mon-setup}). 
  %
  \item A definition formalising what it means for a monitor to correctly analyse a property over multiple runs (\Cref{sec:mon-correct}) and, dually, what it means for a property to be monitorable over multiple runs (\Cref{sec:monitorability}).
  \item The identification of an \emph{extended} logical fragment that is monitorable over the augmented monitoring setup handling multiple runs (\Cref{sec:monitorability}),
  %
  %
  and the establishment that the extended fragment is maximally expressive (\Cref{sec:monitorability}).
  %
  \item An instantiation of the 
  multi-run RV framework to  actor-based 
  systems (\Cref{sec:actor-systems}), a 
  popular concurrency paradigm. 
  %
  %
  \item A method for systematically determining 
  the number of SUS executions required to conduct RV
  from the syntactic structure of the formula being verified (\Cref{sec:lowerbounds}).
\end{enumerate}


\section{Preliminaries}
\label{sec:preliminaries}

We 
assume a set of actions, $\stact,\stactt {\in} \Act {=} \TAct {\uplus} \{\tau\}$, with a distinguished \emph{silent (untraceable) action} $\tau$ and a set of \emph{traceable actions}, $\tact,\tactt {\in} \TAct {=} \EAct {\uplus} \IAct$, that consists of two disjoint sets. 
\emph{External actions}, $\act,\actt{\in}\EAct$, describe 
computation steps observable to an outside entity which are the subject of correctness specifications.  
\emph{Internal actions}, $\iact,\iactt\!\in\!\IAct$, are not of concern to correctness specifications but 
can still be discerned by a monitor with the appropriate instrumentation mechanism.
Notably, silent actions cannot be traced.    

A SUS is modelled as an \emph{Instrumentable Labelled Transition System} (ILTS), a septuple of the form 
$$\langle \Prc, \steq, \EAct, \IAct, \{\tau\}, \traS{\;}, \Det\rangle$$
SUS states are denoted by 
processes, $\pV,\pVV\in\Prc$, with an associated equivalence relation, $\steq\, \subseteq \Prc \times \Prc$. 
The 
transition relation, $\traS{\quad}\,\subseteq (\Prc \times \Act \times \Prc)$, is defined over arbitrary actions (\ie silent, internal and external). 
We write $\pV \traS{\stact} \pVV$ instead of $(\pV,\stact,\pVV) \in \traS{\quad}$, and $\pV \traSN{\stact}$ whenever $\nexists \pVV$ such that $\pV \traS{\stact} \pVV$.
ILTS transitions abstract over equivalent states: 
\begin{center}
  for any $\pV\steq \pVV$, 
  if $\pV\traS{\stact}\pV'$ then there exists $\pVV'$ such that $\pVV \traS{\stact} \pVV'$ where $\pV' \steq \pVV'$.
\end{center}
Instrumentation also can abstract over 
(\emph{non}-traceable) silent transitions  
because they 
are \emph{confluent} \wrt other actions: 
\begin{center}
  for any \pV, whenever $\pV\traS{\tau}\pV'$ and $\pV \traS{\stact} \pV''$ then, \\
  either $\stact = \tau$ and $\pV' \steq \pVV'$, 
  or there exists a state \pVV and transitions $\pV'\traS{\stact}\pVV$ and $\pV''\traS{\tau}\pVV$ joining the diamond.      
\end{center}
An ILTS partitions traceable actions via the predicate $\Det:\TAct \rightarrow \Bool$ where all actions $\tact$ satisfying the predicate, $\Det(\tact) = \btrue$, must be \emph{deterministic}: 
\begin{center}
  if $\pV \traS{\tact} \pV'$ and $\pV \traS{\tact} \pV''$ then $\pV' \steq \pV''$. 
\end{center}
%
\emph{Weak transitions}, $\pV\wtraS{\;\;}\pVV$, abstract over both silent and internal actions whereas \emph{weak traceable transition}, $\pV\wttraS{\;\;}\pVV$, abstract over silent actions only. 
Thus, $\pV\wtraS{\;\;}\pVV$  holds when  $\pV = \pVV$  or $\exists \pV'$ and $\stact \in \bigl(\{\tau\}\cup\IAct\bigr)$  such that $ \pV\traS{\stact}\pV'\wtraS{\;\;}\pVV$. 
Analogously, $\pV\wttraS{\;\;}\pVV$  holds if  $\pV = \pVV$  or $\exists \pV'$   such that $\pV\traS{\tau}\pV'\wttraS{\;\;}\pVV$.
%
We write $\pV\wtraS{\act}\pVV$ when  $\exists \pV',\pV''$ such that $\pV \wtraS{\;\;} \pV' \traS{\act} \pV'' \wtraS{\;\;} \pVV$, and write $\pV\wttraS{\tact}\pVV$ when  $\exists \pV',\pV''$ such that $\pV \wttraS{\;\;} \pV' \traS{\tact} \pV'' \wttraS{\;\;} \pVV$.
Actions can be sequenced to form \emph{traces}, $t,u\in\Trc=\TAct^*$, representing prefixes of system runs.
A trace with action $\tact$ at its head and continuation $t$ is denoted as $\tact t$, whereas a trace with prefix $t$ and action $\tact$ at its end is denoted as $t \tact$.
For $t=\tact_1\cdots\tact_n$, we write $\pV\wttraS{t}\pVV$ instead of the sequence of transitions $\pV \wttraS{\tact_1}\cdots \wttraS{\tact_n}\pVV$.
A system (state) \pV\ \emph{produces} a trace $t$ when $\exists \pVV$ such that $\pV \wttraS{t} \pVV$. 
The set of all the traces produced by the state \pV\ is denoted by $\hstpV$.
\emph{Histories} $\hst\!\in\!\Hst$ where $\Hst\!\subseteq\!\Trc$ are finite sets of traces
%
where $\hst,t$ is shorthand for the disjoint union $\hst \uplus \{ t\}$.

\begin{remark}
  An ILTS provides two (global) views of a 
  SUS: an external one, as viewed by an observer limited to \EAct, and a lower-level view as seen by an instrumented monitor privy to \TAct 
   and \Det. 
  %
  The SUS treatment is still  considered \emph{black-box} since, for 
  any \TAct 
   and \Det, a monitor can \emph{at best} reason about states within the same equivalence class, 
   not 
  specific states.
  Deterministic systems can be modelled by requiring $\detAct{\tact} {=} \btrue$ for all actions, whereas for general systems, we 
  have $\detAct{\tact} {=} \bfalse$. 
  %
  Silent actions capture $\beta$-moves
  \cite{YoshidaHB07,DBLP:books/daglib/0018113}
  and arise naturally 
  as thread-local moves.
  %
  \exqed     
\end{remark}

Properties are formulated for the external SUS view 
in terms of \recHML formulae. 
This logic is 
defined by the negation free grammar in \Cref{fig:syntax-semantics}, which assumes a countably infinite set of formula variables $\XV,\XVV,\ldots{\in}\TVars$.
Apart from the standard constructs for truth, falsity, conjunction and disjunction, the logic includes existential and universal modalities that operate over the 
\emph{external} actions \EAct.
Least and greatest fixed points, \Min{\XV}{\varphi} and \Max{\XV}{\varphi} respectively, bind free instances of variable \XV\ in $\varphi$.
We assume standard definitions for open and closed formulae and work up to $\alpha$-conversion, assuming formulae to be closed and guarded, unless otherwise stated.
%
%
%
For formulae $\varphi$ and $\psi$, and variable \XV,  $\varphi\subS{\psi}{\XV}$ denotes the substitution of all free occurrences of \XV\ in $\varphi$ with $\psi$.

The denotational semantics function \evalE{-} in \Cref{fig:syntax-semantics} 
maps formulae to sets of system states, $\evalE{-}: \recHML \map \mathcal{P}(\Prc)$.
This function is defined with respect to an \emph{environment} $\rho$, which maps formula variables to sets of states, $\rho : \TVars \map \mathcal{P}(\Prc)$.
Given 
a set of states $P$, 
$\rho[\XV \mapsto P]$ 
denotes the environment mapping \XV\ to $P$, mapping as $\rho$ on all other variables.
%
%
%
%
Existential modalities \Em{\act}{\varphi} denote the set of system states that can perform \emph{at least one} $\act$-labelled 
(weak) 
transition and reach a state that satisfies the continuation $\varphi$.  
Conversely, universal modalities \Um{\act}{\varphi} denote the set of systems that reach states satisfying $\varphi$ for \emph{all} (possibly none) their 
$\act$-transitions.
%
%
The set of systems that satisfy least fixed point formulae (\resp greatest fixed point) is given by the intersection (\resp union) of all pre-fixed points (\resp post-fixed points) of the function induced by the corresponding binding formula.
The remaining cases are standard.   
The interpretation of closed formulae is independent of $\rho$; we write \evalE{\varphi} in lieu of \evalE{\varphi, \rho}. 
%
A state $\pV$ \emph{satisfies} $\varphi$ if $\pV\!\in\!\evalE{\varphi}$ and \emph{violates} it if $\pV\!\notin\!\evalE{\varphi}$; 
equivalent states satisfy (\resp\ violate) the same formulae, \Cref{prop:behaviour-equiv}.
Two formulae $\varphi$ and $\psi$ are equivalent, 
$\varphi\equiv\psi$, whenever $\evalE{\varphi}=\evalE{\psi}$.
The negation of a formula can be obtained by duality in the usual way.

\begin{restatable}[Behavioural Equivalence]{proposition}{behaviourEquiv} 
  \label{prop:behaviour-equiv}
  For all (closed) formulae $\varphi\!\in\!\recHML$, if $\pV\!\in\!\evalE{\varphi}$ and $\pV\!\steq\! \pVV$ then $\pVV\!\in\!\evalE{\varphi}$.
  \qed 
\end{restatable}

\begin{figure}[t!]
  \setlength{\abovedisplayskip}{0pt}
  \setlength{\abovecaptionskip}{0pt}
  \textbf{recHML Syntax}
  \smallskip
  \begin{align*}
    & && \hspace{-4em} \varphi,\psi \in \recHML \bnfdef 
    && \quad \XV && \text{(rec. variable)} \\
    & \bnfsepp \T                            && \text{(truth)} 
    && \bnfsepp \Em{\act}{\varphi}           && \text{(existential modality)} \\
    & \bnfsepp \F                            && \text{(falsehood)} 
    && \bnfsepp \Um{\act}{\varphi}           && \text{(universal modality)} \\
    & \bnfsepp \Conj{\varphi}{\psi} \!\!\!\! && \text{(conjunction)} 
    && \bnfsepp \Min{\XV}{\varphi} \!\!\!\!  && \text{(least fixed point)} \\
    & \bnfsepp \Disj{\varphi}{\psi} \!\!\!\! && \text{(disjunction)}  
    && \bnfsepp \Max{\XV}{\varphi} \!\!\!\!  && \text{(greatest fixed point)} 
  \end{align*}  
  \textbf{Branching-Time Semantics} 
  \smallskip
  \begin{align*}
    &\evalE{\T,\rho} \deftxt \Prc
    \hspace{7.5em} \evalE{\F,\rho} \deftxt \emptyset\\  
    & \evalE{\Disj{\varphi}{\psi},\rho} \deftxt \evalE{\varphi,\rho} \cup \evalE{\psi,\rho} 
    \hspace{1.5em} \evalE{\Conj{\varphi}{\psi},\rho} \deftxt \evalE{\varphi,\rho} \cap \evalE{\psi,\rho} \\
    &\evalE{\Um{\act}{\varphi},\rho} \deftxt \big\{ \pV \;|\; \forall \pVV \cdot \pV \wtra{\act} \pVV \text{ implies } \pVV \in \evalE{\varphi, \rho} \big\} \\
    & \evalE{\Em{\act}{\varphi},\rho} \deftxt \big\{ \pV \;|\; \exists \pVV \cdot \pV \wtra{\act} \pVV \text{ and } \pVV \in \evalE{\varphi, \rho} \big\} \\
    &\evalE{\Min{\XV}{\varphi},\rho} \deftxt \bigcap \big\{ P \;|\; \evalE{\varphi, \rho[\XV \mapsto P]} \subseteq P \big\} 
    \hspace{1em}\evalE{\XV, \rho} \deftxt \rho(\XV) \\
    &\evalE{\Max{\XV}{\varphi},\rho} \deftxt \bigcup \big\{ P \;|\; P \subseteq \evalE{\varphi, \rho[\XV \mapsto P]} \big\} 
  \end{align*}
  %
  \vspace{-1em}
  \caption{\recHML in the Branching-Time Setting.}
  \vspace{-1em}
  \label{fig:syntax-semantics}
\end{figure}

Several logical formulae from \Cref{fig:syntax-semantics} are not monitorable \wrt classical RV
limited to one  (partial) execution of the system. 
The safety subset of monitorable \recHML formulae is characterised by the syntactic fragment \sHML~\cite{DBLP:journals/dc/AlpernS87}.

\begin{theorem}[Monitorability~\cite{FrancalanzaAI17:monitorability-branching}]\label{thm:shml}
  Any $\varphi{\in}\recHML$ is monitorable (for violations) iff there exists $\psi \in \sHML$ and $\varphi \equiv \psi$: 
  \begin{align*}
    \varphi,\psi \in \sHML 
    & \bnfdef \T                      
      \bnfsepp \F 
      \bnfsepp \Um{\act}{\varphi}          
      \bnfsepp \Conj{\varphi}{\psi}
      \bnfsepp \Max{\XV}{\varphi} 
      \bnfsepp \XV
      \tag*{\exqed}
  \end{align*}
\end{theorem}

\begin{example}\label{ex:rechml}
  %
  The property \textsl{``after any number of serviced queries, \Um{r}{\Um{s}{\ldots}}, a state that can close a connection, \Em{c}{\T}, \emph{cannot} 
  allocate memory, \Um{a}{\F}''} 
  %
  is \emph{not} monitorable.
  \begin{align*}
    \varphi_4\ &\deftxtS\ \Max{\XV}{\bigl(\Um{r}{\Um{s}{\XV}} \wedge (\Em{c}{\T} \Rightarrow \Um{a}{\F})\bigr)} \\
    &\qquad\;\, \equiv\;\; \Max{\XV}{\bigl(\Um{r}{\Um{s}{\XV}} \wedge (\Um{c}{\F} \vee \Um{a}{\F})\bigr)}
  \end{align*}
  %
  %
  Specifically, a system violates $\varphi_4$ if it 
  is capable of producing \emph{both} actions $a$ and $c$ after an unbounded, but \emph{finite}, sequence of alternating $r$ and $s$ actions.
  %
  \Eg the system  
  $\pV_1 \deftxtS \recX{\bigl(r.s.X + (a.X + c.\nl)\bigr)}$ 
  (see \Cref{def:ccs} for CCS syntax)
  violates this property
  since after zero or more serviced queries, $\pV_1$ reaches a state 
  that can produce both  
  $a$ and $c$. 
  %
  However, no single trace prefix provides enough evidence to detect this. 
  %
  %
  \exqed
\end{example}


\begin{figure*}[t!]
  \small
  \setlength{\abovedisplayskip}{0pt}
  \setlength{\belowcaptionskip}{-10pt}
  \textbf{Monitor Syntax} 
  \vspace{-1.2em} 
  \begin{align*}
    \mV,\mVV \in \Mon ::= 
    \no \;\;
    \bnfsepp \stp \;\;
    \bnfsepp \act.\mV \;\;
    \bnfsepp \recX{\mV} \;\;
    \bnfsepp \XV \;\;
    \bnfsepp \mV\oplus\mVV \;\; 
    \bnfsepp \mV\otimes\mVV \;\; 
    \hspace*{2em} (\odot\!\in\!\{ \oplus, \otimes \}) 
  \end{align*}
  \textbf{Monitor Semantics}
  \smallskip
  \begin{mathpar}
    \inferrule[\rtitSS{mEnd}]
      { }
      { (t, \stp)  \traS{ \act }_\hst (t\act, \stp) }
    \quad
    \inferrule[\rtitSS{mVrP1L}]
      { t\in\hst }
      { (t, \no \odot \mVV)  \traS{ \tau }_\hst (t, \mVV) }
    \quad
    \inferrule[\rtitSS{mVrP2L}]
      { t\notin\hst }
      { (t, \no \odot \mVV)  \traS{ \tau }_\hst (t, \no) }
    \quad
    \inferrule[\rtitSS{mAct}]
      { }
      { (t, \act.\mV)  \traS{ \act }_\hst (t\act, \mV) }
    \quad
    \inferrule[\rtitSS{mRec}]
      {  }
      { (t, \recX{\mV})  \traS{ \tau }_\hst (t, \mV\subS{\recX{\mV}}{\XV}) }
    \and
    \inferrule[\rtitSS{mTauL}]
      { (t, \mV) \traS{\tau}_\hst (t, \mV') }
      { (t, \mV\odot\mVV)  \traS{ \tau }_\hst (t, \mV'\odot\mVV) }
    \and 
    \inferrule[\rtitSS{mPar1}]
      { (t, \mV) \traS{\act}_\hst (t', \mV') \quad (t, \mVV) \traS{\act}_\hst (t', \mVV') }
      { (t, \mV\odot\mVV)  \traS{\act}_\hst (t', \mV'\odot\mVV') }
    \and 
    \inferrule[\rtitSS{mPar2L}]
      { \mVV\neq\no \quad (t, \mV) \traS{\act}_\hst (t', \mV') \quad (t, \mVV) \traSN{\act}_\hst \quad (t, \mVV) \traSN{\tau}_\hst }
      { (t, \mV\odot\mVV)  \traS{\act}_\hst (t', \mV') }
  \end{mathpar}
  \textbf{Instrumentation Semantics}
  \smallskip
  \begin{mathpar}
    \inferrule[\rtitSS{iNo}]
      { }
      {\sysH{\pV}{(t,\no)}{\hst} \traS{\tau} \sysH{\pV}{(t, \stp)}{\hst,t}}
    \and
    \inferrule[\rtitSS{iTer}]
      {\mV\neq\no \quad \pV \traS{\act} \pV' \quad (t, \mV) \traSN{\act}_\hst \quad (t, \mV) \traSN{\tau}_\hst }
      {\sysH{\pV}{(t,\mV)}{\hst} \traS{\act} \sysH{\pV'}{(t\act, \stp)}{\hst} }
    \and
    \inferrule[\rtitSS{iAsS}]
      {\mV\neq\no \quad \pV \traS{\tau} \pV'}
      {\sysH{\pV}{(t, \mV)}{\hst} \tra{\tau} \sysH{\pV'}{(t, \mV)}{\hst}}
    \and 
    \inferrule[\rtitSS{iAsI}]
      {\mV\neq\no \quad \pV \traS{\iact} \pV'}
      {\sysH{\pV}{(t, \mV)}{\hst} \traS{\tau} \sysH{\pV'}{(t\iact, \mV)}{\hst}}
    \and
    \inferrule[\rtitSS{iAsM}]
      {(t,\mV) \traS{\tau}_\hst (t',\mV')}
      {\sysH{\pV}{(t,\mV)}{\hst} \traS{\tau} \sysH{\pV}{(t'\!\!,\mV')}{\hst}}
    \and 
    \inferrule[\rtitSS{iMon}]
      {\pV \traS{\act} \pV' \quad (t, \mV) \traS{\act}_\hst (t', \mV') }
      {\sysH{\pV}{(t,\mV)}{\hst} \traS{\act} \sysH{\pV'}{(t'\!\!,\mV')}{\hst}}
  \end{mathpar}
  \vspace{-1em}
  \caption{Monitors and Instrumentation}
  \vspace{-1em}
  \label{fig:mon}
\end{figure*}

\section{A Framework for Repeated Monitoring}
\label{sec:mon-setup}

Instrumentation permits the monitor to observe the current execution of the SUS until it detects certain behaviour. 
We formalise an \emph{extended} online setup, where monitoring is performed in two steps: history aggregation and history analysis. 
%
During \emph{aggregation}, monitors gather SUS information over \emph{multiple} executions. 
%
Each time a new trace is added to the history, the \emph{analysis} step uses a proof system 
to determine whether the SUS generating such a history is rejected.
If it fails to reject that history, these two steps are repeated until a 
verdict is reached.
SUS instrumentation sits at a lower level of abstraction to the external view used by \recHML which allows monitors to operate with action sequences from \TAct.  

\subsection{History Aggregation}


\vspace{0.5em}
\noindent
\textbf{Monitors.}
Our runtime analysis, 
defined in~\Cref{fig:mon}, \emph{records} the 
traceable actions, \TAct, 
that lead to rejection states.
An \emph{executing-monitor} state 
consists of a tuple $(t,\mV)$,
where $t$ is the trace (\ie sequence of traceable actions) 
collected from the beginning of the run, up to the current execution point, and \mV\ is the current state of the monitor after analysing it.
%
In order to streamline monitor synthesis from formulae (which only mention external actions) the monitor syntax does not reference internal actions, \eg $\act.\mV$ in \Cref{fig:mon}. 
Accordingly, its monitor semantics determines which \emph{external} actions to record, rules \textsc{iMon} and \textsc{mAct}. Internal actions, used to improve the precision of the history analysis, are recorded by the instrumentation semantics, rule \textsc{iAsI}, discussed later.
%

Executing-monitor transitions are defined \wrt a history \hst\ that stores the trace prefixes accumulated in prior executions: 
%
%
$(t,\mV)\traS{\stact}_\hst(t',\mV')$ 
denotes that 
%
$(t,\mV)$ transitions to $(t',\mV')$ either by observing an external action $\act$ produced by the SUS, or by evolving autonomously via the silent action $\tau$. 
A monitor execution can reach one of two final states: a \emph{rejection} verdict, \no, or an \emph{inconclusive} state, \stp.  
The 
 latter
behaves like an identity, transitioning to itself when analysing any external SUS action; see rule \textsc{mEnd}. 
%
Differently, a rejection state indicates to the instrumentation that the (partial) trace analysed thus far should be aggregated to the history. 
After aggregating the trace, it then behaves as \stp; see instrumentation rule \textsc{iNo}, 
discussed later.
%
Rule \textsc{iNo} is the only rule that extends the history to $H,t$.  

The current recorded trace is accrued via monitor sequencing, $\act.\mV$, via rule \textsc{mAct}.
Besides sequencing, \text{(sub-)}monitors can be composed together as a parallel \emph{conjunction},  $\mV\otimes\mVV$, or 
\emph{disjunction}, $\mV\oplus\mVV$.
When analysing SUS actions, parallel monitors, $\mV\!\odot\!\mVV$ where $\odot\!\in\!\{ \oplus, \otimes \}$,  move either autonomously, 
rule \textsc{mTauL}, 
or in unison, rule \textsc{mPar1}.
When a sub-monitor cannot analyse the action proffered by the SUS 
 it is discarded (rule \textsc{mPar2L}); 
%
%
this does not prohibit the former monitor from potentially recording a new trace.
An analogous mechanism is also implemented by the instrumentation rule \textsc{iTer}. 
Four rules determine how a rejection verdict \emph{sub-monitor} is handled. 
%
%
Rule \textsc{mVrP2L} asserts that verdict \no\ 
supersedes its parallel counterpart whenever the accumulated (violating) trace is new, \ie $t \notin \hst$; 
when $\no \odot \mVV$ transitions to \no, it allows the instrumentation rule \textsc{iNo} to add $t$ to the history. 
Dually, if 
$t\!\in\!\hst$, the rejection verdict is discarded, \ie $\no \odot \mVV$ transitions to \mVV, to allow \mVV\  
to potentially collect violating traces with common prefixes, 
rule \textsc{mVrP1L}.
%
The remaining monitor rules are standard and symmetric rules are elided.
Although trace collection does \emph{not} distinguish between parallel conjunction and disjunctions, history analysis does; 
see \Cref{fig:proof-system}.


\vspace{0.5em}
\noindent
\textbf{Instrumentation.}
The 
behaviour of an executing-monitor is connected to that of a 
SUS via the instrumentation 
relation 
in \Cref{fig:mon}. 
It is defined over \emph{monitored systems}, \sysH{\pV}{(t,\mV)}{\hst}, triples consisting of a SUS 
\pV,
an executing-monitor $(t,\mV)$, and a history \hst.  
The transition $\sysH{\pV}{(t,\mV)}{\hst} \traS{\stact} \sysH{\pV'}{(t',\mV')}{\hst'}$ denotes that the executing-monitor $(t,\mV)$ transits to $(t',\mV')$ when analysing a SUS evolving from $\pV$ to $\pV'$ via action $\stact$, while updating the history from $\hst$ to $\hst'$. 
%
Rule \textsc{iMon} formalises the analysis of an external action, whereas rule \textsc{iNo},  previewed earlier, handles the storing of new traces that lead to a rejection verdict.  
Instrumentation also allows the SUS and executing-monitor to (internally) transition independently of one another,
rules \textsc{iAsS} and \textsc{iAsM}.
Rule \textsc{iAsI} allows the SUS to transition with an internal action:  $\iact$ is recorded as part of the aggregated trace while concealing it as a $\tau$ action.
%
%
When $(t,\mV)$ can neither analyse a SUS action, nor perform an internal transition, the instrumentation forces it to terminate prematurely by transitioning to the inconclusive verdict (rule \textsc{iTer}). 
This ensures instrumentation transparency~\cite{DBLP:journals/iandc/Francalanza21,Fra:Concur:17}, where the monitoring infrastructure does not block the behaviour of the SUS whenever the executing monitor cannot analyse an event.
We adopt a similar convention to 
\Cref{sec:preliminaries}; \eg we define weak transitions in a similar manner and write 
$\sysH{\pV}{(t,\mV)}{\hst} \!\wtraS{u}\! \sysH{\pV'}{(t',\mV')}{\hstt}$ in lieu of $\sysH{\pV}{(t,\mV)}{\hst} \!\wtraS{\act_1}\! \!\cdots\! \wtraS{\act_n} \sysH{\pV'}{(t',\mV')}{\hstt}$ for $u\!=\!\act_1\!\cdots\!\act_n$. 

  Our monitor 
  semantics departs from 
  prior work~\cite{FrancalanzaAI17:monitorability-branching,AcetoAFIL19:monitorability-linear};
  it does \emph{not} flag violations but limits itself to aggregating traces. 
  Every 
  monitored execution starts with $t=\epsilon$ and
  can, at most, increase the history the current trace accrued. 
  %
  %
%
%
Our monitors 
work over multiple runs of the \emph{same} SUS.
Starting from an empty history $\hst_0 {=} \emptyset$, traces leading to \no\ states, can be accumulated 
over a sequence of monitored SUS executions by passing 
history $\hst_i$ obtained from the $i^\textsl{th}$ monitored execution on to 
execution $i{+}1$, inducing a (finite) totally-ordered sequence of histories, $\emptyset {=} \hst_0 {\subseteq} \hst_1 {\subseteq} \cdots\!$

\begin{example}\label{ex:inst}
  Monitor 
  $\mV_1 \deftxt \recX{\bigl(r.s.\XV \otimes (a.\no \oplus c.\no)\bigr)}$
  reaches state \no\
  after observing actions $a$ or $c$, following a sequence of serviced queries.
  %
  System $\pV_2 \deftxt \recX{\bigl(r.s.X + (\ut.a.X + \uf.c.\nl)\bigr)}$
  extends $\pV_1$ from \Cref{ex:rechml}, where the decision on whether to allocate memory or close depends on checking whether there is free memory or not, expressed as the internal actions \ut\ and \uf\ respectively. 
  When $\pV_2$ is instrumented with 
  the executing-monitor $(\epsilon,\mV_1)$ and history $\hst_0 = \emptyset$, it can reach 
  state \no\ through the prefix $t_1 = rs \ut a$ as shown below.
  %
  With the augmented history $\hst_1 = \{t_1\}$, 
  \sysH{\pV_2}{(\epsilon,\mV_1)}{\hst_1} can then aggregate $t_2 =rs \uf c$ in a subsequent run, \ie 
  $\sysH{\pV_2}{(\epsilon,\mV_1)}{\hst_1} 
  \wtraS{t_2} \sysH{\pV_2}{(t_2, \stp)}{\hst_2}$
  where $\hst_2 = \{ t_1,t_2 \}$. 
  %
    \setlength{\jot}{0ex}
    \begin{align*}
      & \sysH{\pV_2}{(\epsilon,\mV_1)}{\hst_0} \;
      \traS{\tau} \cdot \traS{\tau}\; 
      \tag*{\scriptsize (\textsc{iAsS},\textsc{iAsM})} \\ 
      & \quad { \sysH{r.s.\pV_2 + (\ut.a.\pV_2 + \uf.c.\nl)}{ (\epsilon, r.s.\mV_1 \otimes (a.\no \oplus c.\no)) }{\hst_0}}
      \\
      & \traS{r}\; \sysH{s.\pV_2}{(r, s.\mV_1)}{\hst_0}  
      \traS{s} \sysH{\pV_2}{(rs, \mV_1)}{\hst_0} 
      \tag*{\scriptsize (\textsc{iMon})} \\ 
      &\traS{\tau} \cdot \traS{\tau} \;
      \tag*{\scriptsize (\textsc{iAsP},\;\textsc{iAsM})} \\
      &\quad \sysH{r.s.\pV_2 + (\ut.a.\pV_2 + \uf.c.\nl)}{(rs, r.s.\mV_1 \otimes (a.\no \oplus c.\no))}{\hst_0} 
      \\
      %
        & \traS{\tau}\; \sysH{a.\pV_2}{(rs\ut, r.s.\mV_1 \otimes (a.\no \oplus c.\no))}{\hst_0} 
        \tag*{\scriptsize (\textsc{iAsI})} 
        \\
      & \traS{a}\; \sysH{\pV_2}{(rs\ut a, \no)}{\hst_0} 
      \tag*{\scriptsize (\textsc{iMon})} \\
      & \traS{\tau}\; \sysH{\pV_2}{(rs\ut a, \stp)}{\hst_0\cup\{rs\ut a\}}
      \tag*{\scriptsize (\textsc{iNo})} 
    \end{align*}
  %
  %
  %
  %
  Note that, since monitors assume a passive role~\cite{DBLP:journals/iandc/Francalanza21}, they cannot steer the behaviour of the SUS, 
  meaning 
  the SUS may \emph{not} exhibit \emph{different} behaviour 
  across multiple executions. 
  \qed
\end{example}


The instrumentation mechanism needs to 
aggregate overlapping trace prefixes 
that lead to rejection states.

\begin{example}\label{ex:inst-comp}
  The SUS $\pV_2$ from \Cref{ex:inst} generates 
  traces of the form $(rs\ut a)^\ast$.
  Monitor 
  $\mV_2 \!\deftxtS\! \recX{\bigl(r.s.\XV \otimes a.X \otimes (a.\no \oplus c.\no)\bigr)}$ 
  revises $\mV_1$ 
  where sequences of $rs$ actions can be interleaved with 
  finite sequences of $a$ actions described by  the sub-monitor $a.X$.
  %
  %
  When $(\epsilon, \mV_2)$ is instrumented on $\pV_2$ with $\hst_0=\emptyset$, it can record the prefix $rs\ut a$ during a first run.
  %
  In a subsequent run with 
  an augmented 
  $\hst_1 {=} \{ rs\ut a \}$, we have:
  %
    \begin{align*}
      &\sysH{\pV_2}{(\epsilon, \mV_2)}{\hst_1} \\
      &\wtraS{rs\ut} 
      \sysH{a.\pV_2}{(rs\ut, r.s.\mV_2 \otimes a.\mV_2 \otimes (a.\no \oplus c.\no))}{\hst_1} \\[-0.1em]
      \label{eq:overlap1} 
      & \traS{a} \sysH{\pV_2}{(rs\ut a, \mV_2 \otimes \no)}{\hst_1} \traS{\tau} \sysH{\pV_2}{(rs\ut a, \mV_2)}{\hst_1} \tag{$*$} 
      \\[-0.1em]
      & \wtraS{rs\ut a} \sysH{\pV_2}{(rs\ut ars\ut a, \mV_2\otimes \no)}{\hst_1} 
      \\    
      &\traS{\tau} \sysH{\pV_2}{(rs\ut ars\ut a, \no)}{\hst_1} 
      \\[-0.1em]
      %
      \label{eq:overlap2}
      &\traS{\tau} \sysH{\pV_2}{(rs\ut ars\ut a, \stp)}{\hst_1 \cup \{rs\ut ars\ut a\}}  \tag{$\dagger$}
    \end{align*}
  %
  Transition \eqref{eq:overlap1} follows rule \textsc{mVrP1R} with $(rs\ut a, \mV_2 \otimes \no) \traS{\tau}_{{\hst_1}} (rs\ut a, \mV_2)$ since $rs\ut a\in\hst_1$:
 the executing-monitor does \emph{not} stop accruing at $rs\ut a$ but continues monitoring until it encounters a \emph{new} rejecting trace, $rs\ut ars\ut a$, which is aggregated to $\hst_1$ in transition \eqref{eq:overlap2} using rule \textsc{iNo}. 
  \exqed
\end{example}

\begin{remark}
  Rule \textsc{iNo} encodes the design decision to stop monitoring (by transitioning to 
  $\stp$) as soon as a new trace is aggregated to the history, 
  providing a clear cut-off point for when to pass the aggregated history to the subsequent run. 
  %
  %
  \exqed  
\end{remark}


\subsection{History Analysis}

\begin{figure}[t!] 
  \small
  \setlength{\abovedisplayskip}{1pt}
  \setlength{\belowcaptionskip}{-10pt}
  \begin{mathpar}
    \inferrule[\rtitSS{no}]
      { \hst \neq \emptyset }
      { \rej{\hst, \odF, \no} }
    \and
    \inferrule[\rtitSS{act}]
    { 
      \hstt{=}\sub{\hst,\act} \quad 
      \rej{\hstt, \bigl(\odF {\wedge} \detAct{\act}\bigr), \mV } 
    }
    { \rej{\hst, \odF, \act.\mV} }
    \and
    \inferrule[\rtitSS{actI}]
      { 
        \hstt=\sub{\hst,\iact} \quad 
        \rej{ \hstt, \bigl(\odF {\wedge} \detAct{\iact}\bigr), \act.\mV } 
      }
      { \rej{ \hst, \odF, \act.\mV } }
    \quad
    \inferrule[\rtitSS{parAL}]
      { \rej{\hst, \odF, \mV} }
      { \rej{\hst, \odF, \mV \otimes \mVV} }
    \and
    \inferrule[\rtitSS{parO}]
      { \rej{\hst, \btrue, \mV} \quad \rej{\hst, \btrue, \mVV} }
      { \rej{\hst, \btrue, \mV \oplus \mVV } }
    \and
    \inferrule[\rtitSS{rec}]
      { \rej{ \hst, \odF, \mV\subS{\recX{\mV}}{\XV} } }
      { \rej{ \hst, \odF, \recX{\mV} } }
  \end{mathpar}
  \vspace{-1em}
  \caption{Proof System}
  \vspace{-1em}
  \label{fig:proof-system}
\end{figure}

We formalise how a history is rejected by a monitor through a proof system. 
Its main judgement is $\rej{\hst, \odF, \mV}$, \ie monitor \mV\ \emph{rejects} history \hst\ using \Det with the boolean flag~\odF.   
It uses internal actions and \Det to calculate whether the traces are produced by the same states (up to $\steq$); the flag value $\btrue$ encodes that all the actions analysed up to this point were deterministic actions. 
This analysis is the least relation defined by the rules in \Cref{fig:proof-system}, 
relying on a helper function $\sub{\hst, \tact} = \{ t \; | \; \tact t{\in} \hst \}$; it returns the continuation of any trace
in $\hst$ that is prefixed by a \tact action; \eg when $\hst = \{rsa, rsc, ars\}$, 
we get $\sub{\hst, r} = \{ sa, sc \}$.
The axiom \textsc{no} states that a \no\ monitor rejects all \emph{non-empty} histories,  
%
\ie a monitor cannot reject a SUS outright, without any observation.
%
%
In rule \textsc{act}, a sequenced monitor $\act.\mV$ rejects \hst\ with flag \odF\ if the (sub-)monitor \mV\ rejects the history returned by \sub{\hst,\act} with updated flag $\bigl(\odF {\wedge} \detAct{\act}\bigr)$.
Alternatively, $\act.\mV$ can reject $\hst$ with \odF\ following rule \textsc{actI}, by considering the suffixes of traces prefixed by an internal action $\iact$, again updating the flag to $\bigl(\odF {\wedge} \detAct{\iact}\bigr)$. 
%
%
Parallel conjunctions $\mV \otimes \mVV$ reject \hst\ with \odF\ if either \emph{one} of the constituent monitors \mV\ and \mVV\ rejects \hst\ with \odF\ (rules \textsc{parAL} and \textsc{parAR}).
Importantly, parallel disjunctions $\mV \oplus \mVV$ reject \hst\ with only when the flag is \btrue\ and \emph{both} monitors 
reject it (rule \textsc{parO}), ensuring that the trace prefix analysed consisted of deterministic actions.   
%
Rule \textsc{rec} states that a recursive monitor rejects a history with some flag if its unfolding does.
%
%
As a shorthand, 
we say that monitor $\mV$ rejects history \hst, denoted \rej{\hst,\mV}, whenever \rej{\hst,\btrue,\mV}.

\begin{example}\label{ex:reject-over-runs}
  Recall $\pV_2$ and $\mV_1$ from \Cref{ex:inst}
  and suppose that $\detAct{r}=\detAct{s}=\btrue$. 
  Instrumentation can record $t_1=rs\ut a$ during a first 
  execution, but $\mV_1$ fails to reject the recorded history, 
  $\neg\rej{\{ t_1 \}, \mV_1}$.  
  When $\pV_2$ is monitored again, the additional trace $t_2= rs\uf c$ can be aggregated, which $\mV_1$ now rejects, $\rej{ \{ t_1,t_2 \}, \mV_1}$ (see \Cref{fig:proof-derivation-1,fig:proof-derivation-2}). 
  %
  \exqed
\end{example}

\Cref{ex:no-reject} shows that rejections are always evidence-based.

\begin{example}
  \label{ex:no-reject}
  Although 
  monitor $\no$ trivially rejects \emph{any} 
  \pV, it 
  does so after observing 
  one execution:
  for 
  $\hst_0=\emptyset$, the semantics in \Cref{fig:mon} immediately triggers rule \textsc{iNo}, \ie  
  %
    $\sysH{\pV}{(\epsilon,\no)}{\emptyset} \traS{\tau} \sysH{\pV}{(\epsilon, \stp)}{\{\epsilon\}}$.
  %
  When 
  $\epsilon$ is added to the history, 
  one can conclude $\rej{\{\epsilon\}, \no}$ by rule \textsc{no}.
  \qed
\end{example}

\section{Monitor Correctness}
\label{sec:mon-correct}

RV establishes a correspondence between the operational behaviour of a monitor and the semantic meaning of the property being monitored for~\cite{FrancalanzaAAAC17:rv,AcetoAFIL21:sosym} which transpires the meaning of the statement
  ``monitor $\mV$ \emph{correctly monitors} for a property $\varphi$.''
Our first correctness result concerns the \emph{history aggregation} mechanism of \Cref{sec:mon-setup}.
%
\Cref{prop:trace-veracity} 
states that traces collected 
are indeed 
generated by the instrumented SUS. 
%
Thus, whenever a history $\hst$ is accumulated over a sequence of 
executions of 
some $\pV$, \ie 
$\emptyset \subseteq \hst_1 \subseteq  \cdots \subseteq \hst$, 
then $\hst \subseteq \hstpV$.
%

\begin{restatable}[Veracity]{proposition}{traceveracity} \label{prop:trace-veracity}%
  For any $\hst$, \mV, $\pV$, and $\stact_1,\ldots,\stact_n$,  
  if $\sysH{\pV}{(\epsilon, \mV)}{\hst} \traS{\stact_1}\ldots\traS{\stact_n} \sysH{\pV'}{(t, \mV')}{\hst'}$ then $\pV \wttraS{t} \pV'$. 
  %
  %
  \qed
\end{restatable}

Another 
criteria for our multi-run monitoring setup is that executing-monitors \emph{behave deterministically}~\cite{Fra:Concur:17,AcetoAFIK20:determinising-mon}. 
%
%
Our monitors are \emph{confluent} \wrt $\tau$-moves, \Cref{prop:diamond-moves}, thus 
equated  up to $\tau$-transitions.
Importantly, for a given history, the 
monitors of \Cref{sec:mon-setup} deterministically reach equivalent states
when analysing a (partial) 
trace exhibited by the SUS, \Cref{prop:det-mon}.





%
%
%
%
\begin{restatable}
  [Determinism]{proposition}{detmon}\label{prop:det-mon}
  If $(t, \mV) \wtraSH{u} (t',\mV')$ and $(t, \mV) \wtraSH{u} (t'',\mV'')$, then $t'=t''$ and 
  there is 
  $\mVV \in \Mon$ such that $(t',\mV') (\traSH{\tau})^* (t',\mVV)$ and $(t'',\mV'')  (\traSH{\tau})^* (t'',\mVV)$.
  \qed
\end{restatable}

\begin{example}\label{ex:monitor-equiv}
  Recall $\mV_2$
  from \Cref{ex:inst-comp}.
  Given 
  $u=rsa$, the executing-monitor $(\epsilon,\mV_2)$ can reach either 
  $(u, \no)$ or $(u, \mV \otimes \no)$  
  %
  which $\tau$-converges to $(u, \no)$ 
  via rule \textsc{mVrP2R}. 
  %
  \exqed
\end{example}

A characteristic sanity check is \emph{verdict irrevocability}~\cite{DBLP:journals/iandc/Francalanza21,Fra:Concur:17, AcetoAFIL21:sosym}.
%
This translates to \Cref{prop:irrevocability} stating that, once a SUS is rejected (using \emph{history analysis} of \Cref{fig:proof-system})
for exhibiting history $\hst$, further observations 
(in terms of longer traces, \emph{length}, or additional traces, \emph{width}) 
do \emph{not} alter 
the
conclusion.
%
%
\begin{restatable}[Irrevocability]{proposition}{irrevocability}
  \label{prop:irrevocability}%
  If \rej{(\hst,t), \mV} then \rej{(\hst,tu), \mV}.
  If \rej{\hst, \mV} then \rej{ \hst {\cup} \hstt, \mV}.
  \qed
\end{restatable}


The least \emph{correctness} requirement expected 
of our (irrevocable) \emph{history analysis} 
is that any rejections imply property violations. 
%
%
Concretely, $\mV$ monitors 
soundly for  
$\varphi$ if, for \emph{any} 
system \pV, 
whenever \mV rejects a history \hst\ produced by \pV, 
\ie  \rej{\hst,\mV} for $\hst {\subseteq} \hstpV$,
then 
\pV also violates the property, \ie $\pV {\notin}  \evalE{\varphi}$. 
%
%
%
The universal quantification over systems 
of \Cref{def:full-sound} manifests a  black-box treatment of the SUS.
%
%

\begin{definition}[Soundness]\label{def:full-sound}
  $\mV$ monitors \emph{soundly} for 
  $\varphi$ 
  when $\forall \pV\in\Prc$, 
  if $\exists\hst\subseteq\hstpV$ such that \rej{\hst,\mV} then $\pV\notin \evalE{\varphi}$.
  \exqed
\end{definition}

\begin{example}\label{ex:sound}
  $\mV_1$ 
  from \Cref{ex:inst} 
  monitors soundly for 
  $\varphi_4$ 
  from \Cref{ex:rechml}.
  %
  %
  \Cref{ex:inst} illustrates how trace prefixes $rs\ut a$ and $rs\uf c$ of $\pV_2$ can be veraciously accumulated as a history 
  and \Cref{ex:reject-over-runs} shows that such a history is rejected.
  Accordingly, $\pV_2$ violates $\varphi_4$. 
  By comparison, monitor $\mV_3 \deftxt r.s.a.\no$ is \emph{not} sound for $\varphi_4$; it can collect and reject histories that contain the trace $rs\ut a$, but systems such as $\recX{r.s.\ut.a.X}$ and $r.s.\ut.a.\nl$ (which can exhibit such a trace) do \emph{not} violate $\varphi_4$.
  \exqed
\end{example}

The dual requirement to soundness is (rejection) completeness: 
%
\mV \emph{monitors completely} for 
$\varphi$ if any $\pV {\notin}  \evalE{\varphi}$
can be rejected based on some history it produces. 
%

\begin{definition}[Completeness]\label{def:full-complete}
  $\mV$ monitors \emph{completely} for 
  $\varphi$ 
  when 
  $\forall \pV\notin\evalE{\varphi}$ 
  implies $\exists \hst\subseteq \hstpV$ such that \rej{\hst, \mV}.
  \exqed
\end{definition}


\begin{example}\label{ex:full-complete}
  $\mV_4 \deftxt s.\no \otimes a.\no \otimes c.\no$ monitors completely for $\varphi_0$ 
  from \Cref{ex:intro}.
  %
  Any violating system 
  can exhibit 
  a trace of the form $ts$, $ta$ or $tc$ for some 
  $t\in\IAct^*$.
  Once exhibited (and aggregated), one can show that $\mV_4$ rejects 
  that history. 
  %
  %
  %
    \exqed
\end{example}

For monitors that are veracious and produce irrevocable verdicts
(\Cref{sec:mon-setup}), 
(rejection) soundness and completeness constitute the basis for our definition of monitor correctness. 

\begin{definition}[Correct Monitoring]\label{def:full-correct}
  Monitor \mV monitors \emph{correctly} for 
  formula $\varphi$ if it can do so \emph{soundly} and \emph{completely}. 
\end{definition}



\section{Monitorability}
\label{sec:monitorability}

Monitorability~\cite{DBLP:journals/sttt/FalconeFM12,FrancalanzaAI17:monitorability-branching,BartocciFFR18,AcetoAFIL21:sosym} delineates between the properties that can be correctly monitored 
and those that cannot, 
%
realised as a correspondence between the declarative semantic of \Cref{sec:preliminaries} and the operational semantics of \Cref{sec:mon-setup}.
%
The chosen approach~\cite{FrancalanzaAAAC17:rv} applies to a variety of settings~\cite{PnueliZaks06:RV-MC,AcetoAFIL19:monitorability-linear,HenzingerS21:LICS,CastanedaR23:PODC,FerrandoCard25:SCP}. 
It fosters a separation of concerns between the specification semantics and the verification method employed,  which is relevant to our investigation on the increase in expressive power when moving from single-run monitoring to multi-runs; see~\cite{AcetoAFIL21:sosym} for a comparison between distinct notions of monitorability.
%
%
%
%
Specifically, \Cref{def:full-mon} (below) is \emph{parametric} \wrt the definition of 
\emph{``$\mV$ monitors correctly for $\varphi$''}; 
prior work~\cite{FrancalanzaAI17:monitorability-branching} formalised this as single-run monitoring whereas \Cref{def:full-correct} redefines it as multi-run monitoring.


\begin{definition}[Monitorability~\cite{FrancalanzaAI17:monitorability-branching}]
  \label{def:full-mon}
   Formula $\varphi {\in} \recHML$ is \emph{monitorable} iff 
  $\exists \mV{\in}\Mon$ monitoring \emph{correctly} for it.
  Sublogic $\mathcal{L} {\subseteq} \recHML$ is monitorable iff 
  $\forall\varphi{\in}\mathcal{L}$ are monitorable.
  \exqed
\end{definition}

Several formulae are 
unmonitorable (for violations) according to \Cref{def:full-mon}, particularly when they include existential modalities and least fixed points.   
%

\begin{example}\label{ex:unmonitorable}
  %
  Assume, towards a contradiction, that there exists a sound and complete monitor $\mV$ for the formula $\Em{\act}\T$.
  Pick 
  some $\pV \notin \evalE{\Em{\act}\T}$, \ie $\pV \traN{\act}$. 
  By \Cref{def:full-complete}, there exists a history $\hst \subseteq T_\pV$ such that $\rej{\hst, \mV}$.
  %
  Using \pV, we can build another 
  system $\pV + \act.\inert$ where $\pV+ \act.\inert \in \evalE{\Em{\act}\T}$ irrespective of the value of $\Det(\act)$.  
  %
  We also know that 
  \hst\ is a history of  $\pV + \act.\inert$ since $\hst \!\subseteq\! T_\pV \!\subseteq\! T_{\pV + \act.\inert}$. 
  %
  This and $\rej{\hst, \mV}$ makes $\mV$ unsound, contradicting our assumption. 

  Similarly, 
  assume, towards a contradiction, 
  that there exists a monitor $\mV$ that can monitor soundly and completely for 
  \Min{\XV}{(\Disj{\Um{\act}{\XV}}{\Um{\actt}{\F}})}.
  The 
  single state system $\pV$ with the sole transition 
  $\pV \traS{\act} \pV$ violates the formula.
  Due to \Cref{def:full-complete}, we must have $\rej{\hst, \mV}$ for some $\hst \subseteq T_\pV$. 
  From 
  the structure of \pV, we also know \hst\ is a \emph{finite} set of the form $\{\act^n\ |\ n \in \Nat\}$.  
  Fix $k$ to be the length of the \emph{longest} trace in \hst\ and then consider the system \pVV consisting of $k+1$ states and the transitions $\pVV=\pVV_0 \traS{\act} \ldots \traS{\act} \pVV_{k}$ exclusively. 
  Clearly, \pVV satisfies \Min{\XV}{(\Disj{\Um{\act}{\XV}}{\Um{\actt}{\F}})}.  
  Since 
  $\hst \subseteq T_\pVV$ as well, 
  $\rej{\hst, \mV}$ contradicts the assumption that \mV is sound.
  %
  \exqed
\end{example}

Disjunctions are the only other \recHML logical constructs 
excluded from \sHML, as restated in \Cref{thm:shml}. 
Formulae containing disjunctions can be monitorable with a few caveats.  

\begin{example}\label{ex:unmonitorable-disj}
  Recall $\varphi_2\deftxt \Um{r}{(\Disj{\Um{s}{\F}}{\Um{a}{\F}})}$ from \Cref{ex:non-det}. 
  When $\Det(r) = \bfalse$,  $\varphi_2$ is \emph{not} monitorable.
  %
  By contradiction, assume 
  a correct 
  $\mV$ exists. 
  Since $\pV_3 \deftxt r.(s.\nl + a.\nl) + r.s.\nl \notin \evalE{\varphi_2}$, 
  then we should have \rej{\hst, \mV} for some $\hst {\subseteq} T_{\pV_{3}}$. 
  %
  But $\hst\subseteq T_{\pV_{4}} {=} T_{\pV_{3}}$ for $\pV_4 \deftxt r.s.\nl + r.a.\nl \in \evalE{\varphi_2}$, and 
  %
  \rej{\hst, \mV} would make \mV\ unsound, contradicting our initial assumption.
   
  However, when $\Det(r) {=} \btrue$, $\varphi_2$ is monitorable: an obvious correct monitor is $\mV_5 \deftxt r.(s.\no \oplus a.\no)$.
  Although systems $\pV_3$ and $\pV_4$ would be ruled out, an ILTS would still allow systems such as 
  $\pV_5 \deftxt r.(s.\nl + a.\nl) + r.(s.\nl + a.\nl + a.\nl)$ 
  that reach the equivalent states $s.\nl + a.\nl$ and $s.\nl + a.\nl + a.\nl$ after an $r$-transition.
  %
  Even if $\hst=\sset{ra,rs}$ is aggregated by passing through \emph{different} intermediary states, \ie $s.\nl + a.\nl$ and $s.\nl + a.\nl + a.\nl$, the monitor analysis would still be sound in rejecting $\pV_5$ via \hst;
  see \Cref{prop:behaviour-equiv}.   

  A trickier formula is $\varphi_4 {\deftxtS} \Max{\XV}{\bigl(\Um{r}{\Um{s}{\XV}} \wedge (\Um{a}{\F} \vee \Um{c}{\F})\bigr)}$ from \Cref{ex:rechml}.
  Although the disjunction is syntactically not prefixed by any universal modality, it can be reached after a recursive unfolding, \ie $\varphi_4 \steq \Um{r}{\Um{s}{\varphi_4}} \wedge (\Um{a}{\F} \vee \Um{c}{\F})$.
  %
  By similar reasoning to that for $\varphi_2$, 
  formula $\varphi_4$ is monitorable whenever $\Det(r) {=} \Det(s) {=} \btrue$ but unmonitorable otherwise.
  %
  %
  \exqed 
\end{example}

\Cref{def:shmlwDet} characterises the extended class of \recHML monitorable formulae for multi-run monitoring, parametrised by \EAct and the associated action determinacy delineation defined by \Det.
It employs a flag to calculate deterministic prefixes via rule \textsc{cUm} along the lines of \Cref{fig:proof-system}. This is then used by rule \textsc{cOr}, which is only defined when the flag is \btrue.  


\begin{definition}\label{def:shmlwDet}
  $\odF \cons \varphi$  
  is defined coinductively as the largest relation of the form $(\Bool \times \recHML)$ satisfying the rules 
  %
  \begin{mathpar}[\small]
    \inferrule[\rtitSS{cA}]
      { \varphi\in\{ \F, \T, \XV \} }
      { \odF \cons \varphi }
    \and
    \inferrule[\rtitSS{cUm}]
      { \odF \wedge \detAct{\act} \cons \varphi}
      { \odF \cons \Um{\act}{\varphi} }
    \and
    \inferrule[\rtitSS{cAnd}]
      { \odF \cons \varphi \quad \odF \cons \psi }
      { \odF \cons \Conj{\varphi}{\psi} }
    \and
    \inferrule[\rtitSS{cOr}]
      { \btrue \cons \varphi \quad \btrue \cons \psi }
      { \btrue \cons \Disj{\varphi}{\psi} }
    \and
    \inferrule[\rtitSS{cMax}]
      { \odF \cons \varphi\subS{\Max{\XV}{\varphi}}{\XV} }
      { \odF \cons \Max{\XV}{\varphi} }
  \end{mathpar}
  %
  $\sHMLwDet \deftxt \setof{\varphi}{\btrue \cons \varphi}$ defines the set of extended monitorable formulae. 
  It extends \sHML with disjunctions as long as these are prefixed by universal modalities of deterministic external actions (up to  largest fixed point unfolding).
  \qed
\end{definition}

\begin{example}
  For $\Det(r) = \Det(s) = \btrue$, we can show $\varphi_{2},\varphi_{4} {\in} \sHMLwDet$.   
  Exhibiting the relation 
  $ 
    \relR = \{
      (\btrue,\Um{r}{(\Disj{\Um{s}{\F}}{\Um{a}{\F}})}),
      (\btrue,\Disj{\Um{s}{\F}}{\Um{a}{\F}}),
      (\btrue,\Um{s}{\F}),
      (\btrue,\Um{a}{\F}),
      (\btrue,\F) 
    \}
  $
  suffices to prove the inclusion of  $\varphi_{2}$ in \sHMLwDet. 
  \exqed
\end{example}

Although the tracing of internal actions as part of the history 
helps with correct monitoring, multi-run RV requires us to limit systems to deterministic internal actions in order to attain violation completeness for monitors $\Mon$ of \Cref{fig:mon}.

\begin{example}
  \label{ex:shmlwDet}
  %
  $\pV_6 \deftxt \ut.r.s.\nl + \uf.r.a.\nl$ and $\pV_7 \deftxt \iact.r.s.\nl + \iact.r.a.\nl$ both \emph{satisfy} $\varphi_2$ from \Cref{ex:unmonitorable-disj} with $\Det(r)=\btrue$.
  In the case of $\pV_6$, 
  $\mV_5$ from \Cref{ex:unmonitorable-disj} does \emph{not} reject the history $\sset{\ut rs, \uf ra}$ because the application of rule \textsc{actI} of \Cref{fig:proof-system} (for either \ut\ or \uf) necessarily reduces the history size of the premise to \emph{one} trace.
  %
  For $\pV_7$, we  must have $\Det(\iact)=\bfalse$; 
  %
  when $\mV_5$ analyses the history $\sset{\iact rs, \iact ra}$ using rule \textsc{actI}, the premise flag can only be $\bfalse$ which prohibits the analysis from using \textsc{parO}.
  Both systems $\pV_8 \deftxt r.(\ut.s.\nl + \uf.a.\nl)$ and $\pV_9 \deftxt r.(\iact.s.\nl + \iact.a.\nl)$ \emph{violate} $\varphi_2$.
  Accordingly, both are rejected by $\mV_5$ via the respective histories $\sset{r\ut s,  r\uf a}$ and $\sset{r\iact s,  r\iact a}$.  
  %
  
  Non-deterministic internal actions 
  hinder completeness. 
  System $\pV_{10} \deftxt \iact.\pV_8 + \iact.\nl$ violates $\varphi_2$ but $\mV_5$ cannot reject the history $\sset{\iact r\ut s,  \iact r\uf a}$:
  again, $\Det(\iact) = \bfalse$ limits the flag premises for \textsc{actI} to \bfalse, prohibiting the use of \textsc{parO}.   
  \exqed
\end{example}

Showing that a logical fragment
is monitorable, \Cref{def:full-mon}, is non-trivial due to the 
universal quantifications 
to be considered, \eg all $\varphi {\in} \mathcal{L}$ and all 
$\pV{\in} \Prc$ 
from \Cref{def:full-sound,def:full-complete}.
We prove the monitorability of $\sHMLwDet$ systematically, by concretising the existential quantification of a correct monitor for every 
$\varphi\in\sHMLwDet$ via the monitor synthesis 
$\synth{\varphi}$. 
We then prove that for any $\varphi {\in} \sHMLwDet$, the synthesised $\synth{\varphi}$ 
monitors correctly for it 
(\Cref{def:full-mon}).
A 
by-product of this 
 proof strategy is that the synthesis function in \Cref{def:synth} can be used directly for tool construction to automatically generate (correct) witness monitors from specifications; see~\cite{DBLP:conf/forte/AttardAAFIL21,DBLP:conf/coordination/AcetoAAEFI22}.

\begin{definition}
  \label{def:synth}
  $\synth{-}\!:\! \sHMLwDet \map \Mon$ is defined as follows:
  {\small
  \begin{align*}
    &\synth{\F} \deftxtS \no 
    &&\synth{\Conj{\varphi }{\varphi}} \deftxtS  \synth{\varphi} \otimes \synth{\varphi}  
    &&\synth{\Um{\act}{\varphi}} \deftxtS \act.\synth{\varphi} \quad \synth{\XV} \deftxtS \XV \\  
    & \synth{\T} \deftxtS \stp  
    &&\synth{\Disj{\varphi }{\varphi }} \deftxtS  \synth{\varphi} \oplus \synth{\varphi} 
    &&\synth{\Max{\XV}{\varphi}} \deftxtS \recX{\synth{\varphi}} 
    \tag*{\exqed}
  \end{align*}
  }%
\end{definition}

If we limit ILTSs to deterministic internal actions, \ie $\Det(\iact) = \btrue$ for all $\iact {\in} \IAct$, we 
can show monitorability for arbitrary ILTSs and the fragment \sHMLwDet.

\begin{restatable}{proposition}{sound} \label{prop:sound}%
  $\synth{\varphi}$ is sound for $\varphi\in\sHMLwDet$. 
  \qed
\end{restatable}

\begin{restatable}{proposition}{complete} \label{prop:complete}%
  If $\detAct{\iact} = \btrue$ for all $\iact\in\IAct$, 
  then $\synth{\varphi}$ is complete for all $\varphi\in\sHMLwDet$. 
  \qed
\end{restatable}

\begin{restatable}[Monitorability]{theorem}{fullMonitorability}
  \label{thm:full-monitorability}
  When $\Det(\iact) = \btrue$ for all $\iact \in \IAct$, all $\varphi \in \sHMLwDet$ are 
  monitorable. 
  \qed
  \end{restatable}
  
  

%
%
We can show an even stronger 
result
which ensures that restricting specifications to \sHMLwDet does not exclude any monitorable properties,
\Cref{thm:maximal-expressivity}.
%
%
Maximality typically relies on a reverse synthesis $\synthRev{-}$ that maps any  $\mV\in\Mon$ to a characteristic formula $\synthRev{m}\in\sHMLwDet$  it monitors correctly for.
This method is however complicated by the occurrence of non-deterministic actions: \eg if $\detAct{r} = \bfalse$ the monitor $r.(s.\no \oplus a.\no)$ does \emph{not} correctly monitor for 
$\Um{r}{(\Disj{\Um{s}{\F}}{\Um{a}{\F}})}$ 
but instead never rejects; to obtain our results we first normalise such a monitor to $r.\stp$; see \Cref{sec:maximality-proof}. 
%
Maximality permits a verification framework to determine if a property is monitorable via 
a simple syntactic check, 
or else employ alternative verification techniques.
%
%
The development of an RV tool can also 
exclusively target 
\sHMLwDet, 
knowing that all monitorable properties are covered.




\begin{restatable}[Maximality]{theorem}{maximality}
  \label{thm:maximal-expressivity}
  If $\Det(\iact) {=} \btrue$ for all $\iact \in \IAct$
  and $\cL\subseteq \recHML$ is monitorable \wrt \Mon, 
  then for all $\varphi \in \cL$, 
  there exists $\psi \in \sHMLwDet$ such that $\evalE{\varphi} = \evalE{\psi}$.  
  \qed
\end{restatable}

\begin{remark}
%
\Cref{sec:implementability} outlines the steps for a full tool automation and gives a corresponding complexity analysis.
%
\exqed
\end{remark}


\section{Actor Systems}
\label{sec:actor-systems}

We validate the utility and applicability of monitoring ILTSs from \Cref{sec:preliminaries} 
via an instantiation to actor systems~\cite{DBLP:conf/ijcai/HewittBS73,DBLP:books/Agha-Actors,DBLP:books/CesariniThompsonErl,goodwin2015AKKA,juric2024elixir,swift:24}   
%
where a set of processes called \emph{actors} interact 
via \emph{asynchronous message-passing}. 
Each actor, $\actrs{\idV}{\eV}{\qV}$, is identified by its unique ID, $\idV,\idVV, \idVVV,\idVVVV \in \Pid$, used by other actors to address messages to it \ie the \emph{single-receiver} property.
Internally, actors consist of a running expression $\eV$ and a mailbox $\qV$, \ie  a list of values denoting a message queue. 
%
%
%
\begin{align*}
  \actV,\actVV \in \Actors & \bnfdef\ \actrs{\idV}{\eV}{\qV} \bnfsepp  \nl  \bnfsepp \actV\actPar\actVV \bnfsepp \actNew{\idV}{\actV}  \bnfsepp \pioutB{\idV}{\vV}
\end{align*}
Parallel actors, $\actV\actPar\actVV$, can also be inactive, \nl, or have IDs that are locally \emph{scoped} to a subset of actors, 
$\actNew{\idV}{\actV}$.
%
There may also be messages in transit, \pioutB{\idV}{\vV}, where value $v$ is addressed to \idV.
The set of all free IDs \idV\ identifying 
actors \actrs{\idV}{\eV}{\qV} in \actV is denoted by \fid{\actV}.

Values, $v\in\Val$, range over $\Pid \cup \Atom$  where $\atV,\atVV \in \Atom$ are uninterpreted tags.
Actor expressions $\eV,\eVV\in\Exp$ can be outputs, $\piout{\idV}{\vV}{\eV}$, or reading inputs from the mailbox through pattern-matching, $\rcv{ \{ \patV_n \rTran \eV_n \}_{n\in I} }$, where each expression $\eV_n$ is guarded by a \emph{disjoint} pattern $\patV_n$.
Actors may also refer to themselves, 
$\slf{\xV}{\eV}$, 
spawn other actors, $\spwn{\eVV}{\xV}{\eV}$,
and recurse, \recX{\eV}.
Receive patterns, spawn and recursion bind expression variables $\xV,\xVV \in \Vars$, and term variables $\XV,\XVV \in \TVar$.
Similarly, $\actNew{\idV}{\actV}$ binds the name ID \idV in \actV. 
We work up to $\alpha$-conversion of bound entities. 
%
The list notation \ccat{\vV}{\qV} denotes the mailbox with \vV as the head and \qV as the tail of the queue, whereas \ccat{\qV}{\vV} denotes the mailbox with \vV at the end of the queue preceded by \qV; queue concatenation is denoted as \ccat{\qV}{\qVV}. 
We may elide empty mailboxes
and write \actrss{\idV}{\eV} 
for \actrs{\idV}{\eV}{\mEmpty}.

The ILTS semantics for our language is defined over system states of the form $\conf{\K\;|\;\Obs}{\actV}\in\Prc$.
The implicit \emph{observers} that \actV\ interacts with when running is represented by the set of IDs $\Obs\subseteq \Pid$; 
to model the single receiver property we have $\fid{\actV}\cap \Obs = \emptyset$.
%
\emph{Knowledge}, $\K\subseteq\Pid$, 
denotes the set of IDs known by both actors in \actV\ and \Obs; it keeps track of bound/free names without the need for  name bindings in actions~\cite{DBLP:books/daglib/0004377} where $(\fid{\actV} \cup \Obs)\subseteq \K$; see \cite{BengstonParrow09:nominalPi}.
%
%
Transitions are of the form 
\begin{equation}
  \label{eq:act-trans}
  \conf{\K\;|\;\Obs}{\actV} \traS{\stact} \conf{\K'\;|\;\Obs'}{\actVV}
\end{equation} 
where \stact\ ranges over $\EAct\cup\IAct\cup\{\tau\}$.
\emph{External} actions $\EAct = \setof{\piinA{\idV}{\vV},\ \pioutA{\idV}{\vV},\ \npioutA{\idV}{\idVV}}{\idV,\idVV \in \Pid, \vV \in \Val}$ include input, \piinA{\idV}{\vV}, output, \pioutA{\idV}{\vV}, and scope-extruding output, \npioutA{\idV}{\idVV}. 
\emph{Internal} actions 
$\IAct = \setof{\commA{\idV}{\vV},\ \ncommA}{\idV \in \Pid, \vV \in \Val}$ include internal communication involving either free names, \commA{\idV}{\vV} or scoped names, \ncommA. 
%
\Cref{eq:act-trans}
is governed by the judgement $\confK{\actV} \traS{\stact} \actVV$ with $K'|\Obs'=\after{\K\,|\,\Obs,\stact}$; the latter function 
determines $K$ and \Obs\  where $\after{\K\,|\,\Obs,\npioutA{\idV}{\idVV}} \deftxt \bigl(\K {\cup} \{\idVV\}\bigr) | \Obs$ and  $\after{\K\,|\,\Obs,\piinA{\idV}{\idVV}} \deftxt (\K {\cup} \{\idVV\}) | \bigl(\Obs{\cup} (\{\idVV\}{\setminus} \K)\bigr)$ (all other cases of \stact\ leave $\K|\Obs$ unchanged).
The generation of \emph{external actions} is defined by the following rules where asynchronous output is conducted in two steps, rules \rtitS{snd1} and \rtitS{snd2}, where the latter rule requires the recipient address \idVV to be in \Obs.
Scope-extruded outputs with its 
name management is described by \rtitS{opn}.   
\begin{mathpar}[\small]
  \inferrule*[lab=\rtitSS{snd1}]
    { }
    { \conf{\K \;|\; \Obs }{ \actrs{\idV}{ \piout{\idVV}{\vV}{\eV} }{\qV} }  
      \traS{ \tau } \actrs{\idV}{\eV}{\qV} \actPar \pioutB{\idVV}{\vV} 
    }
  \quad
  \inferrule*[lab=\rtitSS{snd2},Right=${\scriptsize\idVV{\in}\Obs}$]
    { }
    { \confK{ \pioutB{\idVV}{\vV} }  \traS{ \pioutA{\idVV}{\vV} } \nl }
  \and
  \inferrule*[Left=\rtitSS{rcv}]
  { }
  { \confK{ \actrs{\idV}{\eV}{\qV}}  \traS{ \piinA{\idV}{\vV} }  \actrs{\idV}{\eV}{\ccat{\qV}{\vV}} }
  \and
  \inferrule*[Left=\rtitSS{opn}]
  { \conf{(\K,j) \;|\; \Obs}{ \actV } \traS{\pioutA{\idV}{\idVV}} \actVV}
  { \confK{ \actNew{j}{\actV} }   \traS{\npioutA{\idV}{\idVV}}  \actVV}
  \and 
  \inferrule[\rtitSS{rd}]
  { 
    \forall n\in I \cdot \absent{ \pV_n , \qV}  \quad 
    \exists m\in I \cdot \neg\absent{ \pV_m, \vV} \wedge
    \match{\pV_m,\vV} = \sV 
  }
  { \confK{ \actrs{\idV}{ \rcv{ \{\pV_n \rTran \eV_n \}_{n\in I} } }{ \ccat{\qV}{\ccat{\vV}{\qVV}} } }
        \traS{\tau} \actrs{\idV}{ \eV_m\sV }{\ccat{\qV}{\qVV}} 
  }
\end{mathpar}
Rule \rtitS{rcv} details how input actions append to the recipient mailbox, which are then \emph{selectively} read following rule \textsc{rd}. 
Selection relies on the 
helper functions \absent{-} and \match{-} in \Cref{def:pattern} to find the first message \vV\ in the mailbox that matches one of the 
patterns $\pV_m$ in $\{ \pV_n \!\tra{\;}\! \eV_n\}_{n\in I}$.  
%
%
%
If a match is found, the actor branches to $\eV_m\sigma$, where $\eV_m$ is the expression guarded by the matching pattern $\pV_m$ and $\sV {\in} \Subs\!:\! \Vars \pmap \Val$ substitutes the free variables in $\eV_m$ for the values resulting from the pattern-match.
%
%
\begin{mathpar}[\small]
  \inferrule[\rtitSS{commL}]
      { \begin{tabular}{l}
        \conf{\K\;|\;\fid{\actVV}}{\actV} \traS{\pioutA{\idV}{\vV}} $\actV'$
        \\
        \conf{\K\;|\;\fid{\actV}}{\actVV} \traS{\piinA{\idV}{\vV}}  $\actVV'$
      \end{tabular} 
      }
      { \confK{ \actV \actPar \actVV }
        \traS{\commA{\idV}{\vV}} 
        \actV' \actPar \actVV' 
      } 
    \quad
    \inferrule[\rtitSS{ncommL}]
      { \begin{tabular}{l}
        \conf{\K\;|\;\fid{\actVV}}{\actV} \traS{\npioutA{\idV}{\idVV}} $\actV'$
        \\
        \conf{\K\;|\;\fid{\actV}}{\actVV} \traS{\piinA{\idV}{\idVV}}  $\actVV'$
        \end{tabular} 
      }
      { \confK{ \actV \actPar \actVV }
        \traS{\ncommA} 
        \actNew{\idVV}{(\actV' \actPar \actVV')} 
      }
    \and
    \inferrule*[lab=\rtitSS{scp2},Right=${\scriptsize\idVV \in \{ \idV,\vV \}}$]
      { \conf{\K,\idVV \;|\; \Obs}{ \actV } \traS{\commA{\idV}{\vV}} \actVV   
      }
      { \conf{\K \;|\; \Obs}{ \actNew{\idVV}{\actV} }   
        \traS{\ncommA}   
        \actNew{\idVV}{\actVV} 
      }
      \and\qquad
      \inferrule*[lab=\rtitSS{str}] 
      { \begin{tabular}{l}
        \actV \steq\, $\actV'$ \quad  $\actVV'$ \steq\, \actVV
        \\
        \confK{\actV'} \traS{\stact} $\actVV'$    
      \end{tabular}
      }
      { \confK{\actV} \traS{\stact} \actVV }
    \end{mathpar}
Internal actor interaction is described via internal actions to permit monitors to differentiate these steps from the silent transitions. 
Transitions with \commA{\idV}{\vV} labels are deduced via \rtitS{commL} (above) or the symmetric rule \rtitS{commR}, whereas \ncommA-transitions are generated by the \rtitS{ncommL}, \rtitS{ncommR} and \textsc{scp2} rules. 
Our semantics assumes standard structural equivalence as the ILTS equivalence relation, with axioms such as $\actV \steq \actV \actPar \nl$  and $\actV \actPar \actVV \steq \actVV \actPar \actV$; transitions abstract over such states via rule \rtitS{str}.
The remaining transitions are fairly standard.

\subsection{Actor Structural Equivalence and Silent Actions}

To show that our semantics is indeed an ILTS, we need to prove a few additional properties. 
\Cref{prop:equivalent-states} below shows that transitions abstract over structurally-equivalent states.

\begin{restatable}{proposition}{equivalentStates}
\label{prop:equivalent-states}
  For any $\actV \steq \actVV$, 
  whenever $\confK{\actV} \traS{\stact} \actV'$ then there exists $\actVV'$ 
  such that $\confK{\actVV} \traS{\stact} \actVV'$ and $\actV' \steq \actVV'$.
  \qed
\end{restatable}

As a result of \Cref{thm:confluence} below, we are guaranteed that any actor SUS instrumented via a mechanism that implements the semantics in \Cref{fig:mon}  can safely abstract over (non-traceable) silent transitions because they are confluent \wrt\ other actions.

\begin{restatable}
  {proposition}{actorconfluence}
  \label{thm:confluence}
  If $\confK{\actV} \traS{\tau} \actV'$ and $\confK{\actV} \traS{\stact} \actV''$, then either
  $\stact=\tau$ and $\actV'\steq\actV''$ or
  there exists an actor system $\actVV$ and 
  moves 
  $\confK{\actV'} \traS{\stact} \actVV$ and $\conf{\after{\K\,|\,\Obs,\stact}}{\actV''} \traS{\tau} \actVV$. \qed
\end{restatable}

\subsection{Deterministic and Non-deterministic Traceable Actions}

Our ILTS interpretation treats input, output and internal communication 
as deterministic, justified by \Cref{prop:determinacy}.
\begin{restatable}[Determinacy]{proposition}{det} 
  \label{prop:determinacy}
  For all $\idV,\vV$, we have 
  \begin{itemize}[leftmargin=*]
    \item $\confK{\actV} \traS{\pioutA{\idV}{\vV}} \actV'$ and $\confK{\actV} \traS{\pioutA{\idV}{\vV}} \actV''$ implies $\actV' \steq \actV''$
    \item $\confK{\actV} \traS{\piinA{\idV}{\vV}} \actV'$ and $\confK{\actV} \traS{\piinA{\idV}{\vV}} \actV''$ implies $\actV' \steq \actV''$
    \item $\conf{\K\,|\,\Obs}{\actV} \tra{\!\!\!\commA{\idV}{\vV}\!\!} \actV'$ and $\conf{\K\,|\,\Obs}{\actV} \tra{\!\!\!\commA{\idV}{\vV}\!\!} \actV''$ implies $\actV' \!\steq\! \actV''$
    \!\!\!\!\!\!\qed
  \end{itemize}
\end{restatable}

In contrast, scope-extruding outputs and internal communication involving scoped names are \emph{not} considered 
to be deterministic, 
\ie for all $\idV,\idVV\in\Pid$, we have $\detAct{\npioutA{\idV}{\idVV}} = \detAct{\ncommA} = \bfalse$. 
\Cref{ex:non-det-ncomm,ex:non-det-out} illustrate why they are treated differently from other traceable actions.

\begin{example}
\label{ex:non-det-ncomm}
  Consider the actor state \confK{\actV_1} where $\idVV\in\Obs$ and the running actor is defined as $\actV_1 \deftxt \actNew{\idV}{(\, \actrss{\idV}{\rcv{\xV \rightarrow \piout{\idVV}{\xV}{\nl}}} \actPar \pioutB{\idV}{\vV_1} \actPar \pioutB{\idV}{\vV_2}\,) }$ with $\vV_1 \neq \vV_2$; the actor identified by \idV\ is scoped by the outer construct \actNew{\idV}.
  %
  The actor at $\idV$ can internally receive either value $\vV_1$ or $\vV_1$  via rules \textsc{scp2} and \textsc{commR} as follows:
  %
    \begin{align*}
      &\confK{\actV_1} \traS{\ncommA}  
      \confK{
        \actNew{\idV}{( \actrs{\idV}{\rcv{\xV \rightarrow \piout{\idVV}{\xV}{\nl}}}{\vV_1} 
        \actPar \nl \actPar \pioutB{\idV}{\vV_2} ) }
      } 
      \\
      & \confK{\actV_1} \traS{\ncommA}
      \confK{
        \actNew{\idV}{( \actrs{\idV}{\rcv{\xV \rightarrow \piout{\idVV}{\xV}{\nl}}}{\vV_2} \actPar \pioutB{\idV}{\vV_1} \actPar \nl )}
      }
      %
    \end{align*}
  %
  %
  Since $\vV_1\neq\vV_2$, the systems reached are \emph{not} structurally equivalent:  
  %
  they exhibit a different observational behaviour by sending different payloads to the observer actor at \idVV.
  %
  \exqed 
\end{example}

\begin{example}
  \label{ex:non-det-out}
  Consider the actor system $\confK{\actV_2}$ where $\idVVV\in\Obs$ and 
  the running actor is defined as 
  $\actV_2\deftxtS
  \actNew{\idV}{ \bigl(\,\actrss{\idV}{\eV_1} \actPar \pioutB{\idVVV}{\idV} \,\bigr)} 
  \,\actPar\,\actNew{\idV}{\bigl(\,\actrss{\idV}{\eV_2} \actPar \pioutB{\idVVV}{\idV} \bigr)}$;
  name $\idV$ is locally scoped twice and $\eV_1$ and $\eV_2$ exhibit different behaviour.
  %
  %
  %
  The actor system 
  $\confK{\actV_2}$ can scope extrude name $\idV$ by 
  delivering the message \pioutB{\idVVV}{\idV} in two possible ways 
  using rules \textsc{parL}, \textsc{parR} and \textsc{opn}
  as follows: 
  %
    \begin{align*}
      \confK{\actV_2} 
      & \traS{\npioutA{\idVVV}{\idV}} 
      \conf{K\cup\{\idV\}\;|\;\Obs}
      { 
        \bigl(\, \actrss{\idV}{\eV_1} \;\actPar\; \nl \,\bigr) \;\actPar\; \actNew{\idV}{\bigl(\,\actrss{\idV}{\eV_2} \actPar \pioutB{\idVVV}{\idV} \bigr)}
      }
      \\
      \confK{\actV_2}
      &\traS{\npioutA{\idVVV}{\idV}} 
      \conf{\K\cup\{\idV\}\;|\;\Obs}{
        \actNew{\idV}{ \bigl(\,\actrss{\idV}{\eV_1} \actPar \pioutB{\idVVV}{\idV} \,\bigr)}  
        \;\actPar\; (\, \actrss{\idV}{\eV_2} \;\actPar\; \nl \,) } 
      %
    \end{align*}
  %
  Since the systems reached  
  above are \emph{not} structurally equivalent, 
  they are possibly not behaviourally equivalent either. 
  Particularly, once an observer learns of the new actor address \idV, it could interact with it by sending messages and subsequently observe different behaviour through the different $\eV_1$ and $\eV_2$.
  %
  %
  %
  %
  \exqed
\end{example}

\Cref{ex:mon-actors} below showcases how the properties 
in \Cref{ex:intro,ex:non-det,ex:non-det-det,ex:rechml} can be adapted to monitor for actor systems.


\begin{example}\label{ex:mon-actors}
  With the values 
  $\req,\ans,\alloc,\cls,\init \in \Atom$,
  a server, expressed as actor $\idV$, can receive queries, \piinA{\idV}{\req}, reply to an observer client located at \idVV, \pioutA{\idVV}{\ans}, and send messages to a resource manager, abstracted as an observer actor at 
  address \idVVV, to either allocate more memory, \pioutA{\idVVV}{\alloc}, or close a connection, \pioutA{\idVVV}{\cls}.  
  We can reformulate $\varphi_4$ 
  (\Cref{ex:rechml}) as 
  %
  \begin{align*}
    \varphi_6 &\deftxtS
    \Max{\XV}{\bigl(
      \Conj{
        \Um{\piinA{\idV}{\req}}{\Um{\pioutA{\idVV}{\ans}}{\XV}}
      }{
        (\Disj{\Um{\pioutA{\idVVV}{\cls}}{\F}}{\Um{\pioutA{\idVVV}{\alloc}}{\F}})
      } \bigr) }  
  \end{align*}
  Assuming $\{ \idV, \idVV, \idVVV,k_1,k_2\} \subseteq \K$ and $\{\idVV,\idVVV\} \subseteq \Obs$, 
  consider 
  %
  the server implementation \confK{\actV_\text{srv}} 
  that violates $\varphi_6$.
  %
  %
  \begin{align*}
    \actV_\text{srv} &\deftxt 
      \actrss{\idV}{
        \rcv{\req \rightarrow (\piout{k_1}{\init}{\,\piout{k_2}{\init}{\,\pioutA{\idVV}{\ans}}})
        }
        }  \\
      & \qquad
      \;\actPar\; \actrss{k_1}{
          \rcv{\init \rightarrow \pioutA{\idVVV}{\alloc}}
          } 
      \;\actPar\; 
      \actrss{k_2}{\rcv{\init \rightarrow \piout{\idVVV}{\cls}{\nl}}} 
  \end{align*} 
  This implementation can produce the history $\{ t_1, t_2\}$ where we have
  $t_1 {=} (\piinA{\idV}{\req}).\commA{k_1}{\init}.\commA{k_2}{\init}.(\pioutA{\idVV}{\ans}).(\pioutA{\idVVV}{\alloc})$  
  and 
  $t_2 {=} (\piinA{\idV}{\req}).\commA{k_1}{\init}.\commA{k_2}{\init}.(\pioutA{\idVV}{\ans}).(\pioutA{\idVVV}{\cls})$.
  Since, 
  by \Cref{prop:determinacy}, $\detAct{\piinA{\idV}{\req}} = \detAct{\pioutA{\idV}{\ans}} = \btrue$,
  the visibility of the internal actions $\commA{k_1}{\init}$ and $\commA{k_2}{\init}$ suffices for the representative monitor $\mV_6\deftxt\synth{\varphi_6}$ to reject $\actV_\text{srv}$. 
  This changes for $\conf{\K'|\Obs}{\actNew{k_1,k_2}{(\actV_\text{srv})}}$ where $\K' = \K \setminus \{k_1,k_2\}$. 
  The aforementioned traces would change to 
  $t_3 {=} (\piinA{\idV}{\req}).\ncommA.\ncommA.(\pioutA{\idVV}{\ans}).(\pioutA{\idVVV}{\alloc})$  
  and 
  $t_4 {=} (\piinA{\idV}{\req}).\ncommA.\ncommA.(\pioutA{\idVV}{\ans}).(\pioutA{\idVVV}{\cls})$. The obscured \ncommA events prohibit monitoring from determining whether behaviourally equivalent SUS states are reached after these transitions, thus soundly relate $t_3$ with $t_4$ in history $\{ t_3, t_4\}$. 
  %
  %
  %
  \qed
\end{example}

\section{Establishing Bounds}
\label{sec:lowerbounds}

Despite the guarantees provided by 
\Cref{def:full-correct}, 
\Cref{thm:full-monitorability,thm:maximal-expressivity}
do not 
estimate the \emph{number of monitored runs needed} to reject a violating system.
%
This measure is 
crucial for an efficient implementation where history analysis (\Cref{fig:proof-system}) is not invoked  unnecessarily. 
%
We investigate whether there is a correlation between the syntactic structure of properties expressed in 
$\sHMLwDet$ and the number of partial traces required to conduct the verification.
In particular, we study how this measure can be obtained through a \emph{syntactic analysis} of the \emph{disjunction operators} 
in the formula.
Since we can only monitor for \sHMLwDet\ formulae 
when the relevant internal actions are deterministic (see \Cref{ex:shmlwDet}), internal actions are elided in the subsequent discussion.
%
%

\begin{example}\label{ex:intuition-bounds}
  Assume $\detAct{r}=\detAct{s}=\btrue$ and   
  recall $\varphi_2 \deftxt \Um{r}{\bigl(\Disj{\Um{s}{\F}}{\Um{a}{\F}}\bigr)}$ from \Cref{ex:non-det} and its 
  monitor $\mV_5 \deftxt r.\bigl( s.\no \oplus a.\no \bigr) = \synth{\varphi_2}$ from \Cref{ex:unmonitorable-disj}.
  Violating systems can produce the history $\hst\!=\!\{ rs, ra\}$, which is enough for $\mV_5$ to reject.
  %
  At the same time, no violating system for $\varphi_2$ can be rejected with fewer traces.
  Similarly, all violating systems for the formula $\varphi_5 \deftxt \Um{r}{\bigl(\Disj{\Um{s}{\F}}{\Um{a}{\F}}\bigr)} \vee \Um{a}{\F}$ can be rejected via the 3-size history $\{rs, ra, a \}$ (modulo internal actions).
  \exqed
\end{example}

Although 
\Cref{ex:intuition-bounds} suggests that monitoring for a formula with $n$ disjunctions requires $n{+}1$ executions to detect violations, 
this measure could be imprecise for a number of reasons.
%
%
First, there is no guarantee that the SUS will only produce the trace prefixes required to reject as it might also exhibit other behaviour.  
%
History bounds thus assume 
the \emph{best case scenario} where 
\emph{every} monitored run produces a \emph{relevant} trace prefix.
Second, not all SUS violations are justified by the same number of (relevant) trace prefixes: 
%
since formulae such as \Conj{\varphi_1}{\varphi_2} are violated by systems that either violate $\varphi_1$ or $\varphi_2$ (but not necessarily both), 
the number of relevant trace prefixes required to violate each subformula $\varphi_i$ for $i \in 1..2$ might differ. 
%
Thus lower and upper bounds do not necessarily coincide.

%
%

\begin{example}\label{ex:disj-context}
  Consider 
  $\varphi_7 \deftxtS \Um{r}{\bigl(\Disj{\Um{s}{\F}}{\Um{a}{\F}}\bigr)} \wedge \Um{s}{\F}$, a slight modification on $\varphi_2$.
  A representative monitor for $\varphi_7$ can reject violating systems that exhibit both trace prefixes $ra$ and $rs$, but it can also reject others exhibiting the single prefix $s$ via the subformula \Um{s}{\F}. 
  %
  This is problematic since our violating trace estimation needs to universally quantify over all systems.
  %
  \exqed
\end{example} 

Recursive formulae 
complicate further the calculation of the executions required from the 
disjunctions present in a formula.

\begin{example}\label{ex:no-upperbound} 
  $\varphi_8$ 
  is a variation on $\varphi_4$, stating that 
  \textsl{``if the system can allocate memory, then (i) it cannot also perform a close action and (ii) this property is invariant for all the states reached after servicing received queries.''} 
  \begin{align*} 
    \varphi_8 &\deftxtS 
    \Max{\XV}{\bigl( \Em{a}{\T} \Longrightarrow (\Um{c}{\F} \wedge \Um{r}{\Um{s}{\XV}}) \bigr)} 
    \\
    &\qquad\qquad 
    \ \equiv \ \Max{\XV}{\bigl( \Um{a}{\F} \vee (\Um{c}{\F} \wedge \Um{r}{\Um{s}{\XV}} ) \bigr)}
  \end{align*}
  %
  It contains one disjunction and $\mV_7 \deftxt \recX{\bigl( a.\no \oplus (r.s.\XV \otimes c.\no)\bigr)}  {=} \synth{\varphi_8}$ can correctly monitor for it with no fewer than two trace prefixes.
  %
  \Eg
  $\pV_1 \deftxt \recX{\bigl(r.s.\XV + (a.\XV + c.\nl)\bigr)}$
  from \Cref{ex:rechml} violates $\varphi_8$ and $\mV_7$ can detect this 
  via the size-2 history $\{ a, c \} \subseteq T_{\pV_1}$. 
  %
  %
  But this cannot be said for the violating system
  $\pV_{11} \deftxt a.\nl \!+\! r.s.(a.\nl \!+\! c.\nl)$. 
  %
  Since $\pV_{11}\wtraSN{c}$, 
  monitor $\mV_7$ cannot use the previous size-2 history 
  and instead requires the size-3 history, $\{ a, rsa, rsc\} \subseteq T_{\pV_{11}}$.
  %
  Similarly, the violating system $\pV_{14} \deftxt a.\nl \!+\! r.s.(a.\nl \!+\! r.s.(a.\nl \!+\! c.\nl))$ 
  can only be detected 
  via a history containing the traces $\{a, rsa, rsrsa, rsrsc\}$.
  \exqed
\end{example}

\Cref{ex:no-upperbound} 
illustrates how execution upper bounds cannot be easily determined 
from the structure of a formula. 
However, the calculation of execution lower bounds from the formula structure is attainable. 
For instance, the lower bound for a conjunction \Conj{\varphi_1}{\varphi_2} would be the least bound between the lower bounds of  $\varphi_1$ and $\varphi_2$ respectively.  
Crucially,  history \emph{lower} bounds are invariant \wrt recursive formula unfolding. 

\begin{example}\label{ex:lowerbound-invariant}
Recall $\varphi_8$ from \Cref{ex:no-upperbound} 
with a history lower bound of size 2, which is equal to the number of disjunctions 
in  $\varphi_8$ plus 1 (as argued in \Cref{ex:intuition-bounds}).
By the semantics in \Cref{fig:syntax-semantics}, the same 
systems 
also violate the unfolding of $\varphi_8$, 
\ie
\begin{align*}
  \varphi_8' &\deftxtS 
  \Um{a}{\F} \vee \bigl(\Um{c}{\F} \wedge \Um{r}{\Um{s}{(\Max{\XV}{( \Um{a}{\F} \vee (\Um{c}{\F} \wedge \Um{r}{\Um{s}{\XV}} ))})}} \bigr) \\ 
  &\qquad \qquad =
  \Um{a}{\F} \vee (\Um{c}{\F} \wedge \Um{r}{\Um{s}{\varphi_8}} )
\end{align*}
since 
$\varphi_8\steq\varphi'_8$. 
A naive analysis would conclude that $\varphi_8'$ contains 2 disjunctions, 
thereby requiring histories of 
size 3.
%
But a compositional approach based on 
\Cref{ex:disj-context} 
 would 
allow us to conclude that 
lower bounds of size 2 suffice.  
%
To reject a violating SUS for $\varphi_8'$, trace evidence is needed to determine violations for \emph{both} sub-formulae \Um{a}{\F} and \Conj{\Um{c}{\F}}{\Um{r}{\Um{s}{\varphi_8}}}.
%
Whereas 1 trace suffices to reject \Um{a}{\F}, 
%
determining the lower bounds for rejecting \Conj{\Um{c}{\F}}{\Um{r}{\Um{s}{\varphi_8}}} amounts to calculating the \emph{least} lower bound required to reject either \Um{c}{\F} or \Um{r}{\Um{s}{\varphi_8}}. 
Since rejecting \Um{c}{\F} requires only 1 trace, 
%
the total lower bound is that of $1+1=2$ traces, 
which is equal to that 
of $\varphi_8$.
\exqed 
\end{example}  

The function \lb{-} 
formalises the calculation of history lower bounds based on the syntactical analysis of formulae. 

\begin{definition}\label{def:lb}
  $\lb{-}: \sHMLwDet \map \Nat$ is defined as follows:
  \begin{align*}
    &  \lb{\F} \deftxt 0  
    && \lb{\Max{\XV}{\varphi}} \deftxt \lb{\varphi} 
    \qquad 
    \lb{\Um{\act}{\varphi}} \deftxt \lb{\varphi} \\
    &  \lb{\T} \deftxt \infty 
    && \lb{\Conj{\varphi}{\psi}} \deftxt \textsl{min}\{\lb{\varphi}, \lb{\psi}\} \\
    & \lb{\XV} \deftxt \infty 
    && \lb{\Disj{\varphi}{\psi}} \deftxt \lb{\varphi} + \lb{\psi} + 1    
    \tag*{\exqed}  
  \end{align*}
\end{definition}

There is one further complication 
when calculating the number of trace prefixes required from the syntactic structure of formulae.
%
Our implicit assumption has been that, for disjunctions \Disj{\varphi_1}{\varphi_2}, the incorrect system behaviour described by $\varphi_1$ and $\varphi_2$ is distinct.
Whenever this is not the case, formulae do not observe the lower bound proposed above since $\varphi_1$ and $\varphi_2$ might be violated by 
common trace prefixes. 

\begin{example}\label{ex:not-sparse}
  Although  analysing $\varphi_9\deftxt \Disj{\Um{r}{\F}}{\Um{r}{\Um{s}{\F}}}$ syntactically gives the lower bound 2, $\mV_8\deftxt r.\no \!\oplus\! r.s.\no \!=\! \synth{\varphi_9}$ rejects all violating 
  systems 
  with the single prefix $rs$.
  \hfill$\exqed$
\end{example}

We limit our calculations to a subset of \recHML\ ruling out overlapping violating behaviour across disjunctions. 
%
\sHMLnf (below) combines universal modalities and disjunctions into one construct, $\bigvee_{i \in I}\Um{\act_i}{\varphi_i}$, to represent the formula $\Um{\act_1}{\varphi_1} \vee \cdots \vee \Um{\act_n}{\varphi_n}$ for the finite set index $I {=} \{ 1,\ldots, n\}$.  

\begin{definition}\label{def:shmlnf}
  $\sHMLnf\subseteq\recHML$ is defined as: 
  %
  \begin{align*}
    \varphi,\psi {\in} \sHMLnf 
    &\bnfdef\!  \T 
    \!\bnfsepp\!  \F 
    \!\bnfsepp\!  \Conj{\varphi}{\psi} 
    \!\bnfsepp\!  \bigvee\nolimits_{\!\!i \in I}\Um{\act_i}{\varphi_i} 
    \!\bnfsepp\!  \Max{\XV}{\varphi} 
    \!\bnfsepp\! \XV 
  \end{align*}
  where $\forall i,j \in I$, we have $i \neq j$ implies $\act_i \neq \act_j$. 
  \hfill$\exqed$
\end{definition}

 
%
%
%
%
%


To faciliate the statement and establishment of results on history lowerbounds, we define an \emph{explicit} witness-based violation relation $\hst {\viol} \varphi$ that avoids the existential quantifications over SUS histories of \Cref{def:full-sound,def:full-complete}. 
The new judgement $\hst {\viol} \varphi$  corresponds to $\pV {\notin} \evalE{\varphi}$ whenever $\hst {\subseteq} T_\pV$. 
%
%

\begin{definition}\label{def:viol-relation}
  Given a predicate on \TAct denoted as \Det, the \emph{violation relation}, denoted as \viol, is the least relation 
  of the form 
  $(\Hst\times\Bool \times \sHMLwDet)$ 
  satisfying the 
  rules 
  %
  \begin{mathpar}[\small]
    \inferrule[\rtitSS{vF}]
      { \hst \neq \emptyset }
      { (\hst, \odF) \viol \F }
    \quad
    \inferrule[\rtitSS{vMax}]
      { (\hst, \odF) \viol \varphi\subS{\Max{\XV}{\varphi}}{\XV} }
      { (\hst, \odF) \viol \Max{\XV}{\varphi} }
    \quad
    \inferrule[\rtitSS{vAndL}]
      { (\hst, \odF) \viol \varphi }
      { (\hst, \odF) \viol \Conj{\varphi}{\psi} }
    \and
    \inferrule[\rtitSS{vOr}]
      { (\hst, \btrue) \viol \varphi \quad (\hst, \btrue) \viol \psi }
      { (\hst, \btrue) \viol \Disj{\varphi}{\psi} }
    \and
    \inferrule[\rtitSS{vAndR}]
      { (\hst, \odF) \viol \psi }
      { (\hst, \odF) \viol \Conj{\varphi}{\psi} }
  \end{mathpar}
    \begin{mathpar}[\small]
    \inferrule[\rtitSS{vUmPre}]
      { \hstt{=}\sub{\hst,\iact} \quad \odF' {=} \odF {\wedge} \detAct{\act} \quad (\hstt, \odF') \viol \Um{\act}{\varphi}}
      { (\hst, \odF) \viol \Um{\act}{\varphi} }
    \and
    \inferrule[\rtitSS{vUm}]
      { \hstt{=}\sub{\hst,\act} \quad \odF' {=} \odF {\wedge} \detAct{\act} \quad (\hstt, \odF') \viol \varphi}
      { (\hst, \odF) \viol \Um{\act}{\varphi} }
  \end{mathpar} 
  We read ``\hst\ violates $\varphi$'', $\hst {\viol} \varphi$,  when $(\hst,\btrue) \viol \varphi$.
  %
  \exqed
\end{definition}

\Cref{thm:viol-corr-notin} shows that whenever a system \pV\ produces a history \hst\ that violates a formula $\varphi$, \ie $\hst \viol \varphi$, then 
$\pV$ must also violate it, \ie $\pV\notin\evalE{\varphi}$ (for \emph{arbitrary} ILTSs).  
To show correspondence in the other direction, \Cref{thm:viol-complete}, we need to limit ILTSs to deterministic internal actions.
The reason for this is, once again, the set of systems such as $\pV_{10}$ from \Cref{ex:shmlwDet} for which there is \emph{no} history $\hst \subseteq T_{\pV_{10}}$ such that $\hst \viol \varphi_2$, even though $\pV_{10}\notin \evalE{\varphi_2}$.  
%


\begin{restatable}{theorem}{correspondence}\label{thm:viol-corr-notin}
  For all 
  formulae $\varphi\in\sHMLwDet$, 
  if $\bigl(\exists \hst\subseteq\hstpV$ such that $\hst\viol \varphi\bigr)$ then $\pV\notin\evalE{\varphi}$. 
  \qed   
\end{restatable}

\begin{restatable}{theorem}{correspondenceComplete}\label{thm:viol-complete}
  Suppose $\Det(\iact) {=} \btrue$ for all $\iact {\in} \IAct$. 
  For all 
  $\varphi{\in}\sHMLwDet$, 
  if $\pV{\notin}\evalE{\varphi}$ then $\bigl(\exists \hst{\subseteq}\hstpV$ \sth\ $\hst\viol \varphi\bigr)$. 
  \qed   
\end{restatable}



The new judgment allows us to state and verify that disjunction sub-formulae must be violated by \emph{disjoint} 
histories.   
%


\begin{restatable}{proposition}{shmlnfdisjhst}\label{prop:shmlnf-disj-hst}
  For all 
  $\Disj{\varphi}{\psi}\in\sHMLnf$,
  if $\hst\viol\Disj{\varphi}{\psi}$ then $\hst=\hst'\!\uplus\!\hst''$ such that $\hst'\viol\varphi$ and $\hst''\viol\psi$.
  \qed
\end{restatable}

\Cref{thm:lowerbounds} establishes a lower bound on the 
trace prefixes required to detect violations for \sHMLnf formulae. 
%


\begin{restatable}[Lower Bounds]{theorem}{lowerbounds}\label{thm:lowerbounds}
  For all 
  $\varphi\in\sHMLnf$ and 
  $\hst\in\Hst$,  
  if $\hst \viol \varphi$ then $|\hst| \geq \lb{\varphi}+1$.
  \qed
\end{restatable}

\begin{example}
    Following \Cref{thm:lowerbounds},  $\varphi_2,\varphi_4,\varphi_8\!\in\!\sHMLnf$ cannot be violated with fewer than 2 trace prefixes since $\lb{\varphi_2}\!=\!\lb{\varphi_4}\!=\lb{\varphi_8}\!=\!1$.
  \exqed       
\end{example}

\Cref{thm:lowerbounds} 
also provides 
a simple syntactic check to determine whether 
\sHMLnf\ formulae are worth
monitoring for, according to \Cref{def:full-mon}. 
%
\Cref{cor:lb-inf} 
shows that whenever $\lb{\varphi}\!=\!\infty$,
formula $\varphi$ is always satisfied, \ie 
violations for it can \emph{never} be detected, regardless of the system being runtime verified. 


\begin{restatable}{corollary}{lbinf}\label{cor:lb-inf}
  $\lb{\varphi\in\sHMLnf}\!=\!\infty$ 
  implies 
  $\forall \hst \cdot \hst \nviol \varphi$.
  \qed
\end{restatable}

%

Finally, we note that 
although a minimum of $n$ trace prefixes might be required by \Cref{def:lb} for analysis, the SUS might need to be executed \emph{more} than $n$ times to obtain these prefixes.
%
Intuitively, this is caused by redundancies in the monitors and the manner in which said monitors record trace prefixes, as illustrated in \Cref{ex:overlapping-traces}.
%
%
%

\begin{example}\label{ex:overlapping-traces}
  %
  %
  Assuming $\detAct{a} {=} \btrue$, consider $\varphi_{10}$, describing the property \textsl{``after any number of serviced queries interspersed by 
  sequences of memory allocations, a system that can allocate memory cannot also perform a close action.''}
  \begin{align*}
    \varphi_{10} 
    &\deftxtS 
    \Max{\XV}{\bigl(\Um{r}{\Um{s}{\XV}} \wedge \Um{a}{\XV} \wedge (\Um{a}{\F} \vee \Um{c}{\F})\bigr)}
  \end{align*}
  When synthesising $\synth{\varphi_{10}}$, we get monitor $\mV_2$ from \Cref{ex:inst-comp}.
  %
  The system $\pV_{13} \deftxt \recX{r.s.\XV \!+\! a.X \!+\! a.c.\nl}$ violates $\varphi_{10}$, 
  and $\mV_2$
  can reject it via the history $\hst = \{ rsaa, rsac\} \subseteq T_{\pV_{13}}$,
  in line with \Cref{thm:lowerbounds} 
  since $\lb{\varphi_{10}}\!+\!1 = 2$ trace prefixes.
  However, the incremental manner with which traces are aggregated (\Cref{sec:mon-setup}) requires that, whenever $rsaa \in \hst$, 
  then $rsa \in \hst$ as well.
  This is due to the fact that for the 
  trace $rsa\cdots$, we always have $\sysH{\pV_{14}}{(\epsilon,\mV_2)}{\emptyset} \wtraS{rsa} \sysH{\pV'_{14}}{(rsa,\no)}{\emptyset}$ during the first monitored execution.
  %
  Thus, although 2 prefixes are sufficient to detect a violation, the operational mechanism for aggregating the traces for analysis forces us to observe at least \emph{3} SUS executions 
  to gather the necessary traces for analysis.
  %
  \hfill$\exqed$
\end{example}


\section{Related Work}\label{sec:related-work}

Various bodies of work employ monitors over multiple runs for RV purposes.
%
The most prominent 
target \emph{Hyperproperties}, \ie properties describing sets of traces called \emph{hypertraces}, 
used to describe safety and privacy requirements~\cite{Clarkson2010}.
Finkbeiner \etal~\cite{Finkbeiner2019} investigate the monitorability of hyperproperties expressed in HyperLTL \cite{DBLP:conf/post/ClarksonFKMRS14} and identify three classes for monitoring hypertraces: 
the bounded sequential, the unbounded sequential and the parallel classes.
%
%
They also develop a monitoring 
tool~\cite{DBLP:conf/tacas/FinkbeinerHST18} that analyses 
hypertraces sequentially by converting an alternation-free HyperLTL formula into an alternating automaton 
that is executed over 
permutations of the observed traces.
%
They show that deciding monitorability for alternation-free HyperLTL formulae in this class is PSPACE-complete but undecidable in general.
Our 
setup fits their unbounded sequential class because monitors receive each trace in sequence, and a SUS may exhibit an unbounded number of traces. 
Agrawal \etal~\cite{DBLP:conf/csfw/AgrawalB16} give a semantic characterisation for monitorable HyperLTL hyperproperties called $k$-safety. 
They also identify syntactic HyperLTL fragments and show they are $k$-safety properties, backed up by a monitor synthesis algorithm that generates a combination of petri-nets and $LTL_3$ monitors~\cite{DBLP:journals/tosem/BauerLS11}.     
%
%
Stucki \etal~\cite{SSSB21} show that many properties in HyperLTL 
involving 
quantifier alternation cannot be monitored for.  
%
%
%
%
%
They also present a 
methodology for properties with one alternation by combining static verification and RV: the static part extracts information about the set of traces that the SUS can produce (\ie branching information about the number of traces in the SUS, expressed as a symbolic execution tree) that is used 
by monitors to convert quantifications into $k$-(trace)-quantifications.
%

Despite the similarities of using multi-run monitoring, these works differ from ours in a number of ways. 
For instance, the methods used are very different. 
Our monitor synthesis algorithm is directly based on the formula syntax and does not rely on auxiliary models such as alternating automata or petri-nets, which facilitates syntactic-based proofs. 
%
The results presented are also substantially different.
%
Although~\cite{DBLP:conf/csfw/AgrawalB16,SSSB21} prove that their monitor synthesis algorithm is sound, neither work considers completeness results,
maximality 
or execution lower bound estimation.
More importantly, our target logic, \recHML, is intrinsically different from (linear-time) hyperlogics since it (and other branching-time logics) is interpreted over LTSs, 
whereas hyperlogics are defined over sets of traces. 
%
which inherently coarser than an LTSs. 
For instance, 
the 
systems $a.b.\nl + a.c.\nl$ and $a.(b.\nl + c.\nl)$ are described by different LTSs but have an identical trace-based model, \ie 
$\{ ab,ac\}$;
this was a major source of complication for our technical development.
%
Even 
deterministic LTSs 
where the system $a.b.\nl + a.c.\nl$ is disallowed,
it remains unclear how the two types of logics correspond. 
%
For one, hyperlogics employ existential and universal quantifications over traces, which are absent from our logic. 
%
%
%
If we had to normalise these differences (\ie no trace quantifications), a reasonable mapping would be to take a a linear-time interpretation, $\evalE{\varphi}_\textsc{LT}$~\cite{AcetoAFIL19:monitorability-linear,AcetoAFIL21:sosym} for every \recHML branching-time property $\varphi$, and require it to hold for all of its traces: 
%
For all 
$\varphi$ and deterministic systems \pV, we would 
then expect 
$\pV {\in} \evalE{\varphi}$ iff 
$T_\pV {\in} \evalE{\varphi}_\textsc{LT}$.
But even this correspondence fails,  
%
\eg 
\Disj{\Um{a}{\F}}{\Um{b}{\F}}, describes \emph{systems} that cannot perform both $a$ and $b$ actions and $a.\nl + b.\nl$ clearly violates it.  
However, with a linear-time interpretation, this formula denotes a \emph{tautology}: it is satisfied by \emph{all} traces since they are necessarily either not prefixed with an $a$ action or with a $b$ action. 
There are, however, notable similarities between our history evaluation (\Cref{fig:proof-system}) and team semantics for temporal logics \cite{krebs_et_al:LIPIcs:2018:9592,DBLP:conf/fsttcs/VirtemaHFK021}, and  
%
this relationship is worth further investigation.


The closest work to ours is~\cite{AceAFI:18:FOSSACS}, where Aceto \etal give a framework to extend the capabilities of monitors.
They study monitorability under a grey-box assumption where, at runtime, a monitor has access to additional SUS information, linked to the system's states, in the form of decorated states.
The additional state information is parameterised by a class of \emph{conditions} that represent different situations, such as access to information about that state gathered from previous system executions.
Other works have also examined how to use prior knowledge about the SUS to extend monitorability in the linear-time and branching-time settings, \eg~\cite{DBLP:conf/csl/AcetoAFIL21,HS20}.
In contrast, we treat the SUS as a black-box.
Multiple traces are also used to runtime verify traces with imprecise event ordering~\cite{DBLP:conf/rv/WangASL11,DBLP:conf/tacas/ChenR09,DBLP:conf/fm/BarringerFHRR12,DBLP:conf/sefm/AttardF17} due to interleaved executions of components.
%
%
Parametric trace slicing~\cite{DBLP:conf/tacas/ChenR09,DBLP:conf/fm/BarringerFHRR12} infers additional traces from a trace with interleaved events 
by traversing the original trace and dispatching events to the corresponding slice. 
Attard \etal~\cite{DBLP:conf/sefm/AttardF17} partition the observed trace at the instrumentation level by synthesising monitors 
attached to specific system components; they 
hint at how this could enhance the monitoring expressive power for certain properties but do not prove any monitorability results.
Despite their relevance, 
all traces 
in~\cite{DBLP:conf/rv/WangASL11,DBLP:conf/tacas/ChenR09,DBLP:conf/fm/BarringerFHRR12,DBLP:conf/sefm/AttardF17} 
are 
extracted from a \emph{single} execution.
%

In \cite{ABRAMSKY1987225}, Abramsky studies testing on multiple, yet finite, copies of the same system,
combining the information from multiple runs.
%
%
%
%
%
Our approach differs in three key aspects. 
Firstly, our multiple executions correspond to creating multiple copies of the system from its initial state; Abramsky allows copies to be created at \emph{any point} of the execution.
Secondly, tests are composed using parallel composition, can steer the execution of the SUS and can detect refusals. 
In contrast, our monitors are composed using an instrumentation relation: they are passive and their verdicts are evidence-based (\ie what happened, not what could \emph{not} have happened). 
Third, the visibility afforded by monitor instrumentation considered in this work is larger than that obtained via parallel composition.  
 
Silva \etal~\cite{DaSilvaMelo09:MC-RV} investigate combining traces produced by the same system 
to create temporal models that approximates the SUS's behaviour which can 
then be used to 
model check for branching-time properties.
%
This approach is \emph{not} sound as the generated model may violate properties that are not violated by the actual system.   
%
The authors advise using their 
approach as a 
complement to software testing
to suggest possible problems.

\section{Conclusion}\label{sec:conclusion}

We propose a framework to systematically extend
RV to 
verify branching-time properties.
This is in sharp contrast to most research on RV, which centers around monitoring linear-time properties~\cite{BartocciFFR18,DBLP:journals/jlp/LeuckerS09}.
As 
shown in~\cite{AcetoAFIL19:monitorability-linear}, the class of monitorable linear-time (regular) properties is 
syntactically larger than 
that of monitorable branching-time properties, 
explaining, in part, why the linear-time setting appears less restrictive when 
runtime verified.
For instance, linear-time properties that are monitorable for violations are closed under disjunctions, $\varphi {\vee} \phi$, and existential modalities, $\Em{\act}{\varphi}$, as 
these 
can be 
encoded in an effective, if not efficient, manner~\cite{AcetoAFIL19:monitorability-linear,Aceto2019cost}, albeit in a setting with finite sets of actions.
In contrast, disjunctions and existential modalities in a branching-time setting cannot be encoded in terms of other \recHML\ constructs.
%

We show that these limitations can be mitigated by observing multiple system executions. 
%
Our results demonstrate that monitors can extract sufficient information over multiple runs to correctly detect the violation of a class of branching-time properties that may contain disjunctions (\Cref{thm:full-monitorability}).
We also prove that the monitorable fragment $\sHMLwDet$ (\Cref{def:shmlwDet}) is maximally expressive.
In particular, every property that can be monitored correctly using our monitoring framework can always be expressed as a formula in $\sHMLwDet$.
Such a syntactic characterisation of monitorable properties is useful for tool construction.
It is worth pointing out that an implementation based on our theoretical framework could relax 
assumptions 
used only to attain completeness and maximality results; \eg 
%
instead of assuming that all internal actions 
are deterministic, a tool 
could adopt a pragmatic stance and simply stop monitoring 
as soon as a non-deterministic internal action is encountered, which would still yield a sound (but incomplete) monitor.    
To validate the realisability of our multi-run monitoring RV framework, we outline a possible instantiation to actor-based systems. 
%
We also show that the number of expected runs required to effect the runtime analysis can be calculated from the structure of the formula 
being verified (as opposed to 
other means~\cite{SSSB21}); see \Cref{thm:lowerbounds}. 
We are unaware of similar results in the RV literature.
%


\paragraph*{Future Work}
We plan to investigate how our results can be extended 
by considering more of a grey-box view of the system, in order to combine our machinery with techniques from existing work, such as 
that of Aceto \etal~\cite{AceAFI:18:FOSSACS}. 
We will also study strategies to optimise the collection of relevant SUS traces. 
Depending on the application, one might seek to either maximize the information collected from every execution 
(\eg by continuing to monitor the same execution 
after a trace prefix is added to the history)
or 
minimize the runtime during which the monitor is active. 
%
%
This investigation will be used for tool construction, possibly by extending existing (single-run) open-source monitoring tools for \recHML such as \textsf{detectEr}~\cite{DBLP:conf/forte/AttardAAFIL21,DBLP:conf/coordination/AcetoAAEFI22} that already target actor systems.
%
%
%
We also 
plan to extend our techniques to other graph-based formalisms such as Attack/Fault Trees~\cite{Schneier99:AttackTrees,RuijtersS15:FaultTrees,DBLP:conf/esorics/AudinotPK17,DBLP:conf/icics/Kammuller18} used in cybersecurity, which often necessitate verification at runtime.

\bibliographystyle{IEEEtran}
\bibliography{bibliography}

\appendices


\section{LTS Properties}
\label{sec:lts-properties}

We prove some general results about the LTS of \Cref{sec:preliminaries} and give the standard CCS notation (\Cref{def:ccs}), which is often used to describe systems. 

\begin{definition} \label{def:ccs}
  CCS processes~\cite{DBLP:books/daglib/0067019} are inductively defined by the grammar \Prc\ below:
  \begin{mathpar}
    \pV, \pVV \in \Prc 
    \bnfdef \nl \;\;                       
    \bnfsepp \stact.\pV \;\;
    \bnfsepp \pV + \pVV \;\;
    \bnfsepp \recX{\pV} \;\;
    \bnfsepp \XV  
  \end{mathpar} 
  The transition relation is defined as the least relation satisfying the rules
  \begin{mathpar}[\small]
  \inferrule[\rtitSS{act}]
    { } 
    { \stact.\pV \traS{\stact} \pV }
  \quad 
  \inferrule[\rtitSS{rec}]
    { } 
    { \recX{\pV} \traS{\tau} \pV\subS{\recX{\pV}}{\XV} }
  \quad 
  \inferrule[\rtitSS{selL}]
    { \pV \traS{\stact} \pV' } 
    { \pV + \pVV \traS{\stact} \pV' }
  \quad 
  \inferrule[\rtitSS{selR}]
    { \pVV \traS{\stact} \pVV' } 
    { \pV + \pVV \traS{\stact} \pVV' }
  \end{mathpar} 
  \exqed
\end{definition}

The first result that we show is \Cref{lemma:det-taus-trace}.

\begin{lemma}\label{lemma:det-taus-trace}
  Whenever $\pV\traS{\tau}\pV'$ and $\pV \wttraS{t} \pV''$ where $t\in\TAct^*$ then either 
  \begin{itemize}
    \item $\tau = \epsilon$ and $\pV' \steq \pVV'$; or 
    \item there exist moves $\pV'\wttraS{t}\pVV$ and $\pV''\traS{\tau}\pVV$. 
  \end{itemize}
\end{lemma}

\begin{proof}
  Follows from the confluence property of our ILTSes: silent ($\tau$)-transitions are confluent \wrt\ other actions (\Cref{sec:preliminaries}). 
\end{proof}

\begin{lemma}\label{lemma:aftTau-trcEq}
  If $\pV\traS{\tau} \pVV$ then $\hstpV = \hstpVV$.
\end{lemma}

\begin{proof}
  Let $\pV\traS{\tau} \pVV$.
  We show that $\hstpV = \hstpVV$ in two parts.
  \begin{itemize}
    \item Suppose $t\in\hstpV$, that is $\pV \wttraS{t} \pV'$ for some $\pV'$. 
          We show $t\in\hstpVV$, that is $\pVV \wttraS{t} \pVV'$ for some $\pVV'$.  
          This required matching move follows from \Cref{lemma:det-taus-trace}.
    \item Suppose $t\in\hstpVV$. We show $t\in\hstpV$, that is $\pV \wttraS{t} \pV'$ for some $\pV'$.
          The required matching move is $\pV\traS{\tau} \pVV \wttraS{t} \pVV'$.  \qedhere 
  \end{itemize} 
\end{proof}

\begin{corollary}\label{cor:aftTau-trcEq}
  If $\pV\wttraS{\;\;} \pVV$ then $\hstpV = \hstpVV$.
  \qed
\end{corollary}

\begin{lemma}\label{lemma:steq-trcEq}
  If $\pV\steq \pVV$ then $\hstpV = \hstpVV$.
\end{lemma}

\begin{proof}
  Suppose $\pV\steq \pVV$.
  We show that $\hstpV = \hstpVV$ in two parts.
  \begin{itemize}
    \item Suppose $t\in\hstpV$, \ie $\pV\wttraS{t}\pV'$ for some $\pV'$. 
    By definition of $\steq$, we know $\exists \pVV'$ such that $\pVV\wttraS{t}\pVV'$ and $\pV'\steq\pVV'$, which means $t\in\hstpVV$. 
    \item Suppose $t\in\hstpVV$. The proof for showing $t\in\hstpV$ is analogous. \qedhere
  \end{itemize}
\end{proof}

\begin{lemma}\label{lemma:det-tEq}
  For all $\tact\in\TAct$, if $\pV \wttraS{\tact} \pV'$ and $\pV \wttraS{\tact} \pV''$ and $\detAct{\tact}=\btrue$ then $T_{\pV'}=T_{\pV''}$.    
\end{lemma}

\begin{proof}
  Suppose that $\pV \wttraS{\tact} \pV'$ and $\pV \wttraS{\tact} \pV''$.
  By definition, $\exists \pVV_1,\pVV_2,\pVV_3,\pVV_4$ such that 
  \begin{align*}
    \pV \wttraS{\;\;} \pVV_1 \traS{\tact} \pVV_2 \wttraS{\;\;} \pV' \quad \text{and} \quad
    \pV \wttraS{\;\;} \pVV_3 \traS{\tact} \pVV_4 \wttraS{\;\;} \pV''
  \end{align*}
  We have to show that $T_{\pV'} = T_{\pV''}$.
  Repeatedly using the property of our ILTS that silent actions are confluent \wrt\ other actions (\Cref{sec:preliminaries}) and the assumption that $\detAct{\tact}=\btrue$, we obtain the dashed transitions in the diagram below.
  \begin{center}
    \begin{tikzpicture}[shorten >=1pt,node distance=2cm,on grid,auto,scale=1]
      \node[] (p1)                                {\pV};
      \node[] (q1) [right=of p1]                  {$\pVV_1$};
      \node[] (q2) [right=of q1]                  {$\pVV_2$};
      \node[] (p2) [right=of q2]                  {$\pV'$};
      \node[] (q3) [below=of p1,yshift=0.5cm]     {$\pVV_3$};
      \node[] (q4) [below=of q3,yshift=0.5cm]     {$\pVV_4$};
      \node[] (p3) [below=of q4,yshift=0.5cm]     {$\pV''$};
      \node[] (r1) [right=of q3]                  {$r_1$};
      \node[] (r2) [right=of r1]                  {$r_2$};
      \node[] (r3) [right=of q4]                  {$r_3$};
      \path[->]
        (p1) edge [double]                 node [below,xshift=0.7cm] {\tiny T} (q1)
        (q1) edge                          node                      {$\tact$} (q2)
        (q2) edge [double]                 node [below,xshift=0.7cm] {\tiny T} (p2)
        (p1) edge [double]                 node [left,yshift=-0.45cm] {\tiny T} (q3)
        (q3) edge                          node [left]                {$\tact$} (q4)
        (q4) edge [double]                 node [left,yshift=-0.45cm] {\tiny T} (p3)
        (q1) edge [double,dashed]          node [left,yshift=-0.45cm] {\tiny T} (r1)
        (q3) edge [double,dashed]          node [below,xshift=0.7cm]  {\tiny T} (r1)
        (r1) edge [dashed]                 node                       {$\tact$} (r2)
        (q2) edge [double,dashed]          node [left,yshift=-0.45cm] {\tiny T} (r2)
        (r1) edge [dashed]                 node [left]                {$\tact$} (r3)
        (q4) edge [double,dashed]          node [below,xshift=0.7cm]  {\tiny T} (r3)
        (r2) edge [draw=none,below,sloped] node                       {$\steq$} (r3);
    \end{tikzpicture}
  \end{center}
  By \Cref{lemma:steq-trcEq}, we know $T_{r_2} = T_{r_3}$. 
  By \Cref{cor:aftTau-trcEq}, we also know $T_{\pV'} = T_{\pVV_2} = T_{r_2}$ and $T_{\pV''} = T_{\pVV_4} = T_{r_3}$.
  We can thus conclude that $T_{\pV'} = T_{\pV''}$.
\end{proof}

\Cref{prop:harmony} shows the relation between the two forms of weak transitions, namely $\wtraS{\;\;}$ and $\wttraS{\;\;}$. 

\begin{proposition}\label{prop:harmony}
  For all systems $\pV,\pVV{\in}\Prc$, external actions $\act{\in}\EAct$ and internal actions $\iact{\in}\IAct$,
  \begin{enumerate}
    \item if $\pV \wttraS{\;\;} \pVV$ then $\pV \wtraS{\;\;} \pVV$;
    \item if $\pV \wttraS{\iact} \pVV$ then $\pV \wtraS{\;\;} \pVV$;
    \item if $\pV \wtraS{\;\;} \pVV$ then $\pV \wttraS{t} \pVV$ for some $t\in\IAct^*$;
    \item if $\pV \wtraS{\act} \pVV$ then $\pV \wttraS{t\act t'} \pVV$ for some $t,t'\in\IAct^*$;
    \item if $\pV \wttraS{\act} \pVV$ then $\pV \wtraS{\act} \pVV$.
  \end{enumerate}
\end{proposition}

\begin{proof} We prove the above as follows:
  \begin{description}
    \setlength\itemsep{0.5em}
    \item[To prove (1)] straightforward by definition.
    \item[To prove (2)] suppose $\pV \wttraS{\iact} \pVV$.
      By definition, $\exists \pV',\pV''$ such that $\pV \wttraS{\;\;} \pV' \traS{\iact} \pV'' \wttraS{\;\;} \pVV$.
      By (1), we obtain $\pV \wtraS{\;\;} \pV' \traS{\iact} \pV'' \wtraS{\;\;} \pVV$, \ie $\pV \wtraS{\;\;} \pVV$. 
    \item[To prove (3)] suppose $\pV \wtraS{\;\;} \pVV$. 
      The proof proceeds by induction on the number of (strong) transitions $n$.
      For the base case (\ie $n=0$), then $\pV=\pVV$ and $\pV \wttraS{\epsilon} \pVV$.
      For the inductive case (\ie $n=k+1$), 
      then either $\exists \pV'$ such that $\pV\traS{\tau}\pV'\wtraS{\;\;}\pVV$ or
      $\exists \pV',\iact$ such that $\pV\traS{\iact}\pV'\wtraS{\;\;}\pVV$.
      For the first subcase, by the IH, we obtain $\pV' \wttraS{t} \pVV$ for some $t\in\IAct^*$, which implies $\pV' \wttraS{t} \pVV$.
      For the second subcase, by the IH, we obtain $\pV' \wttraS{t} \pVV$, which implies $\pV \wttraS{\iact t} \pVV$. 
    \item[To prove (4)] suppose $\pV \wtraS{\act} \pVV$.
      By definition, $\exists \pV',\pV''$ such that $\pV \wtraS{\;\;} \pV' \traS{\act} \pV'' \wtraS{\;\;} \pVV$.
      By (3), we obtain $\pV \wttraS{t} \pV' \traS{\act} \pV'' \wttraS{t'} \pVV$ for some $t,t'\in\IAct^*$, which means $\pV \wttraS{t\act t'} \pVV$, as required.
    \item[To prove (5)] straightforward by definition and (1).
      \qedhere 
  \end{description}
\end{proof}

We also prove \Cref{prop:behaviour-equiv} (restated below), stating that equivalent systems satisfy the same formulae.

\behaviourEquiv*

\begin{proof}
  Suppose $\pV\in\evalE{\varphi}$ and $\pV \steq \pVV$. 
  By definition, $\pV$ and $\pVV$ are also strongly bisimilar~\cite{2007Rs:m}.
  Our result, $\pVV\in\evalE{\varphi}$, then follows by the well-known result that strong bisimulation preserves formula satisfactions.
\end{proof}


\section{History Analysis}
\label{sec:proof-trees}

%

\begin{remark}\label{remark:non-det}
  Derivations for $\rej{\hst, \mV}$ are not necessarily unique since \Cref{fig:proof-system} allows a level of non-determinism.
  \Eg when $\detAct{r} = \detAct{s} = \btrue$, the judgement $ \rej{ \{ rsa, rsc\}, r.s.a.\no \otimes r.s.c.\no } $ admits two derivations, shown below: \\
  %
    \begin{displaymath}
      \prftree[r]{\rtitSS{parAL}}
      {
        \prftree[r]{\rtitSS{act}}
        { 
          \prftree[r]{\rtitSS{act}}
          { 
            \prftree[r]{\rtitSS{act}}
            { 
              \prftree[r]{\rtitSS{no}}
              { }
              { \rej{ \{ \epsilon \}, \no} }
            }
            { \rej{ \{ a, c\}, a.\no} }
          }
          { \rej{ \{ sa, sc\}, s.a.\no} }
        }
        { \rej{ \{ rsa, rsc\}, r.s.a.\no} }
      }
      { \rej{ \{ rsa, rsc\}, r.s.a.\no \otimes r.s.c.\no } }
    \end{displaymath}  
    \\
    %
    \begin{displaymath}
      \prftree[r]{\rtitSS{parAR}}
      {
        \prftree[r]{\rtitSS{act}}
        { 
          \prftree[r]{\rtitSS{act}}
          { 
            \prftree[r]{\rtitSS{act}}
            { 
              \prftree[r]{\rtitSS{no}}
              { }
              { \rej{ \{ \epsilon \}, \no} }
            }
            { \rej{ \{ a, c\}, c.\no} }
          }
          { \rej{ \{ sa, sc\}, s.c.\no} }
        }
        { \rej{ \{ rsa, rsc \}, r.s.c.\no } }
      }
      { \rej{ \{ rsa, rsc \}, r.s.a.\no \otimes r.s.c.\no } } 
    \end{displaymath}  
    %
  This, however, does not affect our theory.
\end{remark}


\Cref{fig:proof-derivation-1,fig:proof-derivation-2} give the missing proof derivations from \Cref{ex:reject-over-runs}:
Specifically, \Cref{fig:proof-derivation-1} shows that monitor $\mV_1$
 rejects the history $\{ t_1, t_2 \}$ 
where $t_1 = rs\ut a$ and $t_2 = rs\uf c$, \ie \rej{\mV_1,\{t_1,t_2\}}. 
\begin{figure}[h!]
  \small
  \begin{displaymath}
    \prftree[r]{\rtitSS{rec}} 
    { 
      \prftree[r]{\rtitSS{parAL}}
      {
        \prftree[r]{\rtitSS{act}}
        {
          \prftree[r]{\rtitSS{act}}
          {
            \prftree[r]{\rtitSS{rec}}
            {
              \prftree[r]{\rtitSS{parAL}}
              {
                \prftree[r]{\rtitSS{parO}}
                {
                  \prftree[r]{\rtitSS{actPre}}
                  {
                    \prftree[r]{\rtitSS{act}}
                    {
                      \prftree[r]{\rtitSS{no}}
                      { }
                      { \rej{ \{ \epsilon \}, \no} }
                    }
                    { \rej{ \{ a \}, a.\no} }
                  }
                  {
                    \rej{ \{ \ut a,\; \uf c \}, a.\no}
                  }
                }
                {
                  \prftree[r]{\rtitSS{actPre}}
                  {
                    \prftree[r]{\rtitSS{act}}
                    {
                      \prftree[r]{\rtitSS{no}}
                      { }
                      { \rej{ \{ \epsilon \}, \no} }
                    }
                    { \rej{ \{ c \}, c.\no} }
                  }
                  {
                    \rej{ \{ \ut a,\; \uf c \}, c.\no}
                  }
                }
                { \rej{ \{ \ut a,\; \uf c \}, a.\no \oplus c.\no} }
              }
              { \rej{ \{ \ut a,\; \uf c \}, r.s.\mV_1 \otimes (a.\no \oplus c.\no) } }
            }
            { \rej{ \{ \ut a,\; \uf c \}, \mV_1 } }
          }
          { \rej{ \{ s\ut a,\; s\uf c \}, s.\mV_1 } }
        }
        { \rej{ \{ rs\ut a,\; rsc \}, r.s.\mV_1 } }
      }
      { \rej{ \{ rs\ut a,\; rs\uf c \}, r.s.\mV_1 \otimes (a.\no \oplus c.\no) } }
    } 
    { \rej{ \{ rs\ut a,\; rs\uf c  \}, \mV_1 } } 
  \end{displaymath}
  \caption{Proof derivation showing that $\mV_1$ rejects $\{t_1,t_2\}$} 
  \label{fig:proof-derivation-1}
\end{figure}
 
Conversely, \Cref{fig:proof-derivation-2} shows that monitor $\mV_1$ cannot reject with fewer traces, \ie $\neg\rej{\mV_1, \{ t_1 \}}$, since no rule can justify $\rej{\emptyset, \no}$ at $(**)$. 
\begin{figure}[h!]
  \small
  \begin{displaymath}
    \prftree[r]{\rtitSS{rec}} 
    { 
      \prftree[r]{\rtitSS{parAL}}
      {
        \prftree[r]{\rtitSS{act}}
        {
          \prftree[r]{\rtitSS{act}}
          {
            \prftree[r]{\rtitSS{rec}}
            {
              \prftree[r]{\rtitSS{parAL}}
              {
                \prftree[r]{\rtitSS{parO}}
                {
                  \prftree[r]{\rtitSS{act}}
                  {
                    \prftree[r]{\rtitSS{no}}
                    {  }
                    { \rej{ \{ \epsilon \}, \no } }
                  }
                  { \rej{ \{ a \}, a.\no } }
                }
                {
                  \prftree[r]{\rtitSS{actPre}}
                  {
                    \prftree[r]{\rtitSS{act}}
                    {
                      \rej{ \emptyset, \no } \quad \textit{(**)} 
                    }
                    { \rej{ \{ a \}, c.\no } }
                  }
                  {
                    \rej{ \{ \ut a \}, c.\no }
                  }
                }
                { \rej{ \{ \ut a \}, a.\no \oplus c.\no } }
              }
              { \rej{ \{ \ut a \}, r.s.\mV_1 \otimes (a.\no \oplus c.\no) } }
            }
            { \rej{ \{ \ut a \}, \mV_1 } }
          }
          { \rej{ \{ s\ut a \}, s.\mV_1 } }
        }
        { \rej{ \{ rs\ut a \}, r.s.\mV_1 } }
      }
      { \rej{ \{ rs\ut a \}, r.s.\mV_1 \otimes (a.\no \oplus c.\no) } }
    } 
    { \rej{ \{ rs\ut a  \}, \mV_1 } } 
  \end{displaymath}
  \caption{Proof derivation showing that $\mV_1$ does not reject $\{t_1\}$} 
  \label{fig:proof-derivation-2}
\end{figure}


\section{Monitor Correctness Proofs}
\label{sec:mon-correct-proof}

In this section, we give the proofs for the instrumentation and monitor properties of \Cref{sec:mon-correct}.

\subsection{Instrumentation Properties}

The proof of \Cref{prop:trace-veracity} relies on several technical lemmas that help us reason about the structure of the traces $t$ in executing-monitors $(t,\mV)$.

\begin{lemma}\label{lemma:trace-structure-act}
  $(t,\mV)\traSH{\act}(t',\mV')$ implies $t'{=}t\act$.   
\end{lemma}

\begin{proof}
  By rule induction.
  %
  %
\end{proof}

\begin{lemma}\label{lemma:trace-structure-tau}
  If $(t,\mV) \;\traSH{\tau}\; (t',\mV')$ then $t=t'$.  
\end{lemma}

\begin{proof}
  By case analysis. 
  %
  %
\end{proof}

\begin{lemma}\label{lemma:tau-trace-preserved}
  If $(t,\mV) \;(\traSH{\tau})^*\; (t',\mV')$ then $t=t'$.   
\end{lemma}

\begin{proof}
  Follows from \Cref{lemma:trace-structure-tau}.
\end{proof}



\traceveracity*

\begin{proof}
  The proof proceeds by induction on $n$. \\[0.5em]
  \emph{For the base case}, when $n=0$, the result is immediate.\\[0.5em]
  \emph{For the inductive case}, when 
  $n\!=\!k{+}1$, the transitions are as follows: 
  $$\sysH{\pV}{(\epsilon,\mV)}{\hst} \traS{\stact_1} \cdots \traS{\stact_k} \sysH{\pV'}{(t,\mV')}{\hstt} \traS{\stact_{k{+}1}} \sysH{\pV''}{(t',\mV'')}{\hstt'}$$
  We show that $\pV \wttraS{t'} \pV''$.\\
  By the IH and $\sysH{\pV}{(\epsilon,\mV)}{\hst}\traS{\stact_1} \cdots \traS{\stact_k} \sysH{\pV'}{(t,\mV')}{\hstt}$, 
  we obtain that 
  \begin{align}
    \pV\wttraS{t}\pV' \label{eq:veracity-ih}
  \end{align}
  By case analysis, $\sysH{\pV'}{(t,\mV')}{\hstt} \traS{\!\stact_{k{+}1}\!} \sysH{\pV''}{(t',\mV'')}{\hstt'}$ could have been derived via
  several rules:
  \begin{itemize}
    \setlength\itemsep{0.5em}
    %
    \item Using rule \textsc{iNo}, then $\pV''=\pV'$ and $t=t'$, which implies that $\pV'\wttraS{\epsilon}\pV''$.
    By (\ref{eq:veracity-ih}), we conclude that $\pV\wttraS{t}\pV'\wttraS{\epsilon}\pV''$, \ie $\pV\wttraS{t} \pV''$.
    %
    \item Using rule \textsc{iTer}, then $t'=t\act$ and $\pV'\traS{\act}\pV''$ for some $\act\in\EAct$, which implies that $\pV'\wttraS{\act}\pV''$.
    By (\ref{eq:veracity-ih}), we conclude that $\pV\wttraS{t}\pV'\wttraS{\act}\pV''$, \ie $\pV\wttraS{t\act} \pV''$.
    %
    \item Using rule \textsc{iAsS}, then $t=t'$ and $\pV'\traS{\tau}\pV''$, which implies that $\pV'\wttraS{\epsilon}\pV''$.
    By (\ref{eq:veracity-ih}), we conclude that $\pV\wttraS{t}\pV'\wttraS{\epsilon}\pV''$, \ie $\pV\wttraS{t} \pV''$.
    %
    \item Using rule \textsc{iAsI}, then $t'=t\iact$ and $\pV'\traS{\iact}\pV''$ for some $\iact\in\IAct$, which implies $\pV'\wttraS{\iact}\pV''$.
    By (\ref{eq:veracity-ih}), we conclude that $\pV\wttraS{t}\pV'\wttraS{\iact}\pV''$, \ie $\pV\wttraS{t\iact} \pV''$.
    %
    \item Using rule \textsc{iAsM}, then $\pV'=\pV''$ and $(t,\mV')\traSH{\tau}(t',\mV'')$.
    By \Cref{lemma:trace-structure-tau}, we obtain $t=t'$, and since $\pV'=\pV''$, we obtain $\pV'\wttraS{\epsilon}\pV''$.
    Using (\ref{eq:veracity-ih}), we conclude $\pV\wttraS{t}\pV'\wttraS{\epsilon}\pV''$, \ie $\pV\wttraS{t} \pV''$.
    %
    \item Using rule \textsc{iMon}, then $\pV'\traS{\act}\pV''$ and $(t,\mV')\traSH{\act}(t',\mV'')$ for some $\act\in\EAct$.
    By \Cref{lemma:trace-structure-act}, we obtain that $t'=t\act$, and since $\pV'\traS{\act}\pV''$
    we know that $\pV'\wttraS{\act}\pV''$.
    Using (\ref{eq:veracity-ih}), we conclude $\pV\wttraS{t}\pV'\wttraS{\act}\pV''$, \ie $\pV\wttraS{t\act} \pV''$.
    \qedhere
  \end{itemize}
\end{proof}

\subsection{Monitor Properties}

In this section, we give the proof for \Cref{prop:det-mon,prop:irrevocability} from \Cref{sec:mon-correct}.
%
However, we first give a few useful technical results about the executing-monitors of \Cref{fig:syntax-semantics}.

\begin{lemma}\label{lemma:tauL-rep}
  For all $\mVV\in\Mon$, 
  if $(t,\mV) \;(\traSH{\tau})^*\; (t,\mV')$, then 
  $(t,\mV\odot\mVV) \;(\traSH{\tau})^*\; (t,\mV'\odot\mVV)$. 
\end{lemma}

\begin{proof}
  By induction on the number of $\tau$-transitions.
\end{proof}


\Cref{prop:tau-noAct} below asserts that a monitor that $\tau$-transition cannot transition along other actions. 

\begin{restatable}[$\tau$-Race Absence]{lemma}{taunoact}\label{prop:tau-noAct}
  If $(t, \mV) \traSH{\tau} (t, \mVV)$ then $(t, \mV) \traSHN{\act}$ for all $\act{\in}\EAct$.
  \qed
\end{restatable}

\begin{proof}
  Proof is straightforward by case analysis.
\end{proof}


\Cref{prop:diamond-moves} below assures us that monitor behaviour is confluent \wrt\ $\tau$-moves. 
This allows us to equate monitor states up to $\tau$-transitions.


\begin{restatable}[$\tau$-confluence]{proposition}{diamondmoves}\label{prop:diamond-moves}
  If 
  $(t,\mV) \traSH{\tau} (t,\mV')$ and 
  $(t,\mV) \traSH{\tau} (t,\mV'')$, 
  there exist 
  moves $(t,\mV') \,(\traSH{\tau})^*\, (t,\mVV)$ and $(t,\mV'') \,(\traSH{\tau})^*\, (t,\mVV)$ for some $\mVV\in\Mon$.  
\qed
\end{restatable}

\begin{proof}
  The proof proceeds by induction on $(t,\mV) \traSH{\tau} (t,\mV')$.
  %
  \begin{itemize}
    \setlength\itemsep{0.5em}
    \item Case \textsc{mVrP1}. We have $(t,\no \odot \mVV) \traSH{\tau} (t,\mVV)$ where $t\in\hst$.
    The second transition $(t,\no \odot \mVV) \traSH{\tau} (t,\mV'')$ could have been derived in two ways:
    \begin{itemize}
      \item Using rule \textsc{mVrP1}, \ie $(t,\no \odot \mVV) \traSH{\tau} (t,\mVV)$, which requires 0 matching moves.
      \item Using rule \textsc{mTauR}, \ie $(t,\no \odot \mVV) \traSH{\tau} (t,\no \odot \mVV')$ and $(t,\mVV) \traSH{\tau} (t,\mVV')$. 
      Since $t\in\hst$, we know $(t,\no\odot\mVV') \traSH{\tau} (t,\mVV')$ by rule \textsc{mVrP1}.
      This and $(t,\mVV) \traSH{\tau} (t,\mVV')$ give the required matching moves.
    \end{itemize}  
    \item Case \textsc{mTauL}. We have $(t,\mVV_1 \odot \mVV_2) \traSH{\tau} (t,\mVV_1' \odot \mVV_2)$ because $(t,\mVV_1) \traSH{\tau} (t,\mVV_1')$, which implies $\mVV_1{\neq}\no$.
    The transition $(t,\mVV_1 \odot \mVV_2) \traSH{\tau} (t,\mV'')$ could have been derived using either of the following rules:
    \begin{itemize}
      \item Rule \textsc{mVrP1R}, 
      \ie $(t,\mVV_1 \odot \mVV_2) \traSH{\tau} (t,\mVV_1)$ where $\mVV_2=\no$ and $t\in\hst$.
      By \textsc{mVrP1R}, we deduce $(t,\mVV_1' \odot \mVV_2) \traSH{\tau} (t,\mVV_1')$.
      This and $(t,\mVV_1) \traSH{\tau} (t,\mVV_1')$ are the matching moves.    
      \item Rule \textsc{mVrP2R}, 
      \ie $(t,\mVV_1 \odot \mVV_2) \traSH{\tau} (t,\no)$ where $\mVV_2=\no$ and $t\!\notin\!\hst$.
      By rule \textsc{mVrPR2}, we deduce $(t,\mVV_1' \odot \mVV_2) \traSH{\tau} (t,\no)$.
      This and $(t,\no) (\traSH{\tau})^0 (t,\no)$ give the required matching moves.
      \item Rule \textsc{mTauL}, 
      \ie we have $(t,\mVV_1 \odot \mVV_2) \traSH{\tau} (t,\mVV_1'' \odot \mVV_2)$ and $(t,\mVV_1) \traSH{\tau} (t,\mVV_1'')$. 
      By the IH, there exist moves $(t,\mVV_1') (\traSH{\!\tau\!})^* (t,\mVV)$ and $(t,\mVV_1'') (\traSH{\!\tau\!})^* (t,\mVV)$ for $\mVV{\in}\Mon$. 
      The matching moves, 
      $(t,\mVV_1' \odot \mVV_2) (\traSH{\tau})^* (t,\mVV \odot \mVV_2)$ and 
      $(t,\mVV_1'' \odot \mVV_2) (\traSH{\tau})^* (t,\mVV \odot \mVV_2)$, 
      follow by \Cref{lemma:tauL-rep}. 
      \item Rule \textsc{mTauR}, 
      \ie we have $(t,\mVV_1 \odot \mVV_2) \traSH{\tau} (t,\mVV_1 \odot \mVV_2')$ and $(t,\mVV_2) \traSH{\tau} (t,\mVV_1')$.
      The required matching moves, 
      $(t,\mVV_1' \odot \mVV_2) \traSH{\tau} (t,\mVV_1' \odot \mVV_2')$ and
      $(t,\mVV_1 \odot \mVV_2') \traSH{\tau} (t,\mVV_1' \odot \mVV_2')$, 
      follow by rules \textsc{mTauL} and \textsc{mTauR}. 
      \qedhere
    \end{itemize}
  \end{itemize}
\end{proof}

\begin{corollary}\label{cor:diamond-moves}
  If $(t,\mV) \,(\traSH{\tau})^*\, (t,\mV')$ and $(t,\mV) \,(\traSH{\tau})^*\, (t,\mV'')$, then there must exist moves $(t,\mV') \,(\traSH{\tau})^*\, (t,\mVV)$ and $(t,\mV'') \,(\traSH{\tau})^*\, (t,\mVV)$ for some $n\in\Mon$.
\end{corollary}

\begin{proof}
  Follows by repeatedly applying \Cref{prop:diamond-moves}.
\end{proof}

Since, by \Cref{prop:diamond-moves}, we can equate monitor states up $\tau$-transitions, 
we define what it means for monitors to be equivalent up to $\tau$-moves, \Cref{def:tauEq} below.

\begin{definition}\label{def:tauEq}
  Monitors $\mV$ and $\mV'$ are $\tau$-equivalent, denoted as $\mV \tauEq \mV'$, whenever for all $t\in\Trc$, there exists $\mVV \in \Mon$ such that 
  $(t,\mV) \;(\traS{\tau})^*\; (t,\mVV)$ and $(t,\mV') \; (\traS{\tau})^*\; (t,\mVV)$.
\end{definition}

\begin{lemma}\label{lemma:tauEq-equivalence}
  \tauEq\ is an equivalence relation.  
\end{lemma}

\begin{proof}
  Proving \tauEq\ is symmetric and reflexive is straightforward. 
  To prove that \tauEq\ is transitive, suppose that $(t,\mV_1) \tauEq (t,\mV_2) \tauEq (t,\mV_3)$.
  By \Cref{def:tauEq}, we know that there exist monitors $\mVV_1$ and $\mVV_2$ such that:  
  \begin{align*}
    &(t,\mV_1) \;(\traSH{\tau})^*\; (t,\mVV_1) \text{ and } (t,\mV_2) \;(\traSH{\tau})^*\; (t,\mVV_1) \\ 
    &(t,\mV_2) \;(\traSH{\tau})^*\; (t,\mVV_2) \text{ and } (t,\mV_3) \;(\traSH{\tau})^*\; (t,\mVV_2) 
  \end{align*}
  By \Cref{cor:diamond-moves}, we know that there also exists some monitor $\mVV$ such that 
  $(t,\mVV_1) \;(\traSH{\tau})^*\; (t,\mVV)$ 
  and $(t,\mVV_2) \;(\traSH{\tau})^*\; (t,\mVV)$, 
  which implies that 
  $(t,\mV_1) (\traSH{\tau})^* (t,\mVV)$ and $(t,\mV_3) (\traSH{\tau})^* (t,\mVV)$.
  Our result, namely $(t,\mV_1) \tauEq (t,\mV_3)$, follows by \Cref{def:tauEq}.
\end{proof}



\Cref{lemma:act-det} below shows that two $\tau$-equivalent monitors must be equal if they can transition along the same external actions $\act\in\EAct$.
Moreover, the executing-monitors reached after performing that transition are also equal.

\begin{lemma}\label{lemma:act-det}
  If $(t,\mV) \traSH{\act} (t', \mV')$ and $(t,\mVV) \traSH{\act} (t'', \mVV')$ 
  where $(t,\mV) \tauEq (t,\mVV)$, 
  then $\mV=\mVV$ and $\mV'=\mVV'$ and $t'=t''$.
\end{lemma}

\begin{proof}
  Assume $(t,\mV) \traSH{\act} (t', \mV')$ and $(t,\mVV) \traSH{\act} (t'', \mVV')$ where $(t,\mV) \tauEq (t,\mVV)$.
  By \Cref{def:tauEq}, there exists some $\mVV''$ such that $(t,\mV)\;(\traSH{\tau})^*\;(t,\mVV'')$ and $(t,\mVV)\;(\traSH{\tau})^*\;(t,\mVV'')$. 
  But by 
  \Cref{prop:tau-noAct}, we also know $(t,\mV) \traSHN{\tau}$ and $(t,\mVV) \traSHN{\tau}$, which implies that $(t,\mV)\;(\traSH{\tau})^0\;(t,\mVV'')$ and $(t,\mVV)\;(\traSH{\tau})^0\;(t,\mVV'')$, and thus $\mV=\mVV'=\mVV$.

  To show that if $(t,\mV) \traSH{\act} (t', \mV')$ and $(t,\mV) \traSH{\act} (t'', \mVV')$ then $t'=t''$ and $(t',\mV')\tauEq(t'',\mVV')$, we use rule induction on $(t,\mV) \traSH{\act} (t', \mV')$.
  We outline the main cases:
  \begin{itemize}
    \setlength\itemsep{0.5em}
    \item Case \textsc{mEnd}. We have $(t, \stp) \traSH{\act} (t,\stp)$ where $\mV=\stp$. 
    Result follows immediately since the second transition $(t,\stp) \traSH{\act} (t'', \mVV')$ could have only been derived using the rule \textsc{mEnd}, which implies $t''=t$ and $\mV''=\stp$. 
    %
    %
    \item Case \textsc{mPar1}. We have $(t, \mV_1\odot\mV_2) \traSH{\act} (t',\mV_1'\odot\mV_2')$ where $\mV=\mV_1\odot\mV_2$ because $(t, \mV_1) \traSH{\act} (t',\mV_1')$ and $(t, \mV_2) \traSH{\act} (t',\mV_2')$.
    By \Cref{prop:tau-noAct}, we know $(t, \mV_1) \traSHN{\tau}$ and $(t, \mV_2) \traSHN{\tau}$, which implies $\mV_1\neq\no$ and $\mV_2\neq\no$.
    This means that the second transition $(t, \mV_1\odot\mV_2) \traSH{\act} (t'',\mVV')$ could have only been derived by \textsc{mPar1}. 
    Thus, we infer that $\mVV'\!=\!\mVV_1\odot\mVV_2$, $(t, \mV_1) \traSH{\act} (t'',\mVV_1)$ and $(t, \mV_2) \traSH{\act} (t'',\mVV_2)$.
    Our result, $t'=t''$ and $\mV'=\mV''$, follows by the IH.
    \qedhere
    %
    %
  \end{itemize}
\end{proof}

Similarly, $\tau$-equivalent monitors must be equal if they can (weakly) transition with the same trace $u\in\Trc$, in which case the executing-monitors reached are also equal.

\begin{lemma}\label{lemma:det-mon}
  For all $u\!\in\!\Trc$, if $(t, \mV_1) \wtraSH{u} (t_1,\mVV_1)$ and $(t, \mV_2) \wtraSH{u} (t_2,\mVV_2)$ where 
  $(t,\mV_1) \tauEq (t,\mV_2)$, 
  then $t_1\!=\!t_2$ and $(t_1,\mVV_1) \!\tauEq\! (t_2,\mVV_2)$.
\end{lemma}


\begin{proof}
  The proof proceeds by induction on the length $l$ of transitions in $(t, \mV_1) \wtraSH{u} (t_1, \mVV_1)$.
  \begin{itemize}
    \setlength\itemsep{0.5em}
    \item 
    For the \emph{base case}, suppose $l=0$. 
    Then $u=\epsilon$, $\mV_1 = \mVV_1$ and $(t,\mV_2) (\traSH{\tau})^* (t_2,\mVV_2)$.
    By \Cref{lemma:tau-trace-preserved}, we know $t=t_2$.
    By this and \Cref{def:tauEq}, we also know $(t, \mV_2) \tauEq (t_2, \mVV_2)=(t, \mVV_2)$.
    Since $(t,\mV_1)\tauEq (t,\mV_2)$, our result, $(t, \mV_1) \tauEq (t_2,\mVV_2)$, follows via \Cref{lemma:tauEq-equivalence} (transitivity).
    \item 
    For the \emph{inductive case}, suppose $l=k\!+\!1$.
    The transition sequence $(t, \mV_1) \wtraSH{u} (t_1, \mVV_1)$ can be expanded as 
    $$(t, \mV_1) \traSH{\stact} (v_1,\mVV_1') \wtraSH{u'} (t_1, \mVV_1)$$ 
    where $u',v_1\!\in\!\Trc$, $\mVV_1'\!\in\!\Mon$ and $\stact\!\in\!\Act\cup\{ \tau\}$.
    There are two subcases to consider:
    \begin{itemize}
      \setlength\itemsep{0.5em}
      \item When $\stact = \tau$, we have $(t, \mV_1) \traSH{\tau} (v_1, \mVV_1')$ and $u\!=\!u'$, which implies $t \!=\! v_1$ by \Cref{lemma:tau-trace-preserved} and $(t,\mV_1) \tauEq (v_1,\mVV_1')$ by \Cref{def:tauEq}.
      By $(t,\mV_1) \tauEq (t, \mV_2)$ and $(t,\mV_1) \tauEq (v_1,\mVV_1')$ and \Cref{lemma:tauEq-equivalence}, we obtain  
      $(v_1,\mVV_1') \tauEq (t,\mV_2)$. 
      By $(t,\mVV_1') \wtraSH{u} (t_1, \mVV_1)$, the original assumption $(t,\mV_2) \wtraSH{u} (t_2, \mVV_2)$ and IH, we conclude $t_1=t_2$ and $(t_1,\mVV_1) \tauEq (t_2,\mVV_2)$.
      \item When $\stact \!=\! \act{\in}\EAct$, we have $(t, \mV_1) \traS{\act} (v_1, \mVV_1')$ and $u\!=\!\act u'$ for some $u'\in\Trc$.
      The second sequence $(t, \mV_2) \wtraSH{u} (t_2,\mVV_2)$ can be expanded as 
      \begin{center}
        $(t, \mV_2) \;(\traSH{\tau})^*\; (t, \mVV_2') \traSH{\act} (v_2, \mVV_2'') \wtraSH{u'} (t_2,\mVV_2)$
      \end{center} 
      where $v_2\in\Trc$.
      By \Cref{def:tauEq}, we also know $(t,\mV_2) \tauEq (t,\mVV_2')$.
      From this, the original assumption that $(t,\mV_1)\tauEq(t,\mV_2)$ and \Cref{lemma:tauEq-equivalence}, we deduce $(t,\mV_1) \tauEq (t,\mVV_2')$.
      Since $(t,\mV_1) \traSH{\act} (v_1,\mVV_1')$ and $(t,\mVV_2') \traSH{\act} (v_2,\mVV_2'')$ where $(t,\mV_1){\tauEq}(t,\mVV_2')$, 
      we obtain that $\mV_1 = \mVV_2'$ and $\mVV_1'=\mVV_2''$ and $v_1=v_2$ by \Cref{lemma:act-det}.
      Our result, $t_1=t_2$ and $(t_1,\mVV_1)\tauEq(t_2,\mVV_2)$, follows by the IH. 
      \item When $\stact \!=\! \iact{\in}\TAct$, we must have $(t, \mV_1) \traSH{\iact} (v_1, \mVV_1')$ and $u\!=\!\iact u'$ for some $u'\in\Trc$. 
      However, this gives us a contradiction since by the rules in \Cref{fig:mon}, $(t, \mV_1) \traSHN{\iact}$, meaning that this case never arises. 
      \qedhere
    \end{itemize} 
  \end{itemize}
\end{proof}

We can now prove \Cref{def:monDet} from \Cref{sec:mon-correct}, restated below.

\detmon*

\begin{proof}
  Assume that $(t, \mV) \wtraSH{u} (t',\mV')$ and $(t, \mV) \wtraSH{u} (t'',\mV'')$.
  By \Cref{lemma:tauEq-equivalence}, we know $(t,\mV) \tauEq (t,\mV)$.
  By \Cref{lemma:det-mon}, we obtain that $t'=t''$ and $(t',\mV') \tauEq (t'',\mV'')$.
  Our result then follows by \Cref{def:tauEq}.
  %
\end{proof}



We now show monitor rejections are irrevocable in terms of both additional traces, \emph{width}, and longer traces, \emph{length}.

\irrevocability*

\begin{proof}
  The first part follows from \Cref{lemma:length-irrev} below, letting $\odF=\btrue$.
  The second part follows from \Cref{lemma:width-irrev} below, letting $\odF=\btrue$.
\end{proof}



\begin{lemma}\label{lemma:length-irrev}
  \rej{(\hst,t),\odF,\mV} implies \rej{(\hst,tu),\odF,\mV} 
\end{lemma}

\begin{proof}
  The proof proceeds by induction on \rej{(\hst,t),\odF,\mV}.
  
  \begin{itemize}
    \setlength\itemsep{0.5em}
    \item Case \textsc{no}. Follows immediately because $\rej{\hstt,\odF, \no}$ for all $\hstt\neq\emptyset$. 
    %
      
      \item Case \textsc{act}. 
      We know $\rej{(\hst,t), \odF, \act.\mV}$ because $\rej{\hst', \odF',\mV}$ where $\hst'=\sub{(\hst,t), \act}$ and $\odF'=\od\wedge\detAct{\act}$. 
      There are two subcases to consider:
      
      \begin{itemize}
        \item 
        When $t=\act t'$, then $\hst'= (\hst'',t')=\sub{(\hst,t), \act}$ for some $\hst''$. 
        By the IH, we deduce $\rej{(\hst'',t'u), \odF',\mV}$.
        But by definition, we also know $(\hst'',t'u) = \sub{(\hst,tu),\act}$, meaning that $\rej{\sub{(\hst,tu),\act}, \odF',\mV}$. 
        Our result, $\rej{(\hst,tu),\odF, \act.\mV}$, follows by rule \textsc{act}. 
        %
        \item When $t=\actt t'$, we know by definition that $\sub{(\hst,t),\act} = \sub{\hst,\act} = \sub{(\hst,tu),\act}$.
        Our result, $\rej{(\hst,tu),\odF, \act.\mV}$, follows immediately by applying rule \textsc{act}. 
      \end{itemize}
      \item Case \textsc{actPre}. Proof is similar to that for \textsc{act}.  
      \item Case \textsc{parAL}. 
        We know that $\rej{(\hst,t), \odF, \mV'\otimes\mV''}$ because of $\rej{(\hst,t), \odF, \mV'}$. 
        By the IH, we obtain $\rej{(\hst,tu),\odF, \mV'}$. 
        Using rule \textsc{parAL}, we can conclude $\rej{(\hst,tu), \odF, \mV'\otimes\mV''}$. 
      \item Case \textsc{parAR}. Proof is analogous to that for \textsc{parAR}.
      \item Case \textsc{parO}. 
        We know $\rej{(\hst,t), \btrue, \mV'\otimes\mV''}$ because $\rej{(\hst,t), \btrue, \mV'}$ and $\rej{(\hst,t),\btrue, \mV''}$. 
        By the IH, $\rej{(\hst,tu),\btrue,\mV'}$ and $\rej{(\hst,tu),\btrue,\mV''}$. 
        Applying rule \textsc{parO}, we obtain $\rej{(\hst,tu), \btrue, \mV'{\otimes}\mV''}$.
      %
      \item Case \textsc{rec}. 
        We know $\rej{(\hst,t), \odF, \rec{\XV}{\mV}}$ because $\rej{(\hst,t), \odF, \mV\subS{\recX{\mV}}{\XV}}$. 
        By the IH, we obtain $\rej{(\hst,tu),\odF, \mV\subS{\recX{\mV}}{\XV}}$. 
        Our result, $\rej{(\hst,tu), \odF, \rec{\XV}{\mV}}$, follows by rule \textsc{rec}. 
        \qedhere
  \end{itemize}
\end{proof}



\begin{lemma}\label{lemma:width-irrev}
  \rej{\hst,\odF,\mV} implies \rej{\hst\cup\hstt,\odF,\mV}
\end{lemma}

\begin{proof}
  Straightforward by induction on \rej{\hst,\odF,\mV}. 
\end{proof}


\section{Proving Monitorability}
\label{sec:monitorability-proof}

In this section, we prove \Cref{thm:full-monitorability} from \Cref{sec:monitorability}.
This theorem is proven in two steps; first, we show the monitors generated via the synthesis function \synth{-}
are sound, \Cref{prop:sound}, and then we show they are complete, \Cref{prop:complete}.
These rely on a number of results that use the alternative definition for property violations in \Cref{def:viol-relation} as it is easier to establish results with it.
%
Concretely, \Cref{lemma:hst-sound,lemma:hst-complete} below show there is a tight correspondence between the rejected histories, \rej{\hst,\odF,\mV}, and violating histories, $(\hst,\odF) \viol\varphi$.
%

\begin{lemma}\label{lemma:synth-sub}
  For all $\varphi,\psi\in\sHMLwDet$, $\synth{\varphi\subS{\psi}{\XV}} = \synth{\varphi}\subS{\synth{\psi}}{\XV}$  
\end{lemma}

\begin{proof}
  By induction on the structure of $\varphi$. 
\end{proof}

\begin{lemma}\label{lemma:hst-sound}
  For all $\varphi\in\sHMLwDet$, 
  if \rej{\hst,\odF,\synth{\varphi}} then $(\hst,\odF)\viol\varphi$.
\end{lemma}

\begin{proof}
  The proof proceeds by induction on $\rej{\hst, \odF, \synth{\varphi}}$.
  We outline the main cases:
  \begin{itemize}
    \setlength\itemsep{0.5em}
    \item Case \textsc{act}. 
      We know \rej{\hst, \odF, \act.\mV} because \rej{\hstt, \odF',\mV} where
      $\hstt = \sub{\hst, \act}$ and
      $\odF' = \odF \wedge \detAct{\act}$ and  
      $\varphi = \Um{\act}{\psi}$ and $\mV=\synth{\psi}$.
      By the IH, we obtain $(\hstt, \odF') \viol \psi$.
      Our result, $(\hst,\odF) \viol\Um{\act}{\psi}$, follows by rule \textsc{vUm}.
    \item Case \textsc{rec}. 
      We know that $\rej{\hst, \odF, \recX{\mV}}$ because $\rej{\hst, \odF, \mV\subS{\recX{\mV}}{\XV}}$ where $\varphi=\Max{\XV}{\psi}$ and $\mV=\synth{\psi}$.
      By \Cref{lemma:synth-sub}, we also know that 
      $\mV\subS{\recX{\mV}}{\XV} = \synth{\psi}\subS{\synth{\Max{\XV}{\psi}}}{\XV} = \synth{\psi\subS{\Max{\XV}{\psi}}{\XV}}$.
      Using the IH, we then obtain $(\hst,\odF) \viol \psi\subS{\Max{\XV}{\psi}}{\XV}$.
      Our result, $(\hst,\odF) \viol \Max{\XV}{\psi}$, follows by rule \textsc{vMax}. 
      \qedhere
  \end{itemize}
\end{proof}



\sound*

\begin{proof}
  Expanding \Cref{def:full-sound}, we need to show that
  for all $\varphi\in\sHMLwDet$ and $\pV\in\Prc$,  
  \begin{center}
    if $\bigl( \exists \hst\subseteq\hstpV \text{ such that } \rej{\hst, \synth{\varphi}} \bigr)$ 
    then $\pV \notin \evalE{\varphi}$. 
  \end{center}
  Suppose $\exists\hst\subseteq\hstpV$ such that \rej{\hst, \synth{\varphi}}.
  By \Cref{lemma:hst-sound}, letting $\odF=\btrue$, we get $\hst \viol \varphi$. 
  Our result, $\pV\!\notin\!\evalE{\varphi}$, follows by \Cref{thm:viol-corr-notin}.
\end{proof}

\begin{lemma}\label{lemma:hst-complete}
  For all $\varphi\in\sHMLwDet$, 
  if $(\hst,\odF)\viol\varphi$ then \rej{\hst,\odF,\synth{\varphi}}.
\end{lemma}

\begin{proof}
  Follows with a proof similar to that for \Cref{lemma:hst-sound}. 
\end{proof}


\complete*

\begin{proof}
  Suppose that $\detAct{\iact} = \btrue$ for all $\iact\in\IAct$.
  Expanding \Cref{def:full-complete}, we need to show that for all $\varphi\in\sHMLwDet$ and $\pV\in\Prc$, we have that 
  \begin{center}
    if $\pV\notin\evalE{\varphi}$ 
    then $\bigl( \exists \hst\subseteq\hstpV$ such that $\rej{\hst, \synth{\varphi}} \bigr)$
  \end{center}
  Suppose $\pV\notin\evalE{\varphi}$.
  By 
  \Cref{thm:viol-corr-notin}, we know $\exists \hst\subseteq\hstpV$ such that $\hst\viol\varphi$.
  Our result, $\rej{\hst,\synth{\varphi}}$, follow by \Cref{lemma:hst-complete}, letting $\odF=\btrue$.
\end{proof}

We can now show \sHMLw\ is monitorable, \Cref{thm:full-monitorability}.

\fullMonitorability*

\begin{proof}
  Follows from \Cref{prop:sound,prop:complete}, with \synth{\varphi} as the witness correct monitor.
\end{proof}


\section{Proving Maximal Expressiveness}
\label{sec:maximality-proof}

The first step towards showing that $\sHMLwDet$ is maximally expressive, namely \Cref{thm:maximal-expressivity}, 
is to define expressive-completeness \wrt the monitoring setup \Mon\ of \Cref{sec:mon-setup}.

\begin{definition}[Expressive-complete] \label{def:expressive-complete}
  A subset $\cL \subseteq \recHML$ is \emph{expressive-complete} if for all monitors $\mV \in \Mon$, 
  there exists 
  $\varphi \in \cL$ such that \mV\ monitors correctly for it. 
  \qed
\end{definition}

We prove that the language \sHMLwDet is expressive-complete systematically, by concretising the existential quantification of a formula $\varphi$ in \sHMLwDet\ for every monitor $\mV$ in \Mon\ such that \mV\ monitors correctly for it (\Cref{def:full-correct}). 
\Cref{def:synthRev} below formalises a function $\synthRev{-}$ that maps every monitor in \Mon\ to a corresponding formula.
%


\begin{definition}\label{def:synthRev}
  The function $\synthRev{-}: \Mon \map \recHML$ is defined inductively as follows:
  \begin{align*}
    & \synthRev{\no} \deftxt \F  
    && \synthRev{\mV \oplus \mVV} \deftxt \Disj{\synthRev{\mV}}{\synthRev{\mVV}}
    && \synthRev{\act.\mV} \deftxt \Um{\act}{\synthRev{\mV}}  \\
    & \synthRev{\stp\,} \deftxt \T 
    && \synthRev{\mV \otimes \mVV} \deftxt \Conj{\synthRev{\mV}}{\synthRev{\mVV}} 
    && \\
    & \synthRev{\XV} \deftxt \XV 
    && \synthRev{\recX{\mV}} \deftxt \Max{\XV}{\synthRev{\mV}} 
    &&
    \tag*{\qed}
  \end{align*}
\end{definition}

Note that, $\cod{\synthRev{-}} = \recHML$ as, when given arbitrary monitors, we have \emph{no} guarantee that $\synthRev{\mV} = \varphi$ is in \sHMLwDet.
%

\begin{example}
  Recall monitor $\mV_1 \deftxt \recX{(r.s.\XV \otimes (a.\no \oplus c.\no))}$ from \Cref{ex:inst}.
  %
  %
  When $\detAct{r} = \bfalse$, the formula $\synthRev{\mV_1} = \Max{\XV}{\Conj{\Um{r}{\Um{s}{\XV}}}{(\Disj{\Um{a}{\F}}{\Um{c}{\F}})}} = \varphi_4$ is neither monitorable, according to \Cref{def:full-mon}, nor does it belong to $\sHMLwDet$, as shown in \Cref{ex:unmonitorable-disj}.
  This occurs because parallel disjunction monitors prefixed with non-deterministic actions will generate formulas containing disjunctions prefixed with non-deterministic universal modalities.    
  \qed
\end{example}

\Cref{def:monDet} characterises a subset of monitors from \Mon, parametrised by \EAct and the associated action determinacy delineation defined by \Det.
Similar to \Cref{def:shmlwDet}, it employs a flag to calculate deterministic prefixes via rule \textsc{cAct} along the lines of \Cref{fig:proof-system}. 
This is then used by rule \textsc{cOr}, which is only defined when the flag is \btrue.

\begin{definition}\label{def:monDet}
  The judgement $\odF \cons \mV$  for monitors $\mV\in\Mon$ and flag $\odF\in\Bool$ is defined coinductively as the \emph{largest} relation satisfied by the following rules.
  \begin{mathpar}[\small]
    \inferrule[\rtitSS{cM}]
      { \mV\in\{ \stp, \F, \XV \} }
      { \odF \cons \mV }
    \and
    \inferrule[\rtitSS{cAct}]
      { \odF \wedge \detAct{\act} \cons \mV}
      { \odF \cons \act.\mV }
    \and
    \inferrule[\rtitSS{cParA}]
      { \odF \cons \mV \;\; \odF \cons \mVV }
      { \odF \cons \mV \otimes \mVV }
    \and
    \inferrule[\rtitSS{cParO}]
      { \btrue \cons \mV \;\; \btrue \cons \mVV }
      { \btrue \cons \mV\oplus\mVV }
    \and
    \inferrule[\rtitSS{cRec}]
      { \odF \cons \varphi\subS{\rec{\XV}{\mV}}{\XV} }
      { \odF \cons \rec{\XV}{\mV} }
  \end{mathpar}
  The set $\MonDet \deftxt \setof{\mV}{\btrue \cons \mV}$ defines the set of monitors where all parallel disjunctions are prefixed by deterministic external actions (up to recursion unfolding).
  \exqed
\end{definition}

We can show that whenever we limit systems to deterministic internal actions (see \Cref{ex:shmlwDet}), monitor $\mV\in\MonDet$ monitors correctly for the formula $\synthRev{\mV}$.
This relies on \Cref{prop:monDet-to-shmlwDet}, asserting that $\synthRev{\mV}\in\sHMLwDet$ whenever $\mV\in\MonDet$.

\begin{proposition}\label{prop:monDet-to-shmlwDet}
  If $\mV\in\MonDet$ then $\synthRev{\mV}\in\sHMLwDet$.
  \qed
\end{proposition}

\begin{proposition}\label{prop:expressive-complete-monDet}
  Suppose $\detAct{\iact}=\btrue$ for all $\iact\in\IAct$. 
  For all $\mV\in\MonDet$, monitor $\mV$ monitors correctly for $\synthRev{\mV}$.
\end{proposition}

\begin{proof}
  Pick $\mV\in\MonDet$. 
  By \Cref{prop:monDet-to-shmlwDet}, we know $\synthRev{\mV}\in\sHMLwDet$.
  We show that $\mV$ is sound and complete for $\synthRev{\mV}$:
  \begin{description}
    \setlength\itemsep{0.5em}
    \item[To prove soundness,] suppose \rej{\hst, \mV}.
      Since $\synthRev{\mV}\in\sHMLwDet$, we can use \Cref{prop:sound}, letting $\varphi=\synthRev{\mV}$, to obtain that 
      $\pV\notin\evalE{\synthRev{\mV}}$.
    \item[To show completeness,] suppose $\pV\notin\evalE{\synthRev{\varphi}}$.
      Since $\synthRev{\mV}\in\sHMLwDet$ and $\detAct{\iact}=\btrue$ for all $\iact\in\IAct$, we can use \Cref{prop:sound}, letting $\varphi=\synthRev{\mV}$, to obtain that there exists $\hst\subseteq\hstpV$ such that \rej{\hst,\mV}.  
      \qedhere
  \end{description} 
\end{proof}

While we have demonstrated that a formula $\varphi\in\sHMLwDet$ exists for every monitor $\mV\in\MonDet$, we want to establish a stronger result: that a formula $\varphi\in\sHMLwDet$ exists for every monitor $\mV\in\Mon$.
We show this by generating a monitor $\mV$ in $\MonDet$ for each monitor $\mVV$ in $\Mon$ such that $\mV$ and $\mVV$ reject the same histories.
This is done using the function \trnsfM{-}, formalised in \Cref{def:trnsfM} below.

The function \trnsfM{-} employs a flag to compute deterministic prefixes via rule \textsc{tAct}, which is then used by rule \textsc{tParF} to transform parallel disjunction monitors to the inactive monitor \stp\ when the flag is \bfalse.
%
%
Additionally, this function relies on a mapping $\sV\in \Subs : \TVars \map \Mon \times \Bool$.
When the transformation encounters a recursion monitor \recX{\mV} with the flag \odF, the entry $\XV\map\langle \recX{\mV}, \odF \rangle$ is added to $\sV$. 
Recursion variables are unfolded if there is an entry for them in \sV\ and have not already been visited with the current flag (rule \textsc{tTVar3}).


\begin{definition}\label{def:trnsfM}
  Given a predicate on \TAct denoted as \Det, the function 
  $\mathcal{T} : \Mon \times \Bool \times \Subs \map \MonDet$ 
  is the smallest relation satisfied by the following rules. 
  \begin{mathpar}[\small]
    \inferrule[\rtitSS{tNo}]
      { }
      { \trnsfM{\no,\odF,\sV} = \no }
    \and
    \inferrule[\rtitSS{tEnd}]
      {  }
      { \trnsfM{\stp,\odF,\sV} = \stp }
    \and
    \inferrule[\rtitSS{tTVar1}]
      {  \XV\notin\dom{\sV} }
      { \trnsfM{\XV,\odF,\sV} = \XV }
    \and
    \inferrule[\rtitSS{tTVar2}]
      {  \sV(\XV) = \langle \mV, \odF \rangle }
      { \trnsfM{\XV,\odF,\sV} = \XV }
    \quad
    \inferrule[\rtitSS{tTVar3}]
      {  \sV(\XV) = \langle \mV, \odF' \rangle \quad \odF'\neq\odF \quad \trnsfM{\mV,\odF,\sV} = \mVV }
      { \trnsfM{\XV,\odF,\sV} = \mVV }
    \and
    \inferrule[\rtitSS{tRec}]
      { \trnsfM{\mV,\odF,\sV[\XV \mapsto \langle \recX{\mV}, \odF \rangle]} = \mVV }
      { \trnsfM{\recX{\mV},\odF,\sV} = \recX{\mVV} }
    \and
    \inferrule[\rtitSS{tAct}]
      { \trnsfM{\mV,\odF\wedge \detAct{\act},\sV} = \mVV }
      { \trnsfM{\act.\mV,\odF,\sV} = \act.\mVV }
    \and
    \inferrule[\rtitSS{tParA}]
      { \trnsfM{\mV,\odF,\sV} = \mV' \quad \trnsfM{\mVV,\odF,\sV} = \mVV' }
      { \trnsfM{\mV \otimes \mVV,\odF,\sV} = \mV' \otimes \mVV' }
    \and
    \inferrule[\rtitSS{tParOT}]
      { \trnsfM{\mV,\btrue,\sV} = \mV' \quad \trnsfM{\mVV,\btrue,\sV} = \mVV' }
      { \trnsfM{\mV \otimes \mVV,\btrue,\sV} = \mV' \oplus \mVV' }
    \quad
    \inferrule[\rtitSS{tParOF}]
      { }
      { \trnsfM{\mV \otimes \mVV,\bfalse,\sV} = \stp }
  \end{mathpar} \\
  We write $\trnsfM{\mV, \odF, \sV} = \mVV$ whenever there exists a proof derivation satisfying that judgement.
  %
  As a shorthand, we also write \trnsfM{\mV} in lieu of \trnsfM{\mV,\btrue,\emptyset}.  
  %
  \exqed
\end{definition}

\Cref{prop:tnsfm-rej} establishes a correspondence between the monitors \mV\ and \trnsfM{\mV}.
Specifically, these monitors reject the same histories.

\begin{restatable}{proposition}{tnsfmRej}\label{prop:tnsfm-rej}
  For all $\mV\in\Mon$ and $\hst\in\Hst$, \rej{\hst, \mV} iff \rej{\hst, \trnsfM{\mV}}
\end{restatable}

\begin{proof}
  Since the proof is quite involved and relies on several additional results, we prove it in a separate subsection at the end of this section.
\end{proof}

\begin{example}
  Monitor $\mV_1 \deftxt \recX{r.s.\XV \otimes (a.\no \oplus c.\no)}$
  rejects a history \hst\ with flag \btrue, \ie \rej{\hst,\mV}, if and only if its unfolding does, 
  \ie \rej{r.s.\mV_1 \otimes (a.\no \oplus c.\no), \hst, \btrue}.
  When $\detAct{r}=\bfalse$, we can show that $\neg\rej{\hst, r.s.\mV_1}$, which implies 
  $\rej{\hst, a.\no \oplus c.\no}$.
  This means that $\mV_1$ rejects all histories containing the traces $a$ and $c$, corresponding to the histories rejected by the generated monitor $\mV_6$ below.
  \begin{align*}
    &\mV_6 \deftxt \recX{(r.s.(\recX{r.s.\XV \otimes \stp}) \otimes (a.\no \oplus c.\no))} = \trnsfM{\mV_1} \\
    &\varphi_3' \deftxt \max{\XV}{(\Um{r}{\Um{s}{(\Max{\XV}{\Um{r}{\Um{s}{\XV}} \wedge \T})}} \wedge (\Um{a}{\no} \vee \Um{c}{\no}))}
  \end{align*}  
  Importantly, monitor $\mV_6$ monitors correctly for the $\sHMLwDet$ formula $\varphi_3'$ above.
  \exqed
\end{example}

\begin{corollary}\label{cor:expressive-complete}
  Suppose $\detAct{\iact}=\btrue$ for all $\iact\in\IAct$. 
  For all $\mV\in\Mon$, monitor $\mV$ monitors correctly for \synthRev{\trnsfM{\mV}}.
\end{corollary}

\begin{proof}
  Assume $\detAct{\iact}=\btrue$ for all $\iact\in\IAct$.
  Pick $\pV\in\Prc$.
  \begin{description}
    \setlength\itemsep{0.5em}
    \item[To show soundness,]
      assume there exists $\hst\subseteq\hstpV$ such that \rej{\hst,\mV}.
      By \Cref{prop:tnsfm-rej}, we know \rej{\hst,\trnsfM{\mV}}.
      Since $\cod{\trnsfM{-}} = \MonDet$, then $\trnsfM{\mV}\in\MonDet$. 
      Thus, using \Cref{prop:expressive-complete-monDet}, we conclude $\pV\notin\evalE{\synthRev{\trnsfM{\mV}}}$.
    \item[To show completeness,] 
      assume that $\pV\notin\evalE{\synthRev{\trnsfM{\mV}}}$.
      Since $\cod{\trnsfM{-}} = \MonDet$, we know $\trnsfM{\mV}\in\MonDet$, which by \Cref{prop:monDet-to-shmlwDet}, implies that 
      $\synthRev{\trnsfM{\mV}}\in\sHMLwDet$.
      Thus, using \Cref{prop:expressive-complete-monDet}, we deduce that there exists $\hst\subseteq\hstpV$ such that \rej{\hst,\trnsfM{\mV}}. 
      Our result, \rej{\hst,\mV}, follows by \Cref{prop:tnsfm-rej}.
      \qedhere
  \end{description}
\end{proof}

Equipped with these results, we can now prove \Cref{thm:expressive-complete} 

\begin{theorem} \label{thm:expressive-complete}
  If $\detAct{\act}=\btrue$ for all $\act\in\EAct$, 
  $\sHMLwDet$ is Expressive-Complete \wrt \Mon.
\end{theorem}

\begin{proof}
  Pick $\mV\in\Mon$.
  By \Cref{cor:expressive-complete}, we know monitor \mV\ monitors correctly for the formula $\synthRev{\trnsfM{\mV}}$. 
  Also, since $\cod{\trnsfM{-}} = \MonDet$, we know $\trnsfM{\mV}\in\MonDet$, which by \Cref{prop:monDet-to-shmlwDet}, implies that 
  $\synthRev{\trnsfM{\mV}}\in\sHMLwDet$, as required. 
\end{proof}

We can now show that $\sHMLwDet$ is the largest monitorable subset of \recHML up to logical equivalence, \Cref{thm:maximal-expressivity}, restated below.

\maximality*

\begin{proof}
  Assume $\Det(\iact) {=} \btrue$ for all $\iact \in \IAct$.
  Assume also that $\cL\subseteq \recHML$ is monitorable \wrt \Mon. 
  By \Cref{def:full-mon}, this means that for all $\varphi\in\cL$, 
    $$\exists \mV \in \Mon \text{ such that } \mV \text{ monitors correctly for }\varphi$$
  Pick $\varphi\in\cL$ and assume $\exists \mV\in\Mon$ such that \mV\ monitors correctly for it. 
  By \Cref{def:full-correct}, this means that for all $\pV\in\Prc$, 
  \begin{align}
    \pV\notin\evalE{\varphi} \text{ iff } \bigl( \exists \hst\subseteq \hstpV \text{ such that } \rej{\hst, \mV} \bigr) 
    \label{eq:exp-upper-bound-1}
  \end{align} 
  We need to show that $\exists \psi\in\sHMLwDet$ such that $\evalE{\varphi} = \evalE{\psi}$. \\ 
  Using \Cref{thm:expressive-complete}, for the monitor \mV\ used in (\ref{eq:exp-upper-bound-1}), we also know 
  $\exists \psi \in \sHMLwDet$ where $\psi=\synthRev{\trnsfM{\mV}}$ and \mV\ monitors correctly for $\psi$.
  Expanding \Cref{def:full-mon}, this means that for all $\pV\in\Prc$,
  \begin{align}
    \pV \notin \evalE{\psi} \text{ iff } \bigl( \exists \hst\subseteq \hstpV \text{ such that } \rej{\hst, \mV} \bigr)  
    \label{eq:exp-upper-bound-2}
  \end{align}
  We prove $\evalE{\varphi}=\evalE{\psi}$ in two steps; first, we show $\evalE{\psi} \subseteq \evalE{\varphi}$ and then we show that $\evalE{\varphi}\subseteq\evalE{\psi}$. 
  For the former, assume an arbitrary $\pV\notin\evalE{\varphi}$.
  By~(\ref{eq:exp-upper-bound-1}), we know $\exists \hst\subseteq \hstpV$ such that \rej{\hst, \mV}, which by (\ref{eq:exp-upper-bound-2}) implies that $\pV\notin\evalE{\psi}$.
  We thus have 
  \begin{align}
    \pV\notin\evalE{\varphi} \text{ implies } \pV\notin\evalE{\psi} \label{eq:notin-sub}
  \end{align}
  By the contrapositive of (\ref{eq:notin-sub}), 
  we deduce that $\pV\in\evalE{\psi}$ implies $\pV\in\evalE{\varphi}$, \ie $\evalE{\psi} \subseteq \evalE{\varphi}$.
  Dually, we can show $\evalE{\varphi} \subseteq \evalE{\psi}$. 
  Our result, $\evalE{\varphi}=\evalE{\psi}$, follows.
\end{proof}

\subsection{Proving \Cref{prop:tnsfm-rej}}

The proof for \Cref{prop:tnsfm-rej} relies on several additional results.

\begin{lemma}\label{lemma:subs-shadowing}
  Suppose $\XV\notin\fv{\mV}$.
  Then for all $\mV,\mVV\in\Mon$, $\odF,\odF'\in\Bool$ and $\sV\in\Subs$, we have 
  $$\trnsfM{\mV, \odF, \sV} = \trnsfM{\mV, \odF, \sV[\XV\mapsto\langle\mVV, \odF'\rangle]}$$
\end{lemma}

\begin{proof}
  Suppose $\XV\notin\fv{\mV}$.
  The proof proceeds by induction on 
  \mV. 
  The only interesting case is when $\mV = \recX{\mV'}$. We have
  \begin{align*}
    &\trnsfM{\recX{\mV'}, \odF, \sV} \\
    &\quad = \recX{\trnsfM{\mV', \odF, \sV[\XV \mapsto \langle \recX{\mV}, \odF\rangle]}} \\
    &\quad = \recX{\trnsfM{\mV', \odF, \sV[\XV \mapsto \langle \mVV,\odF'\rangle ][\XV \mapsto \langle \recX{\mV}, \odF\rangle]}} \\
    &\qquad\text{ for some }\mVV \text{ and } \odF' \\
    &\qquad \text{ since }
      \sV[\langle \mVV,\odF'\rangle ][\XV \mapsto \langle \recX{\mV}, \odF\rangle] = \sV[\XV \mapsto \langle \recX{\mV}, \odF\rangle] \\
    &\quad = \trnsfM{\recX{\mV'}, \odF, \sV[\XV \mapsto \langle \mVV,\odF'\rangle ]}
  \end{align*}
  The other cases are straightforward. 
\end{proof}

\begin{lemma}\label{lemma:trnsf-preserves-fv}
  For all $\mV\in\Mon$, $\odF\in\Bool$ and $\sV\in\Subs$, 
  if $\XV\notin\fv{\mV}$ and $\XV\notin\fv{\cod{\sV}}$ then $\XV\notin\fv{\trnsfM{\mV,\odF,\sV}}$.
\end{lemma}

\begin{proof}
  Follows from the contrapositive of \Cref{lemma:trnsf-preserves-fv-contrapositive}. 
\end{proof}

\begin{lemma}\label{lemma:trnsf-preserves-fv-contrapositive}
  For all $\mV\in\Mon$, $\odF\in\Bool$ and $\sV\in\Subs$, 
  if $\XV\in\fv{\trnsfM{\mV,\odF,\sV}}$ then either $\XV\in\fv{\mV}$ or $\XV\in\fv{\cod{\sV}}$.
\end{lemma}

\begin{proof}
  Suppose $\XV\in\fv{\trnsfM{\mV,\odF,\sV}}$.
  The proof proceeds by induction on the derivation of \trnsfM{\mV,\odF,\sV}.
  We only outline the main cases:
  \begin{itemize}
    \setlength\itemsep{0.5em}
    %
    %
    %
    %
    \item 
      Case \textsc{tVar3}, \ie $\trnsfM{\XVV,\odF,\sV}=\mVV$ because $\sV(\XVV) = \langle \mV, \odF' \rangle$ where $\odF'\neq\odF$, and $\trnsfM{\mV,\odF,\sV} = \mVV$.
      Since $\fv{\trnsfM{\XVV,\odF,\sV}} = \fv{\mVV} = \fv{\trnsfM{\mV,\odF,\sV}}$, we can use the IH and obtain that either $\XV\in\fv{\mV}$ or $\XV\in\fv{\cod{\sV}}$.
      Since $\sV(\XVV) = \langle \mV, \odF' \rangle$, then it must be that $\XV\in\fv{\cod{\sV}}$.
    \item Case \textsc{tAct}, \ie $\trnsfM{\act.\mV,\odF,\sV}=\act.\mVV$ because $\trnsfM{\mV,\odF\wedge\detAct{\act},\sV}=\mVV$.
      Since $\fv{\trnsfM{\act.\mV,\odF,\sV}} = \fv{\act.\mVV} = \fv{\mVV} = \fv{\trnsfM{\mV,\odF\wedge\detAct{\act},\sV}}$, we can use the IH and obtain that either $\XV\in\fv{\mV}$ or $\XV\in\fv{\cod{\sV}}$.
      In turn, this implies that either $\XV\in\fv{\act.\mV}$ or $\XV\in\fv{\cod{\sV}}$. 
    %
    %
    %
    %
    \item Case \textsc{tRec}, 
      \ie $\trnsfM{\rec{\XVV}{\mV},\odF,\sV} = \rec{\XVV}{\mVV}$ because $\trnsfM{\mV,\odF,\sV'} = \mVV$ where $\sV'=\sV[\XVV \mapsto \langle \rec{\XVV}{\mV}, \odF \rangle]$.
      Working up to $\alpha$-equivalence, we can assume that $\XV\neq\XVV$.
      Since $\XV\in\fv{\rec{\XVV}{\mVV}} = \fv{\mVV} \setminus \{ \XVV \}$, then $\XV\in\fv{\mVV}$.
      By the IH, we obtain that either $\XV\in\fv{\mV}$ or $\XV\in\fv{\cod{\sV'}}$.
      In case of the former, since $\XV\neq \XVV$, we deduce that $\XV\in\fv{\rec{\XVV}{\mV}}$.
      In case of the latter, there are two subcases.
      If $\XV\in\fv{\cod{\sV}}$, then we are done.  
      Otherwise, if $\XV\in\fv{\cod{\sV'}}$ but $\XV\notin\fv{\cod{\sV}}$, then it must be that $\XV\in\fv{\rec{\XVV}{\mV}}$.
      \qedhere
  \end{itemize} 
\end{proof}

\begin{lemma}\label{lemma:change-flag}
  Given $\mVV\in\Mon$ and $\sV\in\Subs$, suppose that $\XV\notin\fv{\cod{\sV}}$ and $\fv{\mVV}\subseteq\{ \XV\}$.
  Then for all $\mV\in\Mon$,  
  \begin{align*}
    &\trnsfM{\mV, \bfalse, \sV[\XV \mapsto \langle \recX{\mVV}, \btrue \rangle ]} = \\
    &\qquad \trnsfM{\mV, \bfalse, \sV[\XV \mapsto \langle \recX{\mVV}, \bfalse \rangle ]}
    \subS{\trnsfM{\recX{\mVV}, \bfalse, \sV}}{\XV}
  \end{align*}
\end{lemma}

\begin{proof}
  The proof proceeds by induction on the structure of \mV.
  \begin{itemize}
    \setlength\itemsep{0.5em}
    \item Case $\mV=\XV$. We have
      \begin{align*}
        &\trnsfM{\XV, \bfalse, \sV[\XV \mapsto \langle \recX{\mVV}, \btrue \rangle ]} \\
        &= \trnsfM{\recX{\mVV}, \bfalse, \sV[\XV \mapsto \langle \recX{\mVV}, \btrue \rangle ]} \\
        &= \recX{\trnsfM{\mVV, \bfalse, \sV[\XV \mapsto \langle \recX{\mVV}, \bfalse \rangle ]}} \\
        &= \bigl(\recX{\trnsfM{\mVV, \bfalse, \sV[\XV \mapsto \langle \recX{\mVV}, \bfalse \rangle ]}}\bigr) 
        \subS{\trnsfM{\recX{\mVV}, \bfalse, \sV}}{\XV} \\
        &\qquad \text{ since \XV\ is not free} \\
        &= \trnsfM{\recX{\mVV}, \bfalse, \sV[\XV \mapsto \langle \recX{\mVV}, \bfalse \rangle ]} 
        \subS{\trnsfM{\recX{\mVV}, \bfalse, \sV}}{\XV}
      \end{align*}
    \item Case $\mV=\XVV$. 
      There are two subcases to consider.
      When $\XVV\notin\dom{\sV}$, the proof is straightforward.
      When $\sV(\XVV) = \langle \mV', \odF \rangle$ for some $\mV'$ and $\odF$,  
      we have
      \begin{align*}
        &\trnsfM{\XVV, \bfalse, \sV[\XV \mapsto \langle \recX{\mVV}, \btrue \rangle ]} \\
        &= \trnsfM{\mV', \bfalse, \sV[\XV \mapsto \langle \recX{\mVV}, \btrue \rangle ]} \\
        &= \trnsfM{\mV', \bfalse, \sV[\XV \mapsto \langle \recX{\mVV}, \bfalse \rangle ]} 
        \text{ by \Cref{lemma:subs-shadowing}} \\
        & \qquad\text{ because since } \XV\notin\fv{\cod{\sV}} \text{ then } \XV\notin\fv{\mV'} \\
        & = \trnsfM{\mV', \bfalse, \sV[\XV \mapsto \langle \recX{\mVV}, \bfalse \rangle ]} \subS{\trnsfM{\recX{\mVV}, \bfalse, \sV}}{\XV} \\
        &\qquad \text{ since } \XV\notin\trnsfM{\mV', \bfalse, \sV[\XV \mapsto \langle \recX{\mVV}, \bfalse \rangle ]} 
        \text{ by \Cref{lemma:trnsf-preserves-fv}} 
      \end{align*}
      %
      %
      \item Case $\mV = \recX{\mV'}$. We have
        \begin{align*}
          &\trnsfM{\recX{\mV'}, \bfalse, \sV[\XV \mapsto \langle \recX{\mVV}, \btrue \rangle ]} \\
          &=\trnsfM{\recX{\mV'}, \bfalse, \sV[\XV \mapsto \langle \recX{\mVV}, \bfalse \rangle ]} \\
          &\qquad \text{ by \Cref{lemma:subs-shadowing} since } \XV\notin\fv{\recX{\mV'}} \\
          &= \recX{\trnsfM{\mV', \bfalse, \sV[\XV \mapsto \langle \recX{\mVV}, \bfalse \rangle ]}} \\
          &=\bigl(\recX{\trnsfM{\mV', \bfalse, \sV[\XV \mapsto \langle \recX{\mVV}, \bfalse \rangle ]}}\bigr)
          \subS{\trnsfM{\recX{\mVV}, \bfalse, \sV}}{\XV} \\
          &\qquad \text{ since \XV\ is not free} \\
          &= \trnsfM{\recX{\mV'}, \bfalse, \sV[\XV \mapsto \langle \recX{\mVV}, \bfalse \rangle ]}
          \subS{\trnsfM{\recX{\mVV}, \bfalse, \sV}}{\XV} 
        \end{align*} 
      \item Case $\mV = \rec{\XVV}{\mV'}$. We have
        \begin{align*}
          &\trnsfM{\rec{\XVV}{\mV'}, \bfalse, \sV[\XV \mapsto \langle \recX{\mVV}, \btrue \rangle ]} \\
          &=\rec{\XVV}{\trnsfM{\mV', \bfalse, \sV[\XV \mapsto \langle \recX{\mVV}, \btrue \rangle ]}} \\
          &=\rec{\XVV}{
            \bigl( 
            \trnsfM{\mV', \bfalse, \sV[\XV \mapsto \langle \recX{\mVV}, \bfalse \rangle ]} 
            \subS{\trnsfM{\recX{\mVV}, \bfalse, \sV}}{\XV} 
            \bigr)
          } \\
          & \qquad \text{ by the IH}\\
          &=
            \bigl( \rec{\XVV}{ \trnsfM{\mV', \bfalse, \sV[\XV \mapsto \langle \recX{\mVV}, \bfalse \rangle ]}} \bigr)  
            \subS{\trnsfM{\recX{\mVV}, \bfalse, \sV}}{\XV}\\
          &=
            \trnsfM{\rec{\XVV}{\mV'}, \bfalse, \sV[\XV \mapsto \langle \recX{\mVV}, \bfalse \rangle ]}  
            \subS{\trnsfM{\recX{\mVV}, \bfalse, \sV}}{\XV}
        \end{align*}
  \end{itemize}
  The remaining cases are more straightforward. 
\end{proof}

\begin{lemma}\label{lemma:trnsfM-subst}
  Given $\mVV\in\Mon$ and $\sV\in\Subs$, suppose $\XV\notin\fv{\cod{\sV}}$ and $\fv{\mVV}\subseteq\{ \XV\}$.
  For all $\mV\in\Mon$ and $\odF\in\Bool$, 
  $$
    \trnsfM{\mV\subS{\recX{\mVV}}{\XV}, \odF, \sV} = 
    \trnsfM{\mV, \odF, \sV'}
    \subS{\trnsfM{\recX{\mVV}, \odF, \sV}}{\XV}
  $$
  where $\sV'=\sV[\XV \mapsto\langle \recX{\mVV}, \odF\rangle]$.
\end{lemma}

\begin{proof}
  Let $\sV' = \sV[\XV \mapsto\langle \recX{\mVV}, \odF\rangle]$.
  Suppose $\XV\notin\fv{\cod{\sV}}$ and $\fv{\mVV}\subseteq\{ \XV\}$.
  Then $\XV\notin\fv{\recX{\mVV}}$ either.
  The proof proceeds by induction on the structure of \mV.
  We outline the main cases.
  \begin{itemize}[leftmargin=*]
    \setlength\itemsep{0.5em}
    \item Case $\mV = \XV$. There are three subcases to consider. 
      \begin{itemize}
        \item When $\sV(\XV) = \langle \mV', \odF \rangle$ for some $\mV'$,  
        we have
        \begin{align*}
          & \trnsfM{\XV\subS{\recX{\mV}}{\XV}, \odF, \sV} 
          = \trnsfM{\recX{\mVV}, \odF, \sV} \\
          & = \XV\subS{\trnsfM{\recX{\mVV}, \odF, \sV}}{\XV} \\
          & = \trnsfM{\XV, \odF, \sV'}\subS{\trnsfM{\recX{\mVV}, \odF, \sV}}{\XV} \\ 
          &\qquad \text{ where } 
          \sV' = \sV[\XV \mapsto\langle \recX{\mVV}, \odF\rangle] 
          \text{  by \Cref{def:trnsfM}}
        \end{align*} 
        \item When $\sV(\XV) = \langle \mV', \odF' \rangle$ for some $\mV'$ and $\odF'\neq\odF$,  
        the proof is similar.
        \item When $\XV\notin\dom{\sV}$, the proof is similar. 
      \end{itemize}
    \item Case $\mV = \XVV$.
      There are three subcases to consider.
      \begin{itemize}
        \item When $\sV(\XVV) = \langle \mV', \odF \rangle$ for some $\mV'$,  
        we have
        \begin{align*}
          &\trnsfM{\XVV\subS{\recX{\mV}}{\XV}, \odF, \sV} \\
          &= \trnsfM{\XVV, \odF, \sV} 
          = \XVV \text{ by \Cref{def:trnsfM}} \\ 
          & = \XVV \subS{\trnsfM{\recX{\mVV}, \odF, \sV}}{\XV} \\
          & = \trnsfM{\XVV, \odF, \sV'} \subS{\trnsfM{\recX{\mVV}, \odF, \sV}}{\XV} \text{ by \Cref{def:trnsfM}}
        \end{align*}
        \item When $\sV(\XVV) = \langle \mV', \odF' \rangle$ for some $\mV'$ and $\odF' \neq\odF$,  
        we have
        \begin{align*}
          &\trnsfM{\XVV\subS{\recX{\mV}}{\XV}, \odF, \sV} \\
          &= \trnsfM{\XVV, \odF, \sV} 
          = \trnsfM{\mV', \odF, \sV} \text{ by \Cref{def:trnsfM}}\\ 
          & = \trnsfM{\mV', \odF, \sV'} 
          \text{ where } \sV' = \sV[\XV \mapsto\langle \recX{\mVV}, \odF\rangle]  \\
          & \qquad 
          \text{ by \Cref{lemma:subs-shadowing}} \text{ because since } \XV\notin\fv{\cod{\sV}} \\
          & \quad \text{ then } \XV\notin\fv{\mV'} \\
          & = \trnsfM{\mV', \odF, \sV'} \subS{\trnsfM{\recX{\mVV}, \odF, \sV}}{\XV} \\
          &\qquad \text{ since } \XV\notin\trnsfM{\mV', \odF, \sV'} \text{ by \Cref{lemma:trnsf-preserves-fv}} \\
          & = \trnsfM{\XVV, \odF, \sV'} \subS{\trnsfM{\recX{\mVV}, \odF, \sV}}{\XV} \text{ by \Cref{def:trnsfM}}
        \end{align*}
        \item When $\XVV\notin\dom{\sV}$, the proof is straightforward.
      \end{itemize}
    \item Case $\mV = \act.\mVV$.
      We have 
      \begin{align*}
        &\trnsfM{(\act.\mV')\subS{\recX{\mV}}{\XV}, \odF, \sV} \\
        &= \trnsfM{\act.(\mV'\subS{\recX{\mV}}{\XV}), \odF, \sV} \\
        &= \act.\trnsfM{\mV'\subS{\recX{\mV}}{\XV}, \odF', \sV} \text{ where } \odF'=\odF\wedge\detAct{\act} \\
        &= \act.\bigl( \trnsfM{\mV', \odF', \sV'} \subS{\trnsfM{\recX{\mVV}, \odF', \sV}}{\XV} \bigr)  \text{ using the IH } \\
        &\qquad \qquad \text{ where } \sV'=\sV[\XV \mapsto \langle \recX{\mVV}, \odF'\rangle ]
      \end{align*}
      There are two subcases to consider. 
      If $\odF = \odF'$, we have
      \begin{align*}
        &\act.\bigl( \trnsfM{\mV', \odF', \sV'} \subS{\trnsfM{\recX{\mVV}, \odF', \sV}}{\XV} \bigr) \\
        &= \act.\bigl( \trnsfM{\mV', \odF, \sV'} \subS{\trnsfM{\recX{\mVV}, \odF, \sV}}{\XV} \bigr) \\
        &\qquad \text{ where }\sV'=\sV[\XV \mapsto \langle \recX{\mVV}, \odF\rangle ] \\
        &= \bigl( \act.\trnsfM{\mV', \odF, \sV'} \bigr) \subS{\trnsfM{\recX{\mVV}, \odF, \sV}}{\XV} \\
        &= \trnsfM{\act.\mV', \odF, \sV'} \subS{\trnsfM{\recX{\mVV}, \odF, \sV}}{\XV} 
        \text{ as required}
      \end{align*}  
      Otherwise, if $\odF = \btrue$ and $\odF' = \bfalse$, we have
      \begin{align*}
        &\act.\bigl( \trnsfM{\mV', \odF', \sV'} \subS{\trnsfM{\recX{\mVV}, \odF', \sV}}{\XV} \bigr) \\ 
        &= \act.\bigl( \trnsfM{\mV', \bfalse, \sV'} \subS{\trnsfM{\recX{\mVV}, \bfalse, \sV}}{\XV} \bigr) \\
        &\qquad \text{ where }\sV'=\sV[\XV \mapsto \langle \recX{\mVV}, \bfalse \rangle ]  \\
        &= \act.\bigl( \trnsfM{\mV', \bfalse, \sV'} \subS{\trnsfM{\recX{\mVV}, \bfalse, \sV}}{\XV}  \bigr) 
        \subS{\trnsfM{\recX{\mVV}, \btrue, \sV}}{\XV} \\
        & \qquad \text{ since \XV\ is not free}\\
        &= \act.\trnsfM{\mV', \btrue, \sV''}  
        \subS{\trnsfM{\recX{\mVV}, \btrue, \sV}}{\XV} \\ 
        &\quad \text{ by \Cref{lemma:change-flag} where } \sV''=\sV[\XV \mapsto \langle \recX{\mVV}, \btrue \rangle ] \\
        &= \trnsfM{\act.\mV', \btrue, \sV''} \subS{\trnsfM{\recX{\mVV}, \btrue, \sV}}{\XV} 
        \text{ as required}
      \end{align*}  
    \item Case $\mV=\recX{\mV'}$. We have
      \begin{align*}
        &\trnsfM{(\recX{\mV'})\subS{\recX{\mVV}}{\XV}, \odF, \sV} \\  
        &= \trnsfM{\recX{\mV'}, \odF, \sV} \text{ since } \XV\notin\fv{\recX{\mV'}} \\
        &= \trnsfM{\recX{\mV'}, \odF, \sV'} \\ 
        &\qquad \text{ where } \sV' = \sV[\XV \mapsto\langle \recX{\mVV}, \odF\rangle] 
        \text{ using \Cref{lemma:subs-shadowing} } \\ 
        &= \recX{\trnsfM{\mV', \odF, \sV'[\XV\mapsto \langle \recX{\mV'}, \odF \rangle ]}} \\
        &= \bigl(\recX{\trnsfM{\mV', \odF, \sV'[\XV\mapsto \langle \recX{\mV'}, \odF \rangle ]}} \bigr) 
        \subS{\trnsfM{\recX{\mVV}, \odF, \sV}}{\XV} \\ 
        &\qquad \text{ since \XV\ is not free} \\
        &= \trnsfM{\recX{\mV'}, \odF, \sV'} \subS{\trnsfM{\recX{\mVV}, \odF, \sV}}{\XV} 
      \end{align*}
    \item Case $\mV=\rec{\XVV}{\mV'}$.
      We have
      \begin{align*}
        &\trnsfM{(\rec{\XVV}{\mV'})\subS{\recX{\mVV}}{\XV}, \odF, \sV} \\  
        &= \trnsfM{\rec{\XVV}{\bigl(\mV'\subS{\recX{\mVV}}{\XV}\bigr)}, \odF, \sV} \\
        &= \rec{\XVV}{\trnsfM{\mV'\subS{\recX{\mVV}}{\XV}, \odF, \sV'}} \\
        &\qquad \text{ where } \sV' = \sV[\XVV\mapsto \langle \rec{\XVV}{\mV'\subS{\recX{\mVV}}{\XV}}, \odF\rangle] \\
        &= \rec{\XVV}{ \bigl( \trnsfM{\mV', \odF, \sV'' } \subS{\trnsfM{\recX{\mVV},\sV',\odF}}{\XV} \bigr)} \\
        &\qquad \text{ by the IH where } \sV'' = \sV'[X \mapsto \langle \recX{\mVV}, \odF \rangle] \\
        &= \bigl( \rec{\XVV}{ \trnsfM{\mV', \odF, \sV'' } } \bigr)
        \subS{\trnsfM{\recX{\mVV},\sV',\odF}}{\XV} \\
        &= \trnsfM{\rec{\XVV}{\mV', \odF, \sV'' }} \subS{\trnsfM{\recX{\mVV},\sV',\odF}}{\XV} \\
        &= \trnsfM{\rec{\XVV}{\mV', \odF, \sV'' }} \subS{\trnsfM{\recX{\mVV},\sV'',\odF}}{\XV}
        \text{ by \Cref{lemma:subs-shadowing} } \\ 
        &\qquad \text{since } \XVV\notin\fv{\cod{\sV'}} \\
        &\qquad \text{and the assumption } \fv{\mVV} \subseteq \{ \XV\} \text{ implies }\XVV\notin\fv{\recX{\mVV}}  
      \end{align*} 
  \end{itemize}
  The remaining cases are more straightforward.
\end{proof}

\Cref{lemma:trnsfM-preserves-rej} shows that the transformation function \trnsfM{-} preserves history rejections.
However, its proof relies on \Cref{lemma:trnsfF-implies-trnsfT} below. 

\begin{lemma}\label{lemma:trnsfF-implies-trnsfT}
  Suppose that for all $\mV\in\Mon$ and $\sV\in\Subs$, $\XV\notin\fv{\cod{\sV}}$ and $\fv{\mV}\subseteq\{ \XV\}$.
  \begin{center}
    If \rej{\hst, \bfalse, \trnsfM{\mV,\bfalse,\sV}}\\ 
    then \rej{\hst, \bfalse, \trnsfM{\mV,\btrue,\sV}}.
  \end{center} 
\end{lemma}

\begin{proof}
  The proof proceeds by rule induction on the judgement \rej{\hst, \bfalse, \trnsfM{\mV,\bfalse,\sV}}. 
  We outline the main cases.
  \begin{itemize}[leftmargin=*]
    \setlength\itemsep{0.5em}
      %
    \item 
      Case \textsc{act}, \ie \rej{\hst, \bfalse, \trnsfM{\act.\mV,\bfalse,\sV} } where $\trnsfM{\act.\mV,\bfalse,\sV} = \act.\trnsfM{\mV,\bfalse,\sV}$ because \rej{\hstt, \bfalse, \trnsfM{\mV,\bfalse,\sV} } where $\hstt=\sub{\hst,\act}$.
      There are two subcases: \\[0.5em]
        If $\detAct{\act}=\bfalse$, $\trnsfM{\act.\mV,\bfalse,\sV}=\trnsfM{\act.\mV,\btrue,\sV}$, 
        which implies \rej{\hst, \bfalse, \trnsfM{\act.\mV,\btrue,\sV} }. \\[0.5em]
        If $\detAct{\act}{=}\btrue$, by the IH, 
        \rej{\hstt, \bfalse, \trnsfM{\mV,\btrue,\sV} }.
        Applying rule \textsc{act}, we get \rej{\hst, \bfalse, \act.\trnsfM{\mV,\btrue,\sV} }.
        Our result follows by the fact that $\act.\trnsfM{\mV,\btrue,\sV} = \trnsfM{\act.\mV,\btrue,\sV}$.
      %
      %
    \item 
      Case \textsc{rec}, \ie 
      \rej{\hst, \bfalse, \trnsfM{\recX{\mV},\bfalse,\sV} } 
      where, by \Cref{def:trnsfM}, we have 
      $\trnsfM{\recX{\mV},\bfalse,\sV} = \recX{\trnsfM{\mV,\bfalse,\sV'}}$ and $\sV'=\sV[\XV \mapsto \langle \recX{\mV}, \bfalse \rangle]$ 
      because 
      \begin{align}
        \rej{\hst, \bfalse, \trnsfM{\mV,\bfalse,\sV'}\subS{\recX{\trnsfM{\mV,\bfalse,\sV}}}{\XV}} \label{eq:trnsf-rec}
      \end{align}
      By \Cref{lemma:trnsfM-subst}, we also know that
      \begin{align}
        &\trnsfM{\mV,\bfalse,\sV'}\subS{\recX{\trnsfM{\mV,\bfalse,\sV}}}{\XV} \notag \\
        &\qquad = \trnsfM{\mV\subS{\recX{\mV}}{\XV},\bfalse,\sV} \label{eq:rec-subst}
      \end{align}
      By (\ref{eq:trnsf-rec}), (\ref{eq:rec-subst}) and the IH, we deduce \rej{\hst, \bfalse, \trnsfM{\mV\subS{\recX{\mV}}{\XV},\btrue,\sV}}, 
      which implies that \rej{\hst, \bfalse, \trnsfM{\mV,\btrue,\sV''}\subS{\recX{\trnsfM{\mV,\btrue,\sV''}}}{\XV}} 
      where $\sV''=\sV[\XV\mapsto \langle \recX{\mV}, \btrue \rangle]$ by \Cref{lemma:trnsfM-subst}.
      Applying rule \textsc{rec}, we obtain \rej{\hst, \bfalse, \recX{\trnsfM{\mV,\btrue,\sV''}}}.
      Our result, \rej{\hst, \bfalse, \trnsfM{\recX{\mV},\btrue,\sV}}, follows by \Cref{def:trnsfM} since $\recX{\trnsfM{\mV,\btrue,\sV''}} = \trnsfM{\recX{\mV},\btrue,\sV}$.
      \qedhere 
  \end{itemize}
\end{proof}

\begin{lemma}\label{lemma:trnsfM-preserves-rej}
  For all $\mV\in\Mon$, 
  \rej{\hst,\odF,\mV} iff \rej{\hst, \odF, \trnsfM{\mV, \odF, \emptyset}}.
\end{lemma}

\begin{proof}
  For the ``only if'' direction,
  the proof proceeds by rule induction on \rej{\hst,\odF,\mV}. We outline the main cases. 
  \begin{itemize}[leftmargin=*]
    \setlength\itemsep{0.5em}
    %
    %
    \item 
      Case \textsc{act}, \ie \rej{\hst,\odF,\act.\mV} because \rej{\hstt,\odF',\mV} where $\hstt=\sub{\hst,\act}$ and $\odF'=\odF{\wedge}\detAct{\act}$.
      By the IH, we obtain that \rej{\hstt,\odF',\trnsfM{\mV,\odF',\emptyset}}.
      Applying rule \textsc{act}, we get \rej{\hst, \odF, \act.\trnsfM{\mV,\odF',\emptyset}} 
      which, by \Cref{def:trnsfM}, implies that \rej{\hst, \odF, \trnsfM{\act.\mV,\odF,\emptyset}}.
    \item
      Case \textsc{actI}, \ie \rej{\hst,\odF,\act.\mV} because \rej{\hstt,\odF',\act.\mV} where $\hstt{=}\sub{\hst,\iact}$ and $\odF'{=}\odF{\wedge}\detAct{\iact}$ for some $\iact{\in}\IAct$.
      By the IH, we obtain \rej{\hstt,\odF',\trnsfM{\act.\mV,\odF',\emptyset}}.
      There are two subcases to consider. 
      If $\odF = \odF'$, we can apply rule \textsc{actI} and conclude \rej{\hst, \odF, \trnsfM{\act.\mV,\odF,\emptyset}}. 
      If $\odF \neq \odF'$, \ie $\odF=\btrue$ and $\odF'=\bfalse$, then by \Cref{lemma:trnsfF-implies-trnsfT}, we obtain \rej{\hstt,\odF',\trnsfM{\act.\mV,\odF,\emptyset}}.
      Applying rule \textsc{actI}, we conclude that \rej{\hst, \odF, \trnsfM{\act.\mV,\odF,\emptyset}}. 
    %
    %
    \item 
      Case \textsc{rec}, \ie \rej{\hst,\odF,\recX{\mV}} because \rej{\hst,\odF,\mV\subS{\recX{\mV}}{\XV}}.
      By the IH, we obtain \rej{\hst,\odF,\trnsfM{\mV\subS{\recX{\mV}}{\XV}, \odF, \emptyset}}.
      Using \Cref{lemma:trnsfM-subst} and \Cref{def:trnsfM}, we know 
      \begin{align*}
        &\trnsfM{\mV\subS{\recX{\mV}}{\XV}, \odF, \emptyset} \\
        &= \trnsfM{\mV, \odF, \{ \XV \mapsto\langle \recX{\mV}, \odF\rangle\} }
        \subS{\trnsfM{\recX{\mV}, \odF, \emptyset}}{\XV}  \\
        &= \trnsfM{\mV, \odF, \{ \XV \mapsto\langle \recX{\mV}, \odF\rangle\} }\\
        &\qquad\qquad \subS{\recX{\trnsfM{\mV, \odF, \{ \XV \mapsto\langle \recX{\mV}, \odF\rangle\}}}}{\XV}  
      \end{align*}
      Let $\mVV=\trnsfM{\mV, \odF, \{ \XV \mapsto\langle \recX{\mV}, \odF\rangle\} }$.
      We thus have \rej{\hst,\odF,\mVV\subS{\recX{\mVV}}{\XV}}. \\ 
      By rule \textsc{rec}, we obtain \rej{\hst,\odF,\recX{\mVV}}.
      Our result, \rej{\hst,\odF,\trnsfM{\recX{\mV}, \odF, \emptyset }}, follows by \Cref{def:trnsfM}.
  \end{itemize}
  \vspace{0.5em}
  The proof for the ``if'' direction follows similarly by rule induction on \rej{\hst,\odF,\trnsfM{\mV,\odF,\emptyset}}. 
  The only case that differs slightly is that for \textsc{rec}.
  \begin{itemize}[leftmargin=*]
    \item 
      Case \textsc{rec}.
      We know \rej{\hst,\odF,\trnsfM{\recX{\mV},\odF,\emptyset}}, \ie \rej{\hst,\odF,\recX{\mVV}} 
      where 
      $$\recX{\mVV} = \trnsfM{\recX{\mV},\odF,\emptyset} = \recX{\trnsfM{\mV,\odF,\{ \XV \mapsto \langle \recX{\mV},\odF \rangle \}}}$$
      because \rej{\hst,\odF,\mVV\subS{\recX{\mVV}}{\XV}}.
      Using \Cref{def:trnsfM} and \Cref{lemma:trnsfM-subst}, we know 
      \begin{align*}
        &\mVV\subS{\recX{\mVV}}{\XV} \\
        &= \trnsfM{\mV,\odF,\{ \XV \mapsto \langle \recX{\mV},\odF \rangle \}}\subS{\recX{\trnsfM{\mV,\odF,\{ \XV \mapsto \langle \recX{\mV},\odF \rangle \}}}}{\XV} \\
        &= \trnsfM{\mV,\odF,\{ \XV \mapsto \langle \recX{\mV},\odF \rangle \}}\subS{\trnsfM{\recX{\mV},\odF,\emptyset}}{\XV} \\
        &=\trnsfM{\mV\subS{\recX{\mV}}{\XV} , \odF, \emptyset}
      \end{align*}
      We can thus rewrite \rej{\hst,\odF,\mVV\subS{\recX{\mVV}}{\XV}} as the judgement \rej{\hst,\odF,\trnsfM{\mV\subS{\recX{\mV}}{\XV} , \odF, \emptyset}}. 
      By the IH, we obtain that \rej{\hst, \odF, \mV\subS{\recX{\mV}}{\XV}}.
      Applying rule \textsc{rec}, we conclude \rej{\hst, \odF, \recX{\mV}} as required.
      \qedhere
  \end{itemize}  
\end{proof}

We are now in a position to prove \Cref{prop:tnsfm-rej}, restated below. 

\tnsfmRej*

\begin{proof}
  Follows from \Cref{lemma:trnsfM-preserves-rej}, letting $\odF=\btrue$.
\end{proof}


\section{Implementability Aspects}
\label{sec:implementability}

The verification technique presented in this paper lends itself well to the implementation of a tool that runtime verifies systems over multiple runs. 
We outline the steps for a full automation and give a complexity analysis of this technique. 

\paragraph*{Algorithm.}
The first step is to generate executable monitors from properties expressed as $\sHMLwDet$ formulae, following the synthesis algorithm of~\Cref{def:synth}.
These monitors must then be instrumented to execute alongside the SUS \wrt the history of traces observed thus far (initialised to empty) as \emph{outline} monitors~\cite{BartocciFFR18}, which allows us to treat systems as black-boxes.  
%
%
Instrumentation forwards the events generated by the SUS to the monitor, which aggregates traces according to the mechanism in \Cref{fig:syntax-semantics}.  
Prior work~\cite{DBLP:conf/coordination/AcetoAAEFI22,DBLP:conf/forte/AttardAAFIL21} has shown that the synthesis and implementation of similar operational models is almost one-to-one. 
Aceto \etal~\cite{DBLP:conf/fase/AcetoAFI21} rigorously demonstrate their efficiency, which results in a stable tool called \textsf{detectEr} for runtime verifying asynchronous component systems~\cite{DBLP:conf/forte/AttardAAFIL21}.
Whenever instrumentation aggregates a new trace to the history, the monitor is terminated and 
the history analysis in \Cref{fig:proof-system} is invoked; this can be automated following an approach similar to that in~\cite{DBLP:conf/coordination/AchilleosEFLX22}.
Trace aggregation and history analysis are 
repeated until a permanent verdict is reached (\Cref{prop:irrevocability}). 
%


\paragraph*{Complexity Bounds.}
The algorithm's performance depends on:
\begin{enumerate}
  \item The trace aggregation of \Cref{fig:mon}. 
  Monitors analyse system events sequentially and transition accordingly, each monitor component incurring a linear complexity \wrt the length of the processed trace.
  The required number of monitor components and the cost of simulating these with a single monitor component has been studied extensively for similar monitoring systems in~\cite{Aceto2019cost}. 
  There, the authors prove that monitors without parallel components may require up to a doubly-exponential number of states \wrt the size of the formula that they monitor. 
  This means that it may be necessary to maintain an exponentially long description of the monitor configurations along a run. 
  Under the assumption that formulae are generally significantly smaller than execution traces, or that monitors run asynchronously \wrt the SUS, the resulting overhead is acceptable.
  \item The history analysis of \Cref{fig:proof-system}.
  %
  %
  The complexity of derivations for $\rej{\hst,\mV}$ is polynomial \wrt the size of \mV\ and the longest trace in \hst. 
  Effectively, this amounts to $\mu$-calculus model-checking on trees, \ie to modal logic model-checking on acyclic graphs, which requires a bilinear time \wrt the size of the tree and the formula~\cite{halpern1992guide}.
  %
  With the exception of rules \textsc{parAL}, \textsc{parAR}, \textsc{act} and \textsc{actPre}, derivations are mostly syntax-directed and monitors are guarded, \ie rule \textsc{rec} can only be applied a finite number of times before rule \textsc{act} is used. 
  For a similar (but more complex) tableau format,~\cite[Section 5]{DBLP:conf/coordination/AchilleosEFLX22} showed that, in practice, the doubly-exponential worst-case complexity upper bound identified in~\cite{DBLP:conf/csl/AcetoAFIL21} does not represent the average-case complexity.
  \item The number of traces required by the monitor conducting the verification to reject the aggregated history.
  \Cref{thm:lowerbounds} contributes towards this, but formally answering it is hard since for certain formulae, an upper bound does not exist. 
  We revisit this aspect in \Cref{ex:no-upperbound}.
  %
\end{enumerate}


\section{Actor Systems Formalised}

We validate the realisation of ILTSs from \Cref{sec:preliminaries} and how realistic the constraints adopted in \Cref{sec:monitorability} are by considering an instantiation for Actor-based systems~\cite{DBLP:conf/ijcai/HewittBS73,DBLP:books/Agha-Actors}.
This concurrency model has been adopted by numerous programming languages~\cite{DBLP:books/CesariniThompsonErl,goodwin2015AKKA,juric2024elixir,swift:24}.   
Actor systems are characterised by a set of processes called \emph{actors} that interact with one another via \emph{asynchronous message-passing}. 
Every actor is identified by a unique ID, which is used by other actors to send messages to it \ie the \emph{single-receiver} property. 
Actors are \emph{persistently receptive} meaning that they are always able to receive messages addressed to them.


\begin{figure*}[t!]
  \setlength{\abovedisplayskip}{0pt}
  \setlength{\belowcaptionskip}{-10pt}
  %
  \textbf{Erlang Syntax for Actor Systems} 
  \smallskip
  %
  %
  \begin{align*}
    \actV,\actVV \in \Actors & \bnfdef\ \nl \bnfsepp \actrs{\idV}{\eV}{\qV} \bnfsepp \pioutB{\idV}{\vV} \bnfsepp \actV\actPar\actVV \bnfsepp \actNew{\idV}{\actV}  
    \qquad \quad
    \qV, \qVV \in \MBox \bnfdef\ \mEmpty \bnfsepp \ccat{\vV}{\qV}
    \qquad \quad
    \patV, \patVV \in \Pat \bnfdef\ \xV \bnfsepp \idV \bnfsepp \atV \\
    \eV,\eVV \in \Exp & \bnfdef\ \piout{w_1}{w_2}{\eV} \bnfsepp \rcv{ \{ \patV_n \rTran \eV_n \}_{n\in I} } \bnfsepp \spwn{\eVV}{\xV}{\eV} \bnfsepp \slf{\xV}{\eV} \bnfsepp \recX{\eV} \bnfsepp \XV \bnfsepp \nl  
  \end{align*}
  \textbf{Erlang Semantics for Actor Systems} 
  \smallskip
  \begin{mathpar}[\small]
    \inferrule[\rtitSS{snd1}]
      { }
      { \conf{\K \;|\; \Obs }{ \actrs{\idV}{ \piout{\idVV}{\vV}{\eV} }{\qV} }  
        \traS{ \tau } \actrs{\idV}{\eV}{\qV} \actPar \pioutB{\idVV}{\vV} 
      }
    \quad
    \inferrule[\rtitSS{snd2}]
      { \idVV\in\Obs }
      { \confK{ \pioutB{\idVV}{\vV} }  \traS{ \pioutA{\idVV}{\vV} } \nl }
    \quad
    \inferrule[\rtitSS{rcv}]
      { }
      { \confK{ \actrs{\idV}{\eV}{\qV}}  \traS{ \piinA{\idV}{\vV} }  \actrs{\idV}{\eV}{\ccat{\qV}{\vV}} }
    \quad
    \inferrule[\rtitSS{rec}]
      {  }
      { \confK{ \actrs{\idV}{ \recX{\eV} }{\qV} }    \traS{\tau}     \actrs{\idV}{ \eV\subC{\recX{\eV}}{\XV} }{\qV} }
    \and 
    \inferrule[\rtitSS{commL}]
      { \conf{\K\;|\;\fid{\actVV}}{\actV} \traS{\pioutA{\idV}{\vV}} \actV' \quad 
        \conf{\K\;|\;\fid{\actV}}{\actVV} \traS{\piinA{\idV}{\vV}}  \actVV' 
      }
      { \confK{ \actV \actPar \actVV }
        \traS{\commA{\idV}{\vV}} 
        \actV' \actPar \actVV' 
      } 
    \and
    \inferrule[\rtitSS{ncommL}]
      { \conf{\K\;|\;\fid{\actVV}}{\actV} \traS{\npioutA{\idV}{\idVV}} \actV' \quad 
        \conf{\K\;|\;\fid{\actV}}{\actVV} \traS{\piinA{\idV}{\idVV}}  \actVV' 
      }
      { \confK{ \actV \actPar \actVV }
        \traS{\ncommA} 
        \actNew{\idVV}{(\actV' \actPar \actVV')} 
      }
    \and
    \inferrule[\rtitSS{scp1}]
      { \conf{\K,\idVV \;|\; \Obs}{ \actV } \traS{\stact} \actVV   \quad    \idVV \fresh \fn{\stact} }
      { \conf{\K \;|\; \Obs}{ \actNew{\idVV}{\actV} }   \traS{\stact}   \actNew{\idVV}{\actVV} }
    \and
    \inferrule[\rtitSS{scp2}]
      { \conf{\K,\idVV \;|\; \Obs}{ \actV } \traS{\commA{\idV}{\vV}} \actVV   \quad  
      \idVV \in \{ \idV,\vV \} 
      }
      { \conf{\K \;|\; \Obs}{ \actNew{\idVV}{\actV} }   
        \traS{\ncommA}   
        \actNew{\idVV}{\actVV} 
      }
    \and
    \inferrule[\rtitSS{opn}]
      { \conf{\K,j \;|\; \Obs}{ \actV } \traS{\pioutA{\idV}{\idVV}} \actVV     
      }
      { \confK{ \actNew{j}{\actV} }   \traS{\npioutA{\idV}{\idVV}}     \actVV    }
    \and
    \inferrule[\rtitSS{rd}]
      { 
        \forall n\in I \cdot \absent{ \pV_n , \qV}  \quad 
        \exists m\in I \cdot \neg\absent{ \pV_m, \vV}, \;
        \match{\pV_m,\vV} = \sV 
      }
      { \confK{ \actrs{\idV}{ \rcv{ \{\pV_n \rTran \eV_n \}_{n\in I} } }{ \ccat{\qV}{\ccat{\vV}{\qVV}} } }
        \traS{\tau} \actrs{\idV}{ \eV_m\sV }{\ccat{\qV}{\qVV}} 
      }
    \and
    \inferrule[\rtitSS{parL}]
      { \confK{ \actV } \traS{\stact} \actV'  \qquad   \sbj{\stact} \fresh \fid{\actVV}  } 
      { \confK{ \actV \actPar \actVV }    \traS{\stact}      \actV' \actPar \actVV  } 
    \quad
    \inferrule[\rtitSS{spw}]
      { \idVV \fresh  \K }
      { 
        \confK{ \actrs{\idV}{ \spwn{\eVV}{\xV}{\eV} }{\qV} }
        \traS{\tau} 
        \actNew{\idVV}(\actrs{\idV}{\eV\subC{\idVV}{\xV} }{\qV} \actPar \actrs{\idVV}{\eVV}{\mEmpty}) 
      }
    \quad
    \inferrule[\rtitSS{slf}]
      {  }
      { \confK{ \actrs{\idV}{ \slf{\xV}{\eV} }{\qV} }    \traS{\tau}     \actrs{\idV}{ \eV\subC{\idV}{\xV} }{\qV}   }
    \and 
    \inferrule[\rtitSS{sTr}] 
      { \actV \steq \actV' \quad \confK{\actV'} \traS{\stact} \actVV' \quad \actVV' \steq \actVV }
      { \confK{\actV} \traS{\stact} \actVV }
    \and
    \inferrule[\rtitSS{sNil}]
      { }
      { \actV \steq \actV \actPar \nl  }
    \and
    \inferrule[\rtitSS{sCom}]
      { }
      { \actV \actPar \actVV \steq \actVV \actPar \actV  }
    \and
    \inferrule[\rtitSS{sAss}]
      { }
      { (\actV \actPar \actVV) \actPar \actVVV \steq \actV \actPar (\actVV \actPar \actVVV)  }
    \and
    \inferrule[\rtitSS{sCtxP}] 
      { \actV \steq \actVV }
      { \actV \actPar \actVVV \steq \actVV \actPar \actVVV }
    \and
    \inferrule[\rtitSS{sCtxS}] 
      { \actV \steq \actVV }
      { \actNew{\idV}{\actV} \steq \actNew{\idV}{\actVV} }
    \and
    \inferrule[\rtitSS{sSwp}] 
      { }
      { \actNew{\idV}{\actNew{\idVV}{\actV}} \steq \actNew{\idVV}{\actNew{\idV}{\actV}} }
    \and
    \inferrule[\rtitSS{sExt}] 
      { \idV \fresh \fn{\actV} }
      { \actV \actPar \actNew{\idV}{\actVV}  \steq \actNew{\idV}{( \actV \actPar \actVV )} } 
  \end{mathpar}
  \vspace{-1em}
  \caption{Language for Actor Systems}
  \label{fig:erlang-semantics}
\end{figure*}

\Cref{fig:erlang-semantics} presents the syntax of our model actor language. 
%
%
This grammar 
assumes a set of \emph{disjoint} actor names/addresses $\idV,\idVV, \idVVV,\idVVVV \in \Pid$, atoms $\atV,\atVV \in \Atom$, expression variables $\xV,\xVV \in \Vars$, and term variables $\XV,\XVV \in \TVar$.
Values, $v\in\Val$, range over $\Pid \cup \Atom$ and can be sent as messages. 
Identifiers, $w$, are syntactic entities that range over values and variables.  
An actor system, $\actV,\actVV\in\Actors$, consists of multiple parallel actors $\actV\actPar\actVV$, which can either be inactive, \nl, or locally \emph{scoped} to a subsystem of actors, 
$\actNew{\idV}{\actV}$.
A system may also have a number messages in transit; a message in the ether carrying value $v$ addressed to \idV is denoted as \pioutB{\idV}{\vV}.
%
%
%
Individual actors, $\actrs{\idV}{\eV}{\qV}$, are uniquely identifiable by their name, \idV, 
and consist of a running expression $\eV$ and a mailbox $\qV$, \ie  a list of values denoting a message queue. 
Incoming messages are added at the end of the queue, whereas pattern-matched messages are removed from the front of the queue.   
We use \ccat{\qV}{\qVV} to denote queue concatenation, \ccat{\vV}{\qV} for the mailbox with \vV and \qV at the head and tail of the queue, and \ccat{\qV}{\vV} for the mailbox with \vV at the end of the queue.
When the mailbox is empty, \mEmpty, we often elide it from the individual actor and write \actrss{\idV}{\eV} instead of \actrs{\idV}{\eV}{\mEmpty}.    
%
%
Actor expressions $\eV,\eVV\in\Exp$ can be outputs, $\piout{w_1}{w_2}{\eV}$, or reading inputs from the mailbox through pattern-matching, $\rcv{ \{ \patV_n \rTran \eV_n \}_{n\in I} }$, where each expression $\eV_n$ is guarded by pattern $\patV_n$. 
We assume patterns are disjoint, \ie if some $\vV$ matches $\patV_i$, it does not match any other $\patV_j$ for $i\neq j$ and $i,j\in\K$.
Expressions can also consist of self references (to the actor's own name), 
$\slf{\xV}{\eV}$, 
actor spawning, $\spwn{\eVV}{\xV}{\eV}$,
or recursion, \recX{\eV}.

We assume the standard definitions \fn{\actV} and 
\fv{\actV}
for the free names/variables of an actor system \actV
and work up to $\alpha$-conversion of bound names/variables.
We also write \fid{\actV} for the free names \idV\ of the individual actors \actrs{\idV}{\eV}{\qV} in \actV.
%
\Eg for 
$\actV = \actrss{\idV}{\eV_1} \actPar \actrss{\idVV}{\eV_2} \actPar \actNew{\idVVV}{\actrss{\idVVV}{\eV_3}}$, 
we have $\fid{\actV} = \{ \idV, \idVV \}$ and $\fn{\actV} = \{ \idV,\idVV\} \cup \fn{\eV_1} \cup \fn{\eV_2} \cup (\fn{\eV_1} \setminus \sset{\idVVV})$.
Running actor systems are closed, \ie $\fv{\actV}{=}\emptyset$ and respect the single receiver property, 
\ie if $\actV = \actVV_1\actPar\actVV_2$ then $\fid{\actVV_1} \cap \fid{\actVV_2} = \emptyset$. 
For syntactic objects $o,o'$, we write $o \fresh o'$ to mean that the free names of $o$ and $o'$ are disjoint, \eg $\actV \fresh \actVV$ denotes $\fn{\actV}\cap\fn{\actVV} {=} \emptyset$. 
%
%
We also write $\K,d$ 
to mean $\K{\uplus}\{ d\}$, 
where $\uplus$ denotes disjoint union, 
%
Substitutions are partial maps from variables to values, $\sV {\in} \Subs\!:\! \Vars \pmap \Val$. 

The operational semantics of our language 
is given in terms of an ILTS. 
\emph{Knowledge} $\K\subseteq\Pid$ 
denotes the set of names known by an actor system \actV\ and an implicit observer with which it interacts; $K$ is used by the rules in \Cref{fig:erlang-semantics} to keep track of bound/free names and abstract away  from name bindings in actions~\cite{DBLP:books/daglib/0004377,DBLP:books/daglib/0018113}; see \cite{BengstonParrow09:nominalPi}.
The \emph{observer} $\Obs\subseteq\Pid$ is represented by the set of addresses 
that \actV\ interacts with.
Transitions are defined over system states of the form $\conf{\K\;|\;\Obs}{\actV}\in\Prc$ where
$\fn{\actV}\subseteq\K$, $\Obs\subseteq\K$ and $\fid{\actV} \fresh \Obs$ (respecting the single receiver property).
The ILTS transitions of the form 
\begin{math}
  \conf{\K\;|\;\Obs}{\actV} \traS{\stact} \conf{\K'\;|\;\Obs'}{\actVV}
\end{math}
are governed by the judgement
  $\confK{\actV} \traS{\stact} \actVV$
%
defined by the rules in \Cref{fig:erlang-semantics}.
The evolution of $\K$ and $\Obs$ after $\stact$ is left implicit in 
$\conf{\K\;|\;\Obs}{\actV} \traS{\stact} \actVV$
since it is determined by the function $\after{\K\;|\;\Obs, \stact}$ (below).

\begin{definition}
  \label{def:aft}
  %
  $\after{\K\,|\,\Obs,\stact}$ is inductively defined as follows:
  %
  %
    \begin{align*}
      & \after{\K\;|\;\Obs,\pioutA{\idV}{\vV}} \deftxt \K \;|\; \Obs
      \quad \after{\K\;|\;\Obs,\npioutA{\idV}{\vV}} \deftxt \K \cup \fn{\vV} \;|\; \Obs \\
      & \after{\K\;|\;\Obs,\tau} \deftxt \K\;|\;\Obs 
      \quad \after{\K\,|\,\Obs,\piinA{\idV}{\vV}} \deftxt \K {\cup} \fn{\vV} \,|\, \Obs{\cup} (\fn{\vV}{\setminus} \K)  \\
      & \after{\K\;|\;\Obs,\ncommA} \deftxt \K\;|\;\Obs \!\!
      \quad \after{\K\;|\;\Obs,\commA{\idV}{\vV}} \deftxt \K\;|\;\Obs  
      \tag*{\exqed}
    \end{align*}
  %
\end{definition}

Actors communicate through
asynchronous messages, which are sent in two stages: the actor first creates a message \pioutB{\idVV}{\vV} in the ether 
(rule \textsc{snd1}), and then the ether sends value $\vV$ to actor $\idVV$ (rule \textsc{snd2}). 
Once received, messages are appended to the recipient's local mailbox (rule \textsc{rcv})
and \emph{selectively} read following rule \textsc{rd}. 
This relies on the 
helper functions \absent{-} and \match{-} in \Cref{def:pattern} to find the first message \vV\ in the mailbox that matches one of the 
patterns $\pV_m$ in $\{ \pV_n \!\tra{\;}\! \eV_n\}_{n\in I}$.  
%
%
%
If a match is found, the actor branches to $\eV_m\sigma$, where $\eV_m$ is the expression guarded by the matching pattern $\pV_m$ and $\sigma$ substitutes the free variables in $\eV_m$ for the values resulting from the pattern-match.
%
Otherwise, reading blocks.
%
%
%
%
%
Parallel actors, $\actV\!\actPar\!\actVV$, may \emph{internally communicate} via rule \textsc{ncommL} whenever $\actV$ and $\actVV$ can  respectively transition with 
dual output and input actions, \pioutA{\idV}{\vV} and \piinA{\idV}{\vV}, binding all names extruded by $\vV$
in the process, 
or via rule \textsc{commL} 
if all names in \vV\ are already known 
(symmetric rules \textsc{ncommR}, \textsc{commR} elided).
Actors may also transition independently 
with rule \textsc{parL} (symmetric rule \textsc{parR} elided); 
the 
condition $\sbj{\mu}\!\fresh\!\fid{\actVV}$ enforces the \emph{single-receiver property} and checks 
the message is not destined for 
an actor in \actVV.
An actor may also \emph{scope extrude} names by communicating 
bound names to actors outside the scope (rule \textsc{opn}).
%
%
Dually, when bound 
names are not mentioned in the action along which the transition occurs or the action denotes internal communication \ncommA, the names remain bound (rules \textsc{scp1}, \textsc{scp2}).
The remaining rules, \textsc{slf} and \textsc{spw}, are standard. 
%
Our ILTS semantics uses structural equivalence for actor systems $\actV {\steq} \actVV$, 
lifted as the process equivalence relation from \Cref{sec:preliminaries}, \ie $(\conf{\K_1\,|\,\Obs_1}{\actV_1}) \steq (\conf{\K_2\;|\;\Obs_2}{\actV_2})$ whenever $\K_1 = \K_2$, $\Obs_1 = \Obs_2$ and $\actV_1 \steq \actV_2$.  

\subsection{Actor Structural Equivalence and Silent Actions}

To show that our semantics is indeed an ILTS, we need to prove a few additional properties. 
\Cref{prop:equivalent-states} below shows that transitions abstract over structurally-equivalent states.


\equivalentStates*

As a result of \Cref{thm:confluence} below, we are guaranteed that any actor SUS instrumented via a mechanism that implements the semantics in \Cref{fig:mon}  can safely abstract over (non-traceable) silent transitions because they are confluent \wrt\ other actions.


\actorconfluence*


\subsection{Actor Traceable Actions}

Our actor semantics uses three forms of external actions, 
\begin{align*}
  \EAct & = \setof{\piinA{\idV}{\vV},\ \pioutA{\idV}{\vV},\ \npioutA{\idV}{\idVV}}{\idV,\idVV \in \Pid, \vV \in \Val}  
\end{align*} 
Apart from input actions, \piinA{\idV}{\vV}, and output actions, \pioutA{\idV}{\vV}, we identify a specific form of outputs, \npioutA{\idV}{\idVV}, where the payload $\idVV$ is \emph{scope-extruded} to the observer, which manifests itself as $\idVV\notin \K$ in our setting.
See rule \textsc{opn} in \Cref{fig:erlang-semantics}.
Our semantics also employs two forms of internal actions, 
\begin{align*}
  \IAct = \setof{\commA{\idV}{\vV},\ \ncommA}{\idV \in \Pid, \vV \in \Val}
\end{align*}
We model actor communication via \emph{internal communication actions}, \commA{\idV}{\vV}, as opposed to using silent actions as is standard in~\cite{DBLP:books/daglib/0018113,DBLP:books/daglib/0004377}. 
This permits the instrumented monitors to differentiate between different communication steps which can reach states that are not necessarily behaviourally equivalent.
The exception to this strategy 
is internal communication involving scoped names, \ncommA; see rules \textsc{ncommL} and \textsc{scp2} in in \Cref{fig:erlang-semantics}.
We still allow our monitor instrumentation to differentiate these transitions from silent actions, mainly because they do not satisfy properties such as \Cref{thm:confluence}, and thus treat them differently during runtime verification.

Our ILTS interpretation treats input, output and internal communication actions as deterministic;
%
This treatment is justified by 
\Cref{prop:input-determinacy,prop:output-determinacy,prop:comm-determinacy}.
%
For the full proofs, refer to the dedicated sections, \Cref{sec:in-det,sec:out-det,sec:comm-det}.

%

%


\begin{restatable}[Input Determinacy]{proposition}{inDet} 
  \label{prop:input-determinacy}
  If $\confK{\actV} \traS{\piinA{\idV}{\vV}} \actV'$ and $\confK{\actV} \traS{\piinA{\idV}{\vV}} \actV''$ then $\actV' \steq \actV''$.
\end{restatable}

\begin{proof}
  By rule induction on the two transitions, relying also on the single-receiver property.
\end{proof}

\begin{restatable}[Output Determinacy]{proposition}{outDet} 
  \label{prop:output-determinacy}
  If $\confK{\actV} \traS{\pioutA{\idV}{\vV}} \actV'$ and $\confK{\actV} \traS{\pioutA{\idV}{\vV}} \actV''$ then $\actV' \steq \actV''$.
  %
\end{restatable}

\begin{proof}
  By rule induction on the two transitions, relying also on $\steq$ from \Cref{fig:erlang-semantics}.
\end{proof}

\begin{restatable}[Communication Determinacy]{proposition}{commDet} 
  \label{prop:comm-determinacy}
  If $\confK{\actV} \traS{\commA{\idV}{\vV}} \actV'$ and $\confK{\actV} \traS{\commA{\idV}{\vV}} \actV''$ then $\actV' \!\steq\! \actV''$.
\end{restatable}

\begin{proof}
  By rule induction on the two transitions, relying also on \Cref{prop:input-determinacy,prop:output-determinacy}.
\end{proof}

\section{Properties of Actor Systems}
\label{sec:actor-properties}

We formalise (\resp prove) the omitted definitions (\resp results) from \Cref{sec:actor-systems}.
In particular, the \emph{subject} of an action 
is defined as $\sbj{\tau}=\sbj{\commA{\idV}{\vV}} = \sbj{\ncommA} = \emptyset$ 
and $\sbj{\piinA{c}{d}}=\sbj{\pioutA{c}{d}}=\{c\}$.
\Cref{def:pattern} below describes 
the two helper functions \absent{-} and \match{-} in \Cref{fig:erlang-semantics}.

\begin{definition}[Pattern matching]\label{def:pattern}
  We define $\textsl{match}: \Pat \times \Val \traS{\;} \Subs\cup\{\bot\}$ and $\textsl{absent}: \Pat \times \MBox \traS{\;} \Bool$ as 
  %
  {
  \small
  \begin{align*}
    \match{\pV, \vV} &{=}
    \begin{cases}
      \emptyset           &\text{if } \pV=\vV=\idV \text{ or } \pV=\vV=\idV=\atV \\
      \subC{\vV}{\xV}     &\text{if } \pV=\xV \\
      \biguplus_{i = 1}^{n} \sV_i     
      & \text{if } 
      \pV = \{ \pV_1, ..., \pV_n \},\; \vV {=} \{\vV_1, ..., \vV_n \}, \\
      &\quad \forall i{\in}\{1,...,n\} \cdot \match{\pV_i, \vV_i} {=} \sV_i \\
      \bot                &\text{otherwise}
    \end{cases}\\[0.5em]
    \sV_1 \uplus \sV_2 &= 
      \begin{cases}
        \sV_1 \cup \sV_2  &\text{if } \dom{\sV_1} \cap \dom{\sV_2} = \emptyset \\
        \sV_1 \cup \sV_2  &\text{if } \forall \vV \in \dom{\sV_1} \cap \dom{\sV_2}, \\
        & \quad \sV_1(\vV) = \sV_2(\vV) \\
        \bot              &\text{if } \sV_1 = \bot \text{ or } \sV_2 = \bot \\
        \bot              &\text{otherwise}  
      \end{cases}\\[0.5em]
    \absent{\pV, \mEmpty} &= \btrue \\
    \absent{\pV, \ccat{\vV}{\qV}} &=
    \begin{cases}
      \bfalse             &\text{if } \match{\pV, \vV} = \bot \\
      \absent{\pV,\qV}    &\text{otherwise} 
      \tag*{\qed} 
    \end{cases}
  \end{align*} 
  } %
  \label{def:match}
\end{definition}

\subsection{General Results}

We give some general properties of actor systems that will be used in the following sections. 
%

\begin{lemma}\label{lemma:in-inK}
  If $\confK{\actV} \traS{ \piinA{\idV}{\vV} } \actVV$ then $i\in\fid{\actV}$.
\end{lemma}

\begin{proof}
  Straightforward by rule induction.
\end{proof}

\begin{corollary}\label{cor:no-fid-no-in}
  If $\idV\notin\fid{\actV}$ then $\confK{\actV} \traSN{ \piinA{\idV}{\vV} } \actVV$.
\end{corollary}


\begin{proof}
  Straightforward by rule induction.
\end{proof}

\begin{lemma}\label{lemma:fid-tau}
  If $\confK{\actV} \traS{\tau} \actVV$ then $\fid{\actVV} \subseteq \fid{\actV}$.
\end{lemma}

\begin{proof}
  Straightforward by rule induction.
\end{proof}

\Cref{lemma:in-anyO} below states that if a system can perform an input action, then that transition is always possible, regardless of the external observer \Obs\ \wrt which it is executing. 

\begin{lemma}\label{lemma:in-anyO}
  If $\confK{\actV} \traS{\piinA{\idV}{\vV}} \actVV$ then $\conf{\K\;|\;\Obs'}{\actV} \traS{\piinA{\idV}{\vV}} \actVV$ for every observer $\Obs'$.
\end{lemma}

\begin{proof}
  The proof proceeds by induction on $\confK{\actV} \traS{\piinA{\idV}{\vV}} \actVV$.
  \begin{itemize}[leftmargin=*]
    \setlength\itemsep{0.5em}
    \item Case \textsc{rcv}, \ie $\confK{\actrs{\idV}{\eV}{\qV}} \traS{\piinA{\idV}{\vV}} \actrs{\idV}{\eV}{\ccat{\qV}{\vV}}$.
          Our result, $\conf{\K\;|\;\Obs'}{\actrs{\idV}{\eV}{\qV}} \traS{\piinA{\idV}{\vV}} \actrs{\idV}{\eV}{\ccat{\qV}{\vV}}$, follows 
          by rule \textsc{rcv}.
    \item Case \textsc{scp1}, \ie $\confK{\actNew{\idVV}{\actV}} \traS{\piinA{\idV}{\vV}} \actNew{\idVV}{\actVV}$
          because $\conf{\K,\idVV\;|\;\Obs}{\actV} \traS{\piinA{\idV}{\vV}} \actVV$ and $\idVV \fresh \fn{\piinA{\idV}{\vV}}$.
          By the IH, we obtain $\conf{\K,\idVV\;|\;\Obs'}{\actV} \traS{\piinA{\idV}{\vV}} \actVV$.
          Our result, $\conf{\K \;|\; \Obs'}{\actNew{\idVV}{\actV}} \traS{\piinA{\idV}{\vV}} \actNew{\idVV}{\actVV}$, follows by rule \textsc{scp1}.
    \item Case \textsc{parL}, \ie $\confK{\actV \actPar \actVV} \traS{\piinA{\idV}{\vV}} \actV' \actPar \actVV$
          because $\confK{\actV} \traS{\piinA{\idV}{\vV}} \actV'$ and $\idV \fresh \fid{\actVV}$.
          By the IH, we obtain $\conf{\K\;|\;\Obs'}{\actV} \traS{\piinA{\idV}{\vV}} \actV'$.
          Our result, $\conf{\K\;|\;\Obs'}{\actV \actPar \actVV} \traS{\piinA{\idV}{\vV}} \actV' \actPar \actVV$, follows by \textsc{parL}.
    \item Case \textsc{parR}. The proof is analogous to that for \textsc{parL}.  \qedhere
  \end{itemize}
\end{proof}

Similarly, \Cref{lemma:aft-k-independent} 
states that if a system can $\tau$-transition, then that transition is always possible, regardless of the knowledge \K\ and external observer \Obs\ \wrt which it is executing. 

\begin{lemma}\label{lemma:aft-k-independent}
  $\confK{\actV} \traS{\tau} \confK{\actVV}$ implies $\conf{\K'\;|\;\Obs'}{\actV} \traS{\tau} \conf{\K'\;|\;\Obs'}{\actVV}$ for all knowledge $\K,\K'$ and observers $\Obs,\Obs'$.
\end{lemma}

\begin{proof}
  Straightforward by rule induction.
\end{proof}

\subsection{Inversion Lemmas}

We also prove several results that provide insights into the structure and behaviour of actor systems.






\begin{lemma}\label{lemma:inversion-steq}
  If $\actV \steq \actV_1 \actPar \actV_2$ then one of the following statements must hold: 
  \begin{itemize}[leftmargin=*]
    \item $\actV = \actV_1$ and $\actV_2 = \nl$, or $\actV = \actV_2$ and $\actV_1=\nl$
    \item $\actV_1 \steq \actNew{\idLst_1}{\actV_1'} \actPar \actNew{\idLst_2}{\actV_1''}$ and $\actV_2 \steq \actNew{\idLst_3}{\actV_2'} \actPar \actNew{\idLst_4}{\actV_2''}$ and 
    $\actV = \actNew{\idLst_1,\idLst_2,\idLst_3,\idLst_4}{(\actVV_1 \actPar \actVV_2)}$ and 
    $\actVV_1 = \actV_1'\actPar \actV_2'$ and 
    $\actVV_2 = \actV_1''\actPar \actV_2''$.
  \end{itemize}
\end{lemma}

\begin{proof}
  By rule induction on $\actV \steq \actV_1 \actPar \actV_2$. 
\end{proof}

\begin{lemma}\label{lemma:inversion-input}
  If $\actV \steq \actV_1\actPar\actV_2$ and $\confK{\actV} \traS{\piinA{\idV}{\vV}} \actVV$ then 
    \begin{enumerate}[leftmargin=*]
      \item either 
      $\confK{\actV_1} \traS{\piinA{\idV}{\vV}} \actV_1'$ and $\actVV \steq \actV_1'\actPar\actV_2$;
      \item or 
      $\confK{\actV_2} \traS{\piinA{\idV}{\vV}} \actV_2'$ and $\actVV \steq \actV_1\actPar\actV_2'$.
    \end{enumerate}
\end{lemma}

\begin{proof}
  Suppose that $\actV \steq \actV_1\actPar\actV_2$ and $\confK{\actV} \traS{\piinA{\idV}{\vV}} \actVV$.  
  The proof proceeds by rule induction on the latter. 
  \begin{itemize}[leftmargin=*]
    \setlength\itemsep{0.5em}
    \item Case \textsc{rcv}, \ie $\actV=\actrs{\idV}{\eV}{\qV}$ and $\actV'= \actrs{\idV}{\eV}{\ccat{\qV}{\vV}}$.
      Since $\actV = \actV_1\actPar \actV_2$, then by \Cref{lemma:inversion-steq}, we must have either $\actV = \actV_1$ and $\actV_2 = \nl$, or $\actV = \actV_2$ and $\actV_1 = \nl$. 
      In the first case, condition~(1) is satisfied since $\actrs{\idV}{\eV}{\ccat{\qV}{\vV}} \steq \actrs{\idV}{\eV}{\ccat{\qV}{\vV}} \actPar \nl$.
      Otherwise, condition~(2) is satisfied since $\actrs{\idV}{\eV}{\ccat{\qV}{\vV}} \steq \nl\actPar \actrs{\idV}{\eV}{\ccat{\qV}{\vV}}$. 
    \item Case \textsc{parL}, \ie $\actV = \actV_3 \actPar \actV_4$ and $\actVV = \actVV_3 \actPar \actV_4$ because $\confK{\actV_3} \traS{\piinA{\idV}{\vV}} \actVV_3$. 
      By \Cref{lemma:inversion-steq} and $\actV = \actV_3 \actPar \actV_4$, we must have 
      $\actV_1 \steq \actV_1' \actPar \actV_1''$ and $\actV_2 \steq \actV_2' \actPar \actV_2''$ and 
      $\actV_3 = \actV_1'\actPar \actV_2'$ and $\actV_4 = \actV_1''\actPar \actV_2''$. 
      Using the facts that $\actV_3 = \actV_1'\actPar \actV_2'$ and $\confK{\actV_3} \traS{\piinA{\idV}{\vV}} \actVV_3$ and the IH, we know that  
      \begin{align*}
        &\text{either } \confK{\actV_1'} \traS{\piinA{\idV}{\vV}} \actVV_1' \text{ and } \actVV_3 \steq \actVV_1' \actPar \actV_2' \\
        &\text{or } \confK{\actV_2'} \traS{\piinA{\idV}{\vV}} \actVV_2' \text{ and } \actVV_3 \steq \actV_1' \actPar \actVV_2'
      \end{align*}
      Applying rule \textsc{parL} on the transitions and rule \textsc{sCtxP} on the equivalences, these respectively give us that either
      \begin{align*}
        &\confK{\actV_1'\actPar \actV_1''} \traS{\piinA{\idV}{\vV}} \actVV_1'\actPar\actV_1'' 
        \text{ and } \actVV_3 \actPar \actV_4 \steq (\actVV_1' \actPar \actV_2') \actPar \actV_4 \\
        &\qquad \qquad \text{or} \\
        &\confK{\actV_2'\actPar \actV_2''} \traS{\piinA{\idV}{\vV}} \actVV_2'\actPar\actV_2'' 
        \text{ and } \actVV_3\actPar\actV_4 \steq (\actV_1' \actPar \actVV_2' ) \actPar  \actV_4
      \end{align*}
      Using $\actV_4 = \actV_1''\actPar \actV_2''$, $\actV_1 = \actV_1'\actPar \actV_1''$, $\actV_2 = \actV_2'\actPar \actV_2''$, $\actVV = \actVV_3\actPar\actV_4$ and rules for $\steq$, we can rewrite this as 
      \begin{align*}
        &\text{either } \confK{\actV_1} \traS{\piinA{\idV}{\vV}} \actVV_1'\actPar\actV_1'' 
          \text{ and } \actVV \steq (\actVV_1' \actPar  \actV_1'') \actPar  \actV_2 \\
        &\text{or } \confK{\actV_2} \traS{\piinA{\idV}{\vV}} \actVV_2'\actPar\actV_2'' 
          \text{ and } \actVV \steq \actV_1 \actPar (\actVV_2' \actPar \actV_2'')
      \end{align*}
      which correspond to conditions (1) and (2).
    \item Case \textsc{parR}, similar to that for \textsc{parL}.
    \item Case \textsc{scp1}, \ie $\actV = \actNew{\idVV}{\actV'}$ and $\actVV = \actNew{\idVV}{\actV''}$ because 
      $\conf{\K,\idVV\;|\;\Obs}{\actV'} \traS{\piinA{\idV}{\vV}} \actV''$ and $\idVV\fresh \fn{\idV,\vV}$.
      By \Cref{lemma:inversion-steq}, there are two subcases to consider: 
      \begin{itemize}[leftmargin=*]
        \setlength\itemsep{0.5em}
        \item When $\actV = \actV_1$ and $\actV_2 = \nl$, by this and $\confK{\actNew{\idVV}{\actV'}} \traS{\piinA{\idV}{\vV}} \actNew{\idVV}{\actV''}$, we know $\confK{\actV_1} \traS{\piinA{\idV}{\vV}} \actNew{\idVV}{\actV''}$. 
        By rule \textsc{sNil}, we conclude $\actVV = \actNew{\idVV}{\actV''} \steq \actVV \actPar \nl = \actVV \actPar \actV_2$ as required.   
        \item When $\actV_1 \steq \actNew{\idLst_1}{\actV_1'} \actPar \actNew{\idLst_2}{\actV_1''}$ and 
          $\actV_2 \steq \actNew{\idLst_3}{\actV_2'} \actPar \actNew{\idLst_4}{\actV_2''}$ such that  
          \begin{align*}
            & \actV = \actNew{\idLst_1,\idLst_2,\idLst_3,\idLst_4}{(\actV_3 \actPar \actV_4)} \text{ where } \\
            &\actV_3 = \actV_1'\actPar \actV_2'
            \quad \text{and} \quad
            \actV_4 = \actV_1''\actPar \actV_2''
          \end{align*}
          Since $\actV = \actNew{\idVV}{\actV'}$, we must have that $\idVV\in\idLst_i$ for some $i\in\sset{1,2,3,4}$.
          Consider the case for when $i=1$; other cases follow with similar reasoning. 
          Let $\idLst_5 = \idLst_1 \setminus \{ \idVV\}$.
          Then we know 
          \begin{align}
            \actV' = \actNew{\idLst_5,\idLst_2,\idLst_3,\idLst_4}{(\actV_3 \actPar \actV_4)}  \label{eq:inv-in-1}
          \end{align}
          Since we work up to $\alpha$-conversion of bound entities, we can assume that
          $\idLst_1\fresh\fn{\actV_4}$ and $\idLst_3\fresh\fn{\actV_4}$ 
          and $\idLst_2\fresh\fn{\actV_3}$ and $\idLst_4\fresh\fn{\actV_3}$.
          Using the fact that $\idLst_5\subset\idLst_1$ and the rules defining $\steq$ in \Cref{fig:erlang-semantics}, we also know 
          \begin{align}
            &
            \actV' \steq 
            \actNew{\idLst_5,\idLst_3}{\actV_3}\actPar \actNew{\idLst_2,\idLst_4}{\actV_4}
            \label{eq:inv-in-2} 
          \end{align}
          By (\ref{eq:inv-in-1}), the fact that $\conf{\K,\idVV\;|\;\Obs}{\actV'} \traS{\piinA{\idV}{\vV}} \actV''$ and the IH, we obtain that either 
          \begin{align}
            &\conf{\K,\idVV\;|\;\Obs}{\actNew{\idLst_5,\idLst_3}{\actV_3}} \traS{\piinA{\idV}{\vV}} \actVV_1 
            \text{ and } \actV'' \steq \actVV_1 \actPar \actNew{\idLst_2,\idLst_4}{\actV_4}
            \notag \\
            &\qquad 
            \text{ or } \label{eq:inv-in-3} \\
            &\conf{\K,\idVV\;|\;\Obs}{\actNew{\idLst_2,\idLst_4}{\actV_4}} \traS{\piinA{\idV}{\vV}} \actVV_2 
            \text{ and } \actV'' \steq \actNew{\idLst_5,\idLst_3}{\actV_3} \actPar \actVV_2 \label{eq:inv-in-4}
          \end{align} 
          If (\ref{eq:inv-in-3}) holds, applying rules \textsc{scp1},\textsc{sCtxS},\textsc{sExt},\textsc{sCom} gives us that
          \begin{align*}
            &\actNew{\idLst_1,\idLst_3}{\actV_3} \traS{\piinA{\idV}{\vV}} \actNew{\idVV}{\actVV_1}  
            \text{ and } \\ 
            &\actNew{\idVV}{\actV''} \steq \actNew{\idVV}{\actVV_1} \actPar \actNew{\idLst_2,\idLst_4}{\actV_4}
          \end{align*}
          which corresponds to statement $(1)$ as required.
          If (\ref{eq:inv-in-4}) holds, applying rules \textsc{scp1},\textsc{sCtxS},\textsc{sExt},\textsc{sCom} give us  
          \begin{align*}
            &\confK{\actNew{\idVV,\idLst_2,\idLst_4}{\actV_4}} \traS{\piinA{\idV}{\vV}} \actNew{\idVV}{\actVV_2} 
            \text{ and } \\
            &\actNew{\idVV}{\actV''} \steq \actNew{\idLst_5,\idLst_3}{\actV_3} \actPar \actNew{\idVV}{\actVV_2}
          \end{align*}
          which corresponds to statement $(2)$ as required.
        \end{itemize}
    \item Case \textsc{sTrn}, proof is straightforward. \qedhere
  \end{itemize}
\end{proof}

  

\begin{corollary}\label{cor:inversion-input-basic}
  If $\actrs{\idV}{\eV}{\qV} \steq \actV$ and $\confK{\actV} \traS{\piinA{\idV}{\vV}} \actVV$ then $\actVV \steq \actrs{\idV}{\eV}{\ccat{\qV}{\vV}}$.
\end{corollary}
  
\begin{proof}
  Follows from \Cref{lemma:inversion-input} since 
  $\confK{\actrs{\idV}{\eV}{\qV}} \traS{\piinA{\idV}{\vV}} \actrs{\idV}{\eV}{\ccat{\qV}{\vV}}$ and 
  $\actrs{\idV}{\eV}{\qV} \steq \actV \actPar \nl$.
\end{proof}

\begin{lemma}\label{lemma:inversion-output}
  If $\actV \steq \actV_1\actPar\actV_2$ and $\confK{\actV} \traS{\pioutA{\idV}{\vV}} \actVV$ then 
    \begin{enumerate}[leftmargin=*]
      \item either $\confK{\actV_1} \traS{\pioutA{\idV}{\vV}} \actV_1'$ and $\actVV \steq \actV_1'\actPar\actV_2$;
      \item or $\confK{\actV_2} \traS{\pioutA{\idV}{\vV}} \actV_2'$ and $\actVV \steq \actV_1\actPar\actV_2'$.
    \end{enumerate}
\end{lemma}

\begin{proof}
  The proof is similar to that for \Cref{lemma:inversion-input}.
\end{proof}

\begin{lemma}\label{lemma:inversion-scp1}
  If $\actV \steq \actNew{\idV}{\actV'}$ and $\confK{\actV} \traS{\stact} \actVV$ and $\idV\fresh\fn{\stact}$ then 
  $\actVV \steq \actNew{\idV}{\actVV'}$ and $\conf{\K,\idV \;|\; \Obs}{\actV'} \traS{\stact} \actVV'$.
\end{lemma}

\begin{proof}
  Proof is straightforward.
\end{proof}

\begin{lemma}\label{lemma:inversion-par-stact}
  If $\actV \steq \actV_1\actPar \actV_2$ and $\confK{\actV} \traS{\stact} \actVV$ then one of the following statements must hold: 
  \begin{enumerate}[leftmargin=*]
    \item $\confK{\actV_1} \traS{\stact} \actV_1'$ and $\actVV \steq \actV_1'\actPar\actV_2$ and $\sbj{\stact}\fresh\fid{\actV_2}$
    \item $\confK{\actV_2} \traS{\stact} \actV_2'$ and $\actVV \steq \actV_1\actPar\actV_2'$ and $\sbj{\stact}\fresh\fid{\actV_1}$
    \item $\stact = \commA{\idV}{\vV}$ 
          and $\conf{\K\;|\;\fid{\actV_2}}{\actV_1} \traS{\pioutA{\idV}{\vV}} \actVV_1$
          and $\conf{\K\;|\;\fid{\actV_1}}{\actV_2} \traS{\piinA{\idV}{\vV}} \actVV_2$
          and $\actVV \steq \actVV_1 \actPar \actVV_2$
    \item $\stact = \commA{\idV}{\vV}$ 
          and $\conf{\K\;|\;\fid{\actV_2}}{\actV_1} \traS{\piinA{\idV}{\vV}} \actVV_1$
          and $\conf{\K\;|\;\fid{\actV_1}}{\actV_2} \traS{\pioutA{\idV}{\vV}} \actVV_2$
          and $\actVV \steq \actVV_1 \actPar \actVV_2$
    \item $\stact = \ncommA$ 
          and $\conf{\K\;|\;\fid{\actV_2}}{\actV_1} \traS{\npioutA{\idV}{\idVV}} \actVV_1$
          and $\conf{\K\;|\;\fid{\actV_1}}{\actV_2} \traS{\piinA{\idV}{\idVV}} \actVV_2$
          and $\actVV \steq \actNew{\idVV}{(\actVV_1 \actPar \actVV_2)}$
    \item $\stact = \ncommA$ 
          and $\conf{\K\;|\;\fid{\actV_2}}{\actV_1} \traS{\piinA{\idV}{\idVV}} \actVV_1$
          and $\conf{\K\;|\;\fid{\actV_1}}{\actV_2} \traS{\npioutA{\idV}{\idVV}} \actVV_2$
          and $\actVV \steq \actNew{\idVV}{(\actVV_1 \actPar \actVV_2)}$
  \end{enumerate}
\end{lemma}

\begin{proof}
  We omit the proof due to its length. 
  However, it can be proven via rule induction on $\confK{\actV} \traS{\stact} \actVV$, using also \Cref{lemma:inversion-steq}. 
  The method is similar to that for \Cref{lemma:inversion-input}.
\end{proof}

\subsection{Actor Structural Equivalence and Silent Actions}

We prove \Cref{prop:equivalent-states} from \Cref{sec:actor-systems}.
This result states that transitions abstract over structurally-equivalent states.

\equivalentStates*

\begin{proof}
  Straightforward by induction on $\confK{\actV} \traS{\stact} \actV'$.  
\end{proof}

We can also show that (non-traceable) silent transitions are confluent \wrt\ other actions, \Cref{thm:confluence}.



\actorconfluence*
  
\begin{proof}
  Intuitively, this is true because if $\confK{\actV}$ does two \emph{different} moves $\confK{\actV} \traS{\tau} \actV'$ and $\confK{\actV} \traS{\stact} \actV''$, then both moves must have occurred in different components of $\actV$.
  The proof proceeds by induction on the derivation of the first move, $\confK{\actV} \traS{\tau} \actV'$.   

  All axioms are trivial.
  Rules \textsc{commL}, \textsc{commR}, \textsc{ncommL}, \textsc{ncommR}, \textsc{scp2} and \textsc{opn} do not occur in any derivation of a $\tau$ move. 
  For the inductive cases, the only non-straightforward rule is \textsc{parL} (the proof for \textsc{parR} is analogous).
  If $\confK{\actV} \traS{\tau} \confK{\actV'}$ was derived using this rule, then 
  \begin{align}
    &\confK{\actV_1 \actPar \actV_2} \traS{\tau} \confK{\actV_1' \actPar \actV_2} \quad \notag \\ 
    &\text{ because }
    \confK{\actV_1} \traS{\tau} \confK{\actV_1'} \label{eq:beta-comm-1}
  \end{align} 
  We examine the proof for the second move, which must be of the form $\confK{\actV_1 \actPar \actV_2} \traS{\stact} \conf{\after{\K|\Obs, \stact}}{\actV''}$.
  By \Cref{lemma:inversion-par-stact}, there are six possible ways how this could have occurred. 
  We focus on the main cases:
  \begin{itemize}[leftmargin=*]
    \setlength\itemsep{0.5em}
    \item 
    $\actV'' = \actV_1 \actPar \actV_2'$ and $\confK{\actV_2} \traS{\stact} \conf{\K'}{\actV_2'}$ where $\K'\;|\;\Obs'=\after{\K\;|\;\Obs,\stact}$ and $\sbj{\stact}\fresh\fid{\actV_1}$.
    Diagrammatically
      $$
      \begin{tikzpicture}[shorten >=1pt,node distance=1.5cm,on grid,auto,scale=0.9]
        \node[] (A1) {\confK{\actV_1 \actPar \actV_2}};
        \node[] (A2) [right=of A1,xshift=3cm] {$\confK{\actV_1' \actPar \actV_2}$};
        \node[] (B1) [below=of A1]            {$\conf{\K'\;|\;\Obs'}{\actV_1 \actPar \actV_2'}$};
        
        \path[->]
          (A1) edge node {$\tau$}             (A2)
          (A1) edge node [left]{$\stact$}      (B1);
      \end{tikzpicture} 
      $$
      From \Cref{lemma:aft-k-independent} and (\ref{eq:beta-comm-1}), we get 
      $\conf{\K'\;|\;\Obs'}{\actV_1} \traS{\tau} \conf{\K'\;|\;\Obs'}{\actV_1'}$.\\
      Since $\sbj{\tau}=\emptyset$, thus $\sbj{\tau} \fresh \fid{\actV_2}$, we can apply rule \textsc{parL} to get the move     
      $\conf{\K'\;|\;\Obs'}{\actV_1 \actPar \actV_2'} \traS{\tau} \conf{\K'\;|\;\Obs'}{\actV_1' \actPar \actV_2'}$. \\
      By \Cref{lemma:fid-tau}, we also know $\fid{\actV_1'}\subseteq \fid{\actV_1}$, which implies $\sbj{\stact} \fresh \fid{\actV_1'}$.
      Rule \textsc{parR} can thus be applied to $\confK{\actV_2} \traS{\stact} \conf{\K'\;|\;\Obs'}{\actV_2'}$ to obtain the move $\confK{\actV_1' \actPar \actV_2} \traS{\stact} \conf{\K'\;|\;\Obs'}{\actV_1' \actPar \actV_2'}$.  
      This gives us the required commuting diagram 
      %
      $$
      \begin{tikzpicture}[shorten >=1pt,node distance=1.5cm,on grid,auto,scale=0.9]
        \node[] (A1)                          {\confK{\actV_1 \actPar \actV_2}};
        \node[] (A2) [right=of A1,xshift=3cm] {\confK{\actV_1' \actPar \actV_2}};
        \node[] (B1) [below=of A1]            {\conf{\K'\;|\;\Obs'}{\actV_1 \actPar \actV_2'}};
        \node[] (B2) [right=of B1,xshift=3cm] {\conf{\K'\;|\;\Obs'}{\actV_1' \actPar \actV_2'}};
  
        \path[->]
          (A1) edge             node {$\tau$}             (A2)
          (A1) edge             node [left]{$\stact$}      (B1)
          (A2) edge             node [left]{$\stact$}      (B2)
          (B1) edge             node {$\tau$}             (B2);
      \end{tikzpicture}
      $$
      %
    \item 
      $\actV''=\actV_1' \actPar \actV_2$ and $\confK{\actV_1} \traS{\stact} \conf{\K'\;|\;\Obs'}{\actV_1'}$ where $\K'\;|\;\Obs'=\after{\K\;|\;\Obs,\stact}$ and $\sbj{\stact} \fresh \fid{\actV_1}$. 
      In other words, we have to complete the diagram 
      $$
      \begin{tikzpicture}[shorten >=1pt,node distance=1.5cm,on grid,auto,scale=1]
        \node[] (A1) {\confK{\actV_1 \actPar \actV_2}};
        \node[] (A2) [right=of A1,xshift=3cm] {\confK{\actV_1' \actPar \actV_2}};
        \node[] (B1) [below=of A1]            {\conf{\K'\;|\;\Obs'}{\actV_1'' \actPar \actV_2}};
        \node[] (B2) [right=of B1,xshift=3cm] {\conf{\K'\;|\;\Obs'}{\actV_1''' \actPar \actV_2}};
        
        \path[->]
          (A1) edge node                                {$\tau$}       (A2)
          (A1) edge                       node [left]   {$\stact$}      (B1)
          (A2) edge [dashed]              node [left]   {$\stact$}      (B2)
          (B1) edge [dashed]              node          {$\tau$}       (B2);

      \end{tikzpicture} 
      $$
      %
      But note that we also have the diagram 
      $$
      \begin{tikzpicture}[shorten >=1pt,node distance=1.5cm,on grid,auto,scale=1]
        \node[] (A1)                          {\confK{\actV_1}};
        \node[] (A2) [right=of A1,xshift=2cm] {\confK{\actV_1'}};
        \node[] (B1) [below=of A1]            {\conf{\K'\;|\;\Obs'}{\actV_1''}};
        
        \path[->]
          (A1) edge node {$\tau$}              (A2)
          (A1) edge node [left]{$\stact$}       (B1);
      \end{tikzpicture} 
      $$  
      that can be completed by induction as 
      $$
      \begin{tikzpicture}[shorten >=1pt,node distance=1.5cm,on grid,auto,scale=1]
        \node[] (A1)                          {\confK{\actV_1}};
        \node[] (A2) [right=of A1,xshift=2cm] {\confK{\actV_1'}};
        \node[] (B1) [below=of A1]            {\conf{\K'\;|\;\Obs'}{\actV_1''}};
        \node[] (B2) [right=of B1,xshift=2cm] {\conf{\K'\;|\;\Obs'}{\actV_1'''}};
        
        \path[->]
          (A1) edge node {$\tau$}                (A2)
          (A1) edge node [left]{$\stact$}         (B1)
          (A2) edge node [left]{$\stact$}         (B2)
          (B1) edge node {$\tau$}                (B2);
      \end{tikzpicture} 
      $$
      %
      Applying \textsc{parL} twice on 
      $\conf{\K'\;|\;\Obs'}{\actV_1''} \traS{\tau} \conf{\K'\;|\;\Obs'}{\actV_1'''}$ and 
      $\confK{\actV_1'} \traS{\stact} \conf{\K'\;|\;\Obs'}{\actV_1'''}$ give the two required moves.
    \item 
      $\stact=\ncommA$ and $\actV''=\actNew{\idVV}{\bigl(\actV_1'' \actPar \actV_2''\bigr)}$ 
      and 
      $\conf{\K\;|\;\fid{\actV_2}}{\actV_1} \traS{\npioutA{\idV}{\idVV}} \conf{\K'\;|\;\Obs'}{\actV_1''}$ and 
      $\conf{\K\;|\;\fid{\actV_1}}{\actV_2} \traS{\piinA{\idV}{\idVV}} \conf{\K''\;|\;\Obs''}{\actV_2''}$ 
      for some name $\idV,\idVV\in\Pid$
      where $\K'\;|\;\Obs' = \after{\K\;|\;\fid{\actV_2},\npioutA{\idV}{\idVV}}$
      and $\K''\;|\;\Obs'' = \after{\K\;|\;\fid{\actV_1},\piinA{\idV}{\idVV}}$.
      In other words, we have to complete the diagram 
      %
      $$
      \begin{tikzpicture}[shorten >=1pt,node distance=1.5cm,on grid,auto,scale=1]
        \node[] (A1)                          {\confK{\actV_1 \actPar \actV_2}};
        \node[] (A2) [right=of A1,xshift=3cm] {\confK{\actV_1' \actPar \actV_2}};
        \node[] (B1) [below=of A1]            {\confK{\actNew{\idVV}{\bigl(\actV_1'' \actPar \actV_2''\bigr)}}};
        \node[] (B2) [right=of B1,xshift=3cm] {\confK{\actNew{\idVV}{\bigl(\actV_1'''\actPar \actV_2''\bigr)}}};

        \path[->]
          (A1) edge           node        {$\tau$}                 (A2)
          (A1) edge           node [left] {$\ncommA$}    (B1)
          (A2) edge [dashed]  node [left] {$\ncommA$}    (B2)
          (B1) edge [dashed]  node        {$\tau$}                 (B2);
      \end{tikzpicture} 
      $$  
      %
      But note that we also have the diagram 
      $$
      \begin{tikzpicture}[shorten >=1pt,node distance=1.5cm,on grid,auto,scale=1]
        \node[] (A1)                          {\conf{\K\;|\;\fid{\actV_2}}{\actV_1}};
        \node[] (A2) [right=of A1,xshift=2.5cm] {\conf{\K\;|\;\fid{\actV_2}}{\actV_1'}};
        \node[] (B1) [below=of A1]            {\conf{\K'\;|\;\Obs'}{\actV_1''}};
        
        \path[->]
          (A1) edge node        {$\tau$}                      (A2)
          (A1) edge node [left] {$\npioutA{\idV}{\idVV}$}     (B1);
      \end{tikzpicture} 
      $$  
      that can be completed by induction as
      $$
      \begin{tikzpicture}[shorten >=1pt,node distance=1.5cm,on grid,auto,scale=1]
        \node[] (A1)                            {\conf{\K\;|\;\fid{\actV_2}}{\actV_1}};
        \node[] (A2) [right=of A1,xshift=2.5cm] {\conf{\K\;|\;\fid{\actV_2}}{\actV_1'}};
        \node[] (B1) [below=of A1]              {\conf{\K'\;|\;\Obs'}{\actV_1''}};
        \node[] (B2) [right=of B1,xshift=2.5cm] {\conf{\K'\;|\;\Obs'}{\actV_1'''}};
        
        \path[->]
          (A1) edge node {$\tau$}                             (A2)
          (A1) edge node [left]{$\npioutA{\idV}{\idVV}$}      (B1)
          (A2) edge node [left]{$\npioutA{\idV}{\idVV}$}      (B2)
          (B1) edge node {$\tau$}                             (B2);
      \end{tikzpicture} 
      $$  
      Applying \textsc{ncommL} to 
      $\conf{\K\;|\;\fid{\actV_2}}{\actV_1'} \traS{\npioutA{\idV}{\idVV}} \conf{\K'\;|\;\Obs'}{\actV_1'''}$ and 
      $\conf{\K\;|\;\fid{\actV_1}}{\actV_2} \traS{\piinA{\idV}{\idVV}} \conf{\K''\;|\;\Obs''}{\actV_2''}$ gives the first required move
      $$
        \confK{\actV_1' \actPar \actV_2} \traS{\ncommA} \confK{\actNew{\idVV}{( \actV_1''' \actPar \actV_2'' )}}
      $$
      By \Cref{lemma:aft-k-independent} 
      and $\conf{\K;\;|\;\Obs'}{\actV_1''} \traS{\tau} \conf{\K'\;|\;\Obs'}{\actV_1'''}$, 
      we also know $\confK{\actV_1''} \traS{\tau} \confK{\actV_1'''}$.
      By rule \textsc{parL}, we get $\confK{\actV_1''\actPar \actV_2''} \traS{\tau} \confK{\actV_1'''\actPar \actV_2''}$.
      Then applying \textsc{scp1}, we obtain the second required move
      $$
        \confK{\actNew{\idVV}}{(\actV_1''\actPar \actV_2'')} \traS{\tau} \confK{\actNew{\idVV}{(\actV_1'''\actPar \actV_2'')}}
      $$
      %
      %
  \end{itemize} 
  The remaining cases follow with similar reasoning.
\end{proof}

\subsection{Actor Traceable Actions}

We show that input actions, output actions and communication actions are deterministic, 
\Cref{prop:determinacy}.
%

\Cref{prop:determinacy}, restated below, can be decomposed into three parts; namely, input determinacy, output determinacy and communication determinacy.  

\det*

\begin{proof}
  Follows from \Cref{prop:input-determinacy,prop:output-determinacy,prop:comm-determinacy}, proven in the dedicated sections below.
\end{proof}


\subsection{Proving Input Determinacy.}
\label{sec:in-det}

We show that input actions are deterministic, \Cref{prop:input-determinacy}. 
Its proof relies on \Cref{lemma:input-det} below.

\begin{lemma}\label{lemma:input-det}
  For any $\actV\steq\actV'$, 
  if $\confK{\actV} \traS{\piinA{\idV}{\vV}} \actVV$ and $\confK{\actV'} \traS{\piinA{\idV}{\vV}} \actVV'$ then $\actVV \steq \actVV'$.
\end{lemma}

\begin{proof}
  Suppose $\actV\steq\actV'$ and 
  $\confK{\actV} \traS{\piinA{\idV}{\vV}} \actVV$ and 
  $\confK{\actV'} \traS{\piinA{\idV}{\vV}} \actVV'$.
  We show $\actVV \steq \actVV'$.
  The proof proceeds by induction on the first move. 
  \begin{itemize}[leftmargin=*]
    \setlength\itemsep{0.5em}
    \item Case \textsc{rcv}, \ie $\confK{ \actrs{\idV}{\eV}{\qV}}  \traS{ \piinA{\idV}{\vV} }  \actrs{\idV}{\eV}{\ccat{\qV}{\vV}}$.
          For the second move, we thus have $\confK{\actV'}  \traS{ \piinA{\idV}{\vV} }  \actVV'$
          where $\actrs{\idV}{\eV}{\qV} \steq \actV'$. 
          By \Cref{cor:inversion-input-basic}, we obtain $\actVV'\steq \actrs{\idV}{\eV}{\ccat{\qV}{\vV}}$.
          Our result, $\actrs{\idV}{\eV}{\ccat{\qV}{\vV}}\steq\actVV'$, follows by symmetry.     
    \item Case \textsc{scp1}, \ie $\confK{\actNew{\idVV}{\actV}} \traS{\piinA{\idV}{\vV}} \actNew{\idVV}{\actVV}$ 
          because $\conf{\K,\idVV\;|\;\Obs}{\actV} \traS{\piinA{\idV}{\vV}} \actVV$ because $\idVV \fresh \fn{\pioutA{\idV}{\vV}}$.
          For the second move, we have $\confK{\actV'} \traS{\piinA{\idV}{\vV}} \actVV'$ where $\actV'\steq\actNew{\idVV}{\actV}$. 
          By \Cref{lemma:inversion-scp1}, we know $\conf{\K,\idVV\;|\;\Obs}{\actV} \traS{\piinA{\idV}{\vV}} \actVV''$ and $\actVV' \steq \actNew{\idVV}{\actVV''}$.
          By the IH and the fact that $\actV\steq\actV$, we obtain $\actVV \steq \actVV''$.
          Our result, $\actNew{\idVV}{\actVV} \steq \actNew{\idVV}{\actVV''}$, follows by rule \textsc{sCtxS}.
    \item Case \textsc{parL}, \ie $\confK{\actV_1 \actPar \actV_2} \traS{\piinA{\idV}{\vV}} \actVV_1 \actPar \actV_2$ 
          because $\confK{\actV_1} \traS{\piinA{\idV}{\vV}} \actVV_1$ and $\sbj{\piinA{\idV}{\vV}} \fresh \fid{\actV_2}$.
          For the second move, we have $\confK{\actV'} \traS{\piinA{\idV}{\vV}} \actVV'$ where $\actV'\steq\actV_1\actPar\actV_2$.
          By \Cref{lemma:inversion-input}, \Cref{cor:no-fid-no-in} and $\idV\fresh\fid{\actV_2}$, we know $\confK{\actV_1} \traS{\piinA{\idV}{\vV}} \actVV_1'$ and $\actVV' \steq \actVV_1' \actPar \actV_2$ for some $\actVV_1'$.
          By the IH, 
          $\actVV_1 \steq \actVV_1'$.
          Our result, $\actVV_1 \actPar \actV_2 \steq \actVV_1' \actPar \actV_2$, follows by \textsc{sCtxP}.
    \item Case \textsc{parR}, proof is analogous to that for case \textsc{parL}.   
    \item Case \textsc{str}, \ie $\confK{\actV} \traS{\piinA{\idV}{\vV}} \actVV$ 
          because $\actV\steq\actV''$ and $\confK{\actV''} \traS{\piinA{\idV}{\vV}} \actVV''$ and $\actVV''\steq\actVV$.
          For the second move, we have $\confK{\actV'} \traS{\piinA{\idV}{\vV}} \actVV'$ where $\actV'\steq\actV''$.
          By transitivity, we know $\actV''\steq\actV'$ as well.
          Using the IH, we thus obtain that $\actVV''\steq\actVV'$.
          Our result, $\actVV\steq\actVV'$, follows by symmetry and transitivity. \qedhere
  \end{itemize}
\end{proof}

\inDet*

\begin{proof}
  Follows from \Cref{lemma:input-det} since $\actV\steq\actV$.
\end{proof}

\subsection{Proving Output Determinacy}
\label{sec:out-det}

%
%
The proof showing that output actions are deterministic, \Cref{prop:output-determinacy}, relies on \Cref{lemma:od-free,lemma:output-det}.
We start with the former, which describes the structure of actors capable of performing an output action $\pioutA{\idV}{\vV}$.

\begin{lemma}\label{lemma:od-free}
  If $\confK{\actV} \traS{ \pioutA{\idV}{\vV}} \actVV$
  then $\actV \steq \actV' \actPar \pioutB{\idV}{\vV}$ and $\actVV \steq \actV'$.
  %
\end{lemma}

\begin{proof}
  Suppose $\confK{\actV} \traS{ \pioutA{\idV}{\vV}} \actVV$.
  The proof proceeds by induction on $\confK{\actV} \traS{ \pioutA{\idV}{\vV}} \actVV$. 
  \begin{itemize}[leftmargin=*]
    \setlength\itemsep{0.5em}
    \item Case \textsc{snd2}, \ie $\confK{\pioutB{\idV}{\vV}} \traS{\pioutA{\idV}{\vV}} \nl$ where $\idV\in\Obs$. 
          Result is immediate by rules \textsc{sNil}, \textsc{sCom}.
    %
    \item Case \textsc{scp1}, \ie $\confK{ \actNew{\idVV}{\actV}}  \traS{\pioutA{\idV}{\vV}} \actNew{\idVV}{\actVV}$ 
          because $\conf{\K,\idVV \;|\; \Obs}{ \actV } \traS{\pioutA{\idV}{\vV}} \actVV$ and $\idVV \fresh \fn{\pioutA{\idV}{\vV}}$.  
          By the IH, we obtain that 
          $\actV \steq \actV'\actPar \pioutB{\idV}{\vV}$ and 
          $\actVV \steq \actV'$ for some $\actV'$.
          Applying rule \textsc{sCtxt}, we get $\actNew{\idVV}{\actV} \steq \actNew{\idVV}{(\actV' \actPar \pioutB{\idV}{\vV})}$ and $\actNew{\idVV}{\actVV} \steq \actNew{\idVV}{\actV'}$.\\
          There only remains to show that $\actNew{\idVV}{\actV} \steq (\actNew{\idVV}{\actV'}) \actPar \pioutB{\idV}{\vV}$.
          Since $\idVV \fresh \fn{\pioutA{\idV}{\vV}}$, we know $\idVV\fresh \fn{\pioutB{\idV}{\vV}}$.
          Thus, by rule \textsc{sExt}, $\actNew{\idVV}{(\actV' \actPar \pioutB{\idV}{\vV})} \steq (\actNew{\idVV}{\actV'}) \actPar \pioutB{\idV}{\vV}$. 
          Our result, $\actNew{\idVV}{\actV} \steq (\actNew{\idVV}{\actV'}) \actPar \pioutB{\idV}{\vV}$, follows by transitivity.
    \item Case \textsc{parL}, \ie $\confK{\actV_1 \actPar \actV_2} \traS{ \pioutA{\idV}{\vV} } \actVV_1 \actPar \actV_2$
          because $\confK{\actV_1} \traS{ \pioutA{\idV}{\vV} } \actVV_1$ and $\idV \fresh \fid{\actV_2}$.
          By the IH, we obtain $\actV_1 \steq \actV_1' \actPar \pioutB{\idV}{\vV}$ 
          and $\actVV_1 \steq \actV_1' $ for some $\actV_1'$. 
          Applying rule \textsc{sCtxP}, we get 
          $\actV_1 \actPar \actV_2 \steq (\actV_1' \actPar \pioutB{\idV}{\vV}) \actPar \actV_2$ and 
          $\actVV_1 \actPar \actV_2 \steq \actV_1' \actPar \actV_2$.\\
          There remains to show $\actV_1 \actPar \actV_2 \steq (\actV_1' \actPar \actV_2) \actPar \pioutB{\idV}{\vV}$; this follows by rules \textsc{sAss}, \textsc{sCom}. 
    \item Case \textsc{parR}, analogous to that of \textsc{parL}.
    %
    %
    \item Case \textsc{sTrn}, \ie $\confK{\actV} \traS{\pioutA{\idV}{\vV}} \actVV$
      because $\actV \steq \actV''$, $\actVV \steq \actVV'$ and $\confK{\actV''} \traS{\pioutB{\idV}{\vV}} \actVV'$.
      By the IH, $\actV'' \steq \actV' \actPar \pioutB{\idV}{\vV}$ and $\actVV' \steq \actV'$. 
      By transitivity and symmetry of $\steq$, we can thus conclude $\actV \steq \actV' \actPar \pioutB{\idV}{\vV}$ and $\actVV \steq \actV'$.
      \qedhere  
  \end{itemize}
\end{proof}

\begin{lemma}\label{lemma:output-det}
  For any $\actV\steq\actV'$, 
  if $\confK{\actV} \traS{\pioutA{\idV}{\vV}} \actVV$ and $\confK{\actV'} \traS{\pioutA{\idV}{\vV}} \actVV'$ then $\actVV \steq \actVV'$.  
\end{lemma}

\begin{proof}
  Suppose $\actV\steq\actV'$ and $\confK{\actV} \traS{\pioutA{\idV}{\vV}} \actVV$ and $\confK{\actV'} \traS{\pioutA{\idV}{\vV}} \actVV'$.
  We show $\actVV \steq \actVV'$.
  The proof proceeds by induction on the first move. 
  \begin{itemize}[leftmargin=*]
    \setlength\itemsep{0.5em}
    \item Case \textsc{snd2}, \ie $\confK{\pioutB{\idV}{\vV}} \traS{\pioutA{\idV}{\vV}} \nl$ where $\idV\in\Obs$. 
        For the second move, we have $\confK{\actV'} \traS{\pioutA{\idV}{\vV}} \actVV'$ where $\actV'\steq\pioutB{\idV}{\vV}$.
        By \Cref{lemma:od-free}, we know $\actV'\steq\actV''\actPar\pioutB{\idV}{\vV}$ and $\actVV'\steq\actV''$ for some $\actV''$.
        But since $\actV'\steq\pioutB{\idV}{\vV}$, we can show $\actV''\steq\nl$. 
        Our result, $\nl\steq\actVV'$, follows.
    \item Case \textsc{scp1}, \ie $\confK{ \actNew{\idVV}{\actV}}  \traS{\pioutA{\idV}{\vV}} \actNew{\idVV}{\actVV}$ 
        because $\conf{\K,\idVV\;|\;\Obs}{ \actV } \traS{\pioutA{\idV}{\vV}} \actVV$ and $\idVV \fresh \fn{\vV}\cup\{i\}$.
        Consider the second move, $\confK{\actV'} \traS{\pioutA{\idV}{\vV}} \actVV'$ where $\actNew{\idVV}{\actV} \steq\actV'$.
        By \Cref{lemma:inversion-scp1}, we know 
        $\actVV' \steq \actNew{\idVV}{\actVV''}$ 
        and $\conf{\K,\idVV\;|\;\Obs}{\actV} \traS{\pioutA{\idV}{\vV}} \actVV''$.
        By the IH, we obtain that $\actVV \steq \actVV''$.
        By rule \textsc{sCtxS}, $\actNew{\idVV}{\actVV} \steq \actNew{\idVV}{\actVV''}$.
        Our result, $\actNew{\idVV}{\actVV} \steq \actVV'$, follows by transitivity/symmetry. 
    \item  
        Case \textsc{parL}, \ie $\confK{\actV_1 \actPar \actV_2} \traS{ \pioutA{\idV}{\vV} } \actVV_1 \actPar \actV_2$
        because $\confK{\actV_1} \traS{ \pioutA{\idV}{\vV} } \actVV_1$ and $\idV \fresh \fid{\actV_2}$.
        Consider the second move, $\confK{\actV'} \traS{ \pioutA{\idV}{\vV} } \actVV'$ where $\actV_1\actPar \actV_2\steq\actV'$.
        By \Cref{lemma:inversion-output}, we have two sub-cases:
        \begin{itemize}
          \setlength\itemsep{0.5em}
          \item 
                For the first subcase, $\confK{\actV_1} \traS{ \pioutA{\idV}{\vV} } \actVV_1'$ and 
                $\actVV' \steq \actVV_1' \actPar \actV_2$.
                By the IH, we obtain $\actVV_1 \steq \actVV_1'$.
                By rule \textsc{sCxtP}, $\actVV_1 \actPar \actV_2 \steq \actVV_1' \actPar \actV_2$.
                Our result, $\actVV_1 \actPar \actV_2 \steq \actVV'$, follows by transitivity/symmetry.
          \item 
                For the second subcase, $\confK{\actV_2} \traS{ \pioutA{\idV}{\vV} } \actVV_2$ and 
                $\actVV' \steq \actV_1 \actPar \actVV_2$.
                \Cref{lemma:od-free}, we obtain 
                $\actV_1 \steq \actV_1' \actPar \pioutB{\idV}{\vV}$ and 
                $\actV_2 \steq \actV_2' \actPar \pioutB{\idV}{\vV}$ and 
                $\actVV_1 \steq \actV_1'$ and $\actVV_2 \steq \actV_2'$.  
                This implies that 
                \begin{align*}
                  \actVV_1 \actPar \actV_2 &\steq \actV_1' \actPar (\actV_2' \actPar \pioutB{\idV}{\vV}) \\
                  &\steq (\actV_1' \actPar \pioutB{\idV}{\vV}) \actPar \actV_2'  \text{ using rules } \textsc{sCom} \text{ and } \textsc{sAss} \\
                  &\steq \actV_1 \actPar \actVV_2 \steq \actVV'
                \end{align*}
                Our result, $\actVV_1 \actPar \actV_2 \steq \actVV'$, follows by transitivity.
          \end{itemize}
    \item Case \textsc{parR}, analogous to that of \textsc{parL}.
    %
    \item Case \textsc{str}, \ie $\confK{\actV} \traS{\pioutA{\idV}{\vV}} \actVV$ 
      because $\actV\steq\actV''$ and $\confK{\actV''} \traS{\pioutA{\idV}{\vV}} \actVV''$ and $\actVV''\steq\actVV$.
      For the second move, we have $\confK{\actV'} \traS{\pioutA{\idV}{\vV}} \actVV'$ where $\actV'\steq\actV''$.
      By transitivity, we know $\actV''\steq\actV'$ as well.
      Using the IH, we thus obtain that $\actVV''\steq\actVV'$.
      Our result, $\actVV\steq\actVV'$, follows by symmetry and transitivity. \qedhere
  \end{itemize}
\end{proof}


\outDet*

\begin{proof}
  Follows from \Cref{lemma:output-det} since $\actV\steq\actV$.
\end{proof}


\subsection{Proving Communication Determinacy}
\label{sec:comm-det}

The proof for \Cref{prop:comm-determinacy} relies on \Cref{lemma:comm-st-basic} below, which describes the structure of an actor system capable of performing an internal communication action \commA{\idV}{\vV}.  

\begin{lemma}\label{lemma:comm-st-basic}
  If $\confK{\actV} \traS{\commA{\idV}{\vV}} \actVV$ then
  \begin{enumerate}[label=(\roman*)]
    \item $\actV \steq \actV' \actPar \pioutB{\idV}{\vV}$;
    \item $\confK{\actV'} \traS{\piinA{\idV}{\vV}} \actVV'$ for some $\actVV'$;
    \item $\actVV \steq \actVV'$. 
  \end{enumerate} 
\end{lemma}

\begin{proof}
  Assume $\confK{\actV} \traS{\commA{\idV}{\vV}} \actVV$.
  We proceed by rule induction, outlining only the main cases; the remaining follow similarly.
  \begin{itemize}[leftmargin=*]
    \setlength\itemsep{0.5em}
    \item Case \textsc{commL}, \ie 
          $\actV = \actV_1 \actPar \actV_2$ and $\actVV= \actVV_1 \actPar \actVV_2$
          because $\conf{\K\;|\;\fid{\actV_2}}{\actV_1} \traS{\pioutA{\idV}{\vV}} \actVV_1$ 
          and $\conf{\K\;|\;\fid{\actV_1}}{\actV_2} \traS{\piinA{\idV}{\vV}} \actVV_2$. \\
          By \Cref{lemma:od-free}, we know $\actV_1 \steq \actV_1' \actPar \pioutB{\idV}{\vV}$
          and $\actVV_1 \steq \actV_1'$.
          By rules \textsc{sCtxP,\;sAss,\;sCom} and transitivity, this implies 
          $\actV_1 \actPar \actV_2 \steq (\actV_1' \actPar \actV_2) \actPar \pioutB{\idV}{\vV}$, giving us $(i)$. \\
          Applying rule \textsc{parR} on $\conf{\K\;|\;\fid{\actV_1}}{\actV_2} \traS{\piinA{\idV}{\vV}} \actVV_2$, we obtain 
          $\conf{\K\;|\;\fid{\actV_1}}{\actV_1' \actPar \actV_2} \traS{\piinA{\idV}{\vV}} \actV_1' \actPar \actVV_2$.
          Using \Cref{lemma:in-anyO}, we get $\confK{\actV_1' \actPar \actV_2} \traS{\piinA{\idV}{\vV}} \actV_1' \actPar \actVV_2$,
          giving us $(ii)$.\\
          The result in $(iii)$, namely $\actVV_1 \actPar \actVV_2 \steq \actV_1' \actPar \actVV_2$, follows from $\actVV_1 \steq \actV_1'$ and rule \textsc{sCtxP}. 
    \item Case \textsc{parL}, \ie 
          $\actV = \actV_1 \actPar \actV_2$ and $\actVV = \actVV_1 \actPar \actV_2$ because
          $\confK{\actV_1} \traS{\commA{\idV}{\vV}} \actVV_1$.
          %
          %
          By the IH, we obtain that 
          \begin{align}
            &\actV_1 \steq \actV_1' \actPar \pioutB{\idV}{\vV} \label{eq:stb-pl-1}\\
            &\confK{\actV_1'} \traS{\piinA{\idV}{\vV}} \actVV_1' \label{eq:stb-pl-3}\\
            &\actVV_1 \steq \actVV_1'  \label{eq:stb-pl-4}
          \end{align} 
          The result in $(i)$, namely $\actV_1\actPar\actV_2 \steq (\actV_1' \actPar \actV_2) \actPar \pioutB{\idV}{\vV}$, follows by (\ref{eq:stb-pl-1}), rules \textsc{sCtxP}, \textsc{sAss}, \textsc{sCom} and transitivity. \\
          The result in $(ii)$, namely $\confK{\actV_1' \actPar \actV_2} \traS{\piinA{\idV}{\vV}} \actVV_1' \actPar \actV_2$, follows by (\ref{eq:stb-pl-3}) and rule \textsc{parL}. \\
          The result in $(iii)$, namely $\actVV_1 \actPar \actV_2 \steq \actVV_1' \actPar \actV_2$, follows from (\ref{eq:stb-pl-4}) and rule \textsc{sCtxP}. 
    \item Case \textsc{scp1}, 
          $\actV=\actNew{\idVV}{\actV'}$ and $\actVV = \actNew{\idV}{\actVV'}$
          because $\conf{\K,\idVV}{\actV'} \traS{\commA{\idV}{\vV}} \actVV'$ and $\idVV\fresh\fn{\commA{\idV}{\vV}}$ 
          where $\fn{\commA{\idV}{\vV}} = \{\idV,\vV\}$.
          By the IH, we obtain
          \begin{align}
            &\actV' \steq \actVV'' \actPar \pioutB{\idV}{\vV} \label{eq:stb-scp-1}\\
            &\conf{\K,\idVV\;|\;\Obs}{\actV''} \traS{\piinA{\idV}{\vV}} \actVV'' \label{eq:stb-scp-3}\\
            &\actVV' \steq \actVV'' \label{eq:stb-scp-4}
          \end{align} 
          Applying rule \textsc{sCtxS} on (\ref{eq:stb-scp-1}), we get $\actNew{\idVV}{\actV'} \steq \actNew{\idVV}{(\actV'' \actPar \pioutB{\idV}{\vV})}$.
          Since $\idVV \fresh \{ \idV,\vV\}$, we can use rule \textsc{sExt} to obtain $\actNew{\idVV}{\actV'} \steq (\actNew{\idVV}{\actV''}) \actPar \pioutB{\idV}{\vV}$, giving us the result in $(i)$. \\
          The result in $(ii)$, namely $\confK{\actNew{\idVV}{\actV''}} \traS{\piinA{\idV}{\vV}} \actNew{\idVV}{\actVV''}$, follows by (\ref{eq:stb-scp-3}) and rule \textsc{scp1}. \\
          The result in $(iii)$, namely $\actNew{\idVV}{\actVV'} \steq \actNew{\idVV}{\actVV''}$, follows by (\ref{eq:stb-scp-4}) and rule \textsc{sCtxS}.
          \qedhere
  \end{itemize}
\end{proof}

We are now in a position to prove that communication actions lead to structurally equivalent actor systems, as stated in \Cref{prop:comm-determinacy} below.


\commDet*

\begin{proof}
  Suppose $\confK{\actV} \traS{\commA{\idV}{\vV}} \actVV$ and $\confK{\actV} \traS{\commA{\idV}{\vV}} \actVV'$. 
  We need to show $\actVV \steq \actVV'$.
  By \Cref{lemma:comm-st-basic} and $\confK{\actV} \traS{\commA{\idV}{\vV}} \actVV$, we know that there exist some actor system $\actVVV$ such that
  \begin{align*}
    & \actV \steq \actVVV \actPar \pioutB{\idV}{\vV} \quad \text{and} \quad 
    \confK{\actVVV} \traS{\piinA{\idV}{\vV}} \actVVV' \quad \text{and} \quad
    \actVV \steq \actVVV'
  \end{align*}
  Similarly, by \Cref{lemma:comm-st-basic} and $\confK{\actV} \traS{\commA{\idV}{\vV}} \actVV'$, we know that there exist some actor system \actVVVV\ such that  
  \begin{align*}
    &\actV \steq \actVVVV \actPar \pioutB{\idV}{\vV} \quad \text{and} \quad
    \confK{\actVVVV} \traS{\piinA{\idV}{\vV}} \actVVVV' \quad \text{and} \quad
    \actVV' \steq \actVVVV'
  \end{align*}
  Since $\actV \steq \actVVV \actPar \pioutB{\idV}{\vV}$ and $\actV \steq \actVVVV \actPar \pioutB{\idV}{\vV}$,
  we also know that $\actVVV \actPar \pioutB{\idV}{\vV} \steq \actVVVV \actPar \pioutB{\idV}{\vV}$.
  By case analysis, this could have only been derived using rule \textsc{sCtxP}, which gives us $\actVVV \steq \actVVVV$.
  Thus, by rule \textsc{sTrn} and $\confK{\actVVVV} \traS{\piinA{\idV}{\vV}} \actVVVV'$, we know $\confK{\actVVV} \traS{\piinA{\idV}{\vV}} \actVVVV'$ as well. 
  Using the facts that $\confK{\actVVV} \traS{\piinA{\idV}{\vV}} \actVVVV'$ and $\confK{\actVVV} \traS{\piinA{\idV}{\vV}} \actVVV'$ and \Cref{prop:input-determinacy}, we obtain $\actVVV' \steq \actVVVV'$. 
  Since $ \actVV \steq \actVVV' $ and $ \actVV' \steq \actVVVV'$, using transitivity/symmetry, we can conclude $\actVV \steq \actVV'$.
\end{proof}

\section{Properties of the Violation Relation}
\label{sec:viol-relation-results}

We prove some properties about the violation relation $\viol$ of \Cref{def:viol-relation}, including \Cref{thm:viol-corr-notin,thm:viol-complete} from \Cref{sec:monitorability}.
In this section and the ones that follow, we work up to $\alpha$-equivalence.

\begin{lemma}\label{lemma:viol-hst-notempty}
  If $(\hst,\odF) \viol \varphi$ then $\hst\neq\emptyset$.
\end{lemma}

\begin{proof}
  Straightforward by rule induction.  
\end{proof}

We show that the violation relation $\viol$ observes sanity checks akin to those for the history analysis of \Cref{fig:proof-system}. 
In particular, \Cref{prop:viol-l-irrevoc,prop:viol-w-irrevoc} below guarantee that once a system violates a formula via a history, it will persistently violate that formula, regardless of any other behaviour it might exhibit (described in terms of additional traces added to the history, width, or longer trace prefixes, length).

\begin{proposition}[Width Irrevocability]\label{prop:viol-w-irrevoc}
  If $(\hst,\odF) \viol \varphi$ then $ (\hst\cup\hstt,\odF) \viol \varphi$.   
\end{proposition}

\begin{proof}
  The proof is similar to that for \Cref{prop:irrevocability}.
\end{proof}

\begin{proposition}[Length Irrevocability]\label{prop:viol-l-irrevoc}
  If $(\hst\cup{t},\odF) \viol \varphi$ then $ (\hst\cup\{tu\},\odF) \viol \varphi$.   
\end{proposition}

\begin{proof}
  The proof is similar to that for \Cref{prop:irrevocability}.
\end{proof}

\begin{corollary}\label{cor:smaller-hst-nviol}
  If $(\hst\cup\hstt, \odF) \nviol \varphi$ then $(\hst, \odF) \nviol \varphi$. 
\end{corollary}

We also lift the function \sub{-} to traces: 
$\sub{\hst, \epsilon} =\hst$ and $\sub{\hst, \tact t} = \sub{\sub{\hst,\tact}, t}$.
Similarly, $\detAct{\epsilon} =\btrue$ and $\detAct{\tact t} = \detAct{\tact} \wedge \detAct{t}$. 



\begin{lemma}\label{lemma:inversion}
  For all $\varphi\in\sHMLwDet$ 
  and $t\in\IAct^*$, 
  if $\hstt\!=\!\sub{\hst,t\act}$ and $\odF'\!=\!\odF\wedge\detAct{t\act}$ and $(\hstt,\odF')\viol\varphi$
  then $(\hst, \odF) \viol \Um{\act}{\varphi}$. 
\end{lemma}

\begin{proof}
  The proof proceeds by induction on the length of $t$, \ie $n=|t|$. 
  \begin{itemize}
    \setlength\itemsep{0.5em}
    \item When $n=0$, then $t=\epsilon$, $\hstt=\sub{\hst,\act}$, $\odF'=\odF\wedge\detAct{\act}$ and $(\hstt,\odF')\viol\varphi$.
          Our result, $(\hst,\odF)\viol\Um{\act}{\varphi}$, follows immediately by rule \textsc{vUm}. 
    \item When $n=k+1$, 
          then $t=\iact_1\cdots\iact_n\in\IAct^*$ 
          and $\hstt=\sub{\hst,t\act}$ 
          and $\odF'=\odF\wedge\detAct{t\act}$ 
          and $(\hstt,\odF')\viol\varphi$.
          By definition of \sub{-}, we know there exists some $\hstt'$ such that 
          \begin{align}
            \hstt'=\sub{\hst,\iact_1} \quad\text{ and }\quad \hstt=\sub{\hstt,\iact_2\cdots\iact_n \act}
            \label{eq:decomposed-hst}
          \end{align}
          By definition of \detAct{-}, we also know there exists some $\odF''$ such that 
          \begin{align}
            \odF''= \odF \wedge \detAct{\iact_1} \quad\text{ and } \quad \odF'= \odF'' \wedge \detAct{\iact_2\cdots\iact_n}  
            \label{eq:decomposed-det}
          \end{align}   
          Using (\ref{eq:decomposed-hst}), (\ref{eq:decomposed-det}) and the IH, we obtain $(\hstt',\odF'')\viol\Um{\act}{\varphi}$.
          Our result, $(\hst,\odF)\viol\Um{\act}{\varphi}$, follows by applying rule \textsc{vUmPre}.
          \qedhere
  \end{itemize} 
\end{proof}



We prove that whenever a system \pV\ produces a history \hst\ that violates a formula $\varphi$, \ie $\hst \viol \varphi$, then $\pV$ must also violate it, \ie $\pV\notin\evalE{\varphi}$, namely \Cref{thm:viol-corr-notin} from \Cref{sec:monitorability}.
This proof relies on an additional result. 
Specifically, \Cref{lemma:viol-false-t} below states that if a history violates a formula with the flag set to \bfalse, then a single trace $t$ suffices to violate that formula.

\begin{lemma} \label{lemma:viol-false-t}
  If $(\hst,\bfalse) \viol \varphi$ then $\exists t\in\hst$ such that $(\{t\},\bfalse) \viol \varphi$.
\end{lemma}

\begin{proof}
  Straightforward by rule induction.
\end{proof}

\Cref{lemma:viol-t-notin} below then states that whenever a single trace violates a formula, then the system producing that trace also violates the formula.

\begin{lemma} \label{lemma:viol-t-notin}
  For all $t\in\hstpV$, 
  if $(\{t\},\odF) \viol \varphi$ then $\pV \notin \evalE{\varphi}$.
\end{lemma}

\begin{proof}
  Straightforward by rule induction.
\end{proof}

We are now in a position to prove \Cref{thm:viol-corr-notin}, restated below.

\correspondence*

\begin{proof}
  Suppose that $\exists \hst \subseteq \hstpV$ such that $\hst \viol \varphi$.
  Our result, $\pV\notin\evalE{\varphi}$, follows from \Cref{lemma:viol-implies-notin} below, by letting $\odF=\btrue$.  
\end{proof}

\begin{lemma}\label{lemma:viol-implies-notin}
  For all $\varphi\in\sHMLwDet$ and $\hst\subseteq \hstpV$, 
  if $(\hst,\odF) \viol \varphi$ then $\pV \notin \evalE{\varphi}$. 
\end{lemma}

\begin{proof}
  The proof proceeds by induction on $(\hst,\odF)\viol\varphi$ where $\hst\subseteq \hstpV$.
  \begin{itemize}
    \setlength\itemsep{0.5em}
    \item Case \textsc{vF}, \ie $(\hst,\odF) \viol \F$ where $\hst\neq\emptyset$. Our result, $\pV\notin\evalE{\F}$, is immediate since $\evalE{\F} = \emptyset$.
    \item Case \textsc{vUm}, \ie $(\hst,\odF) \viol \Um{\act}{\varphi}$ because $(\hstt,\odF') \viol \varphi$ 
      where $\hstt = \sub{\hst,\act}$ and $\odF'=\odF\wedge \detAct{\act}$.  
      By \Cref{lemma:viol-hst-notempty}, we also know $\hstt\neq\emptyset$. 
      This means that $\exists n\geq 1$ such that    
      \begin{align}
        &\hstt = \bigcup_{i=1}^{n}\hst_{i}'
        \text{ such that } 
        \pV \wttraS{\act} \pVV_i 
        \text{ and } \hst_i'\subseteq T_{\pVV_i} 
        \notag \\[-0.5em]
        &\qquad\qquad\qquad \text{ for each } i\in\{1,\ldots,n\}
        \label{eq:decompose-hst} 
      \end{align}
      There are two subcases to consider:
      \begin{itemize}
        \setlength\itemsep{0.5em}
        \item If $\odF'=\bfalse$, 
          we know by \Cref{lemma:viol-false-t} that $\exists t\in\hstt$ such that $(\{ t \}, \odF') \viol \Um{\act}{\varphi}$.
          From (\ref{eq:decompose-hst}), we also know $t\in\hstt_k$ for some $k\in\{1,\ldots,n\}$ and that $\pV \wttraS{\act} \pVV_k$ and $\hstt_k \subseteq T_{\pVV_k}$. 
          Using \Cref{lemma:viol-t-notin}, we deduce that $\pVV_k \notin \evalE{\varphi}$, and by \Cref{prop:harmony}, we obtain $\pV \wtraS{\act} \pVV_k$.
          Thus, we can conclude that $\pV\notin \{ \pVV \;|\; \pV \wtraS{\act} \pVV \text{ implies } \pVV\notin\evalE{\varphi} \} = \evalE{\Um{\act}{\varphi}}$, as required.
        \item If $\odF'=\btrue$, then \detAct{\act}.
          From \Cref{lemma:det-tEq}, we know $T_{\pVV_i} = T_{\pVV_j}$ for all $i,j\in\{1,\ldots,n\}$, 
          which implies $\hstt\subseteq T_{\pVV_k}$ for all $k\in\{1,\ldots,n\}$.
          We can thus use the IH and obtain $\pVV_k \notin \evalE{\varphi}$.  
          By \Cref{prop:harmony} and the fact that $\pV\wttraS{\act} \pVV_k$, we also know $\pV\wtraS{\act} \pVV_k$.
          Thus, we can conclude that $\pV\notin \{ \pVV \;|\; \pV \wtraS{\act} \pVV \text{ implies } \pVV\notin\evalE{\varphi} \} = \evalE{\Um{\act}{\varphi}}$, as required.      
        \end{itemize}
      \item Case \textsc{vUmPre}, \ie $(\hst,\odF) \viol \Um{\act}{\varphi}$ because $(\hstt,\odF') \viol \Um{\act}{\varphi}$ 
        where $\hstt = \sub{\hst,\iact}$ and $\odF'=\odF\wedge \detAct{\iact}$.  
        Due to our assumption that all internal actions $\IAct$ are deterministic, \ie \detAct{\iact}, then $\odF'=\btrue$.     
        By \Cref{lemma:viol-hst-notempty}, we also know $\hstt\neq\emptyset$. 
        This means that $\exists n\geq 1$ such that    
        \begin{align*}
          &\hstt = \bigcup_{i=1}^{n}\hst_{i}'
          \text{ such that } 
          \pV \wttraS{\iact} \pVV_i 
          \text{ and } \hst_i'\subseteq T_{\pVV_i} \\[-0.5em]
          &\qquad\qquad \qquad \text{ for each } i\in\{1,\ldots,n\} 
        \end{align*}
        From \Cref{lemma:det-tEq}, we know $T_{\pVV_i} = T_{\pVV_j}$ for all $i,j\in\{1,\ldots,n\}$, 
        which implies $\hstt\subseteq T_{\pVV_k}$ for all $k\in\{1,\ldots,n\}$.
        We can thus use the IH and obtain $\pVV_k \notin \evalE{\Um{\act}{\varphi}}$.  
        By \Cref{prop:harmony} and $\pV\wttraS{\iact} \pVV_k$, we also know $\pV\wtraS{\;\;} \pVV_k$.
        Our result, $\pV\notin\evalE{\Um{\act}{\varphi}}$, follows by definition of \evalE{-}.      
    \item Case \textsc{vAndL}.
      We know $(\hst,\odF) \viol \Conj{\varphi}{\psi}$ because $(\hst,\odF)\viol\varphi$. 
      By the IH, we obtain $\pV\notin\evalE{\varphi}$, which implies 
      that $\pV \notin \evalE{\varphi} \cap \evalE{\psi} = \evalE{\Conj{\varphi}{\psi}}$.
    \item Case \textsc{vAndR}.
      Proof is analogous to that for \textsc{vAndL}.
    \item Case \textsc{vOr}.
      We know $(\hst,\btrue) \viol \Disj{\varphi}{\psi}$ because $(\hst,\btrue)\viol\varphi$ and $(\hst,\btrue)\viol\psi$. 
      By the IH, we obtain $\pV\notin\evalE{\varphi}$ and $\pV\notin\evalE{\psi}$, which implies 
      that $\pV \notin \evalE{\varphi} \cup \evalE{\psi} = \evalE{\Disj{\varphi}{\psi}}$.
    \item Case \textsc{vMax}.
      We know $(\pV,\odF)\viol\Max{\XV}{\varphi}$ because $(\hst,\odF)\viol\varphi\subS{\Max{\XV}{\varphi}}{\XV}$.
      By the IH, we obtain $\pV \notin \evalE{\varphi\subS{\Max{\XV}{\varphi}}{\XV}}=\evalE{\Max{\XV}{\varphi}}$.  
      \qedhere
  \end{itemize}
\end{proof}

We now prove \Cref{thm:viol-complete}, restated below. 

\correspondenceComplete*


\begin{proof}
  Follows from \Cref{lemma:notin-implies-viol-expanded} below, by letting $\odF = \btrue$.
\end{proof}

\begin{lemma}\label{lemma:notin-implies-viol-expanded}
  If $\Det(\iact) = \btrue$ for all $\iact \in \IAct$, then  
  for all $\pV\in\Prc$, $\varphi\in\sHMLwDet$ and $\odF\in\Bool$, 
  if $\odF \cons \varphi$ and $\pV\notin\evalE{\varphi}$ 
  then $\bigl(\exists \hst\subseteq\hstpV$ such that $(\hst,\odF) \viol \varphi \bigr)$.
\end{lemma}

\begin{proof}
  Suppose $\pV\in \evalE{\varphi}$. 
  Since $\hst \subseteq \hstpV$, it suffices to show $(\hstpV, \btrue) \viol \varphi$.
  This follows from \Cref{lemma:notin-implies-viol-hstpV} below.
\end{proof}

\begin{lemma}\label{lemma:notin-implies-viol-hstpV}
  If $\Det(\iact) = \btrue$ for all $\iact \in \IAct$, then  
  for all $\pV\in\Prc$, $\varphi\in\sHMLwDet$ and $\odF\in\Bool$, 
  if $\odF \cons \varphi$ 
  and $\pV\notin\evalE{\varphi}$ then $(\hstpV,\odF) \viol \varphi$.
\end{lemma}

\begin{proof}
  The proof proceeds by rule induction on $\odF \cons \varphi$.
  \begin{itemize}
    \setlength \itemsep{0.5em}
    \item Case \textsc{cA}, \ie $\odF \cons \varphi$ where $\varphi\in\{ \F, \T, \XV \}$. 
      Assume that $\pV\notin\evalE{\varphi}$. 
      When $\varphi=\F$, we immediately obtain that $(\hstpV, \odF) \viol \F$ by rule \textsc{vF}.
      When $\varphi=\T$, the statement is vacuously true since $\evalE{\T}=\Prc$ and thus $\pV\in\evalE{\varphi}$.
      Also, we cannot have that $\varphi = \XV$; we are assuming $\varphi$ is closed.
    \item Case \textsc{cUm}, \ie $\odF \cons \Um{\act}{\varphi}$ because $\odF \wedge \detAct{\act} \cons \varphi$.
      Assume $\pV\notin\evalE{\Um{\act}{\varphi}}$.
      This means that there exists $\pVV$ such that $\pV \wtraS{\act} \pVV$ and $\pVV\notin\evalE{\varphi}$.
      By \Cref{prop:harmony}, we know $\pV \wttraS{t} \pV' \wttraS{\act} \pV'' \wttraS{t'} \pVV$ for some $\pV',\pV''$ and $t,t'\in\IAct^*$, 
      which implies $\pV''\notin\evalE{\varphi}$.
      There are three subcases to consider: 
      \begin{itemize}
        \setlength\itemsep{0.5em}
        \item When $\odF {=} \btrue {=} \detAct{\act}$, we have $\btrue \cons \varphi$.
        By the IH, we deduce $(T_{\pV''}, \btrue) \viol \varphi$, \ie $(T_{\pV''}, \btrue \wedge \detAct{\act}) \viol \varphi$. 
        Applying rule \textsc{vUm}, we obtain $(T_{\pV'}, \btrue) \viol \Um{\act}{\varphi}$.
        Applying rule \textsc{vUmPre} $|t|$ times, we conclude $(\hstpV, \btrue) \viol \Um{\act}{\varphi}$.
        \item When $\odF = \bfalse$, the proof is analogous to that for the previous case.
        \item When $\odF=\btrue$ and $\detAct{\act}=\bfalse$, we have $\bfalse \cons \varphi$.
        By the IH, we get $(T_{\pV''}, \bfalse) \viol \varphi$ \ie $(T_{\pV''}, \btrue \wedge \detAct{\act}) \viol \varphi$.
        Applying rule \textsc{vUm}, we obtain $(T_{\pV'}, \btrue) \viol \Um{\act}{\varphi}$.
        By the assumption that $\detAct{\iact}=\btrue$, then we also know that for $t=\iact_1\cdots\iact_n$, we have $\detAct{\iact_i}=\btrue$.
        We can thus apply rule \textsc{vUmPre} $n$ times and conclude $(\hstpV, \btrue) \viol \Um{\act}{\varphi}$.
      \end{itemize}

    \item Case \textsc{cAnd}, \ie $\odF \cons \Conj{\varphi}{\psi}$ because $\odF \cons \varphi$ and $\odF \cons \psi$.
      Assume $\pV\notin\evalE{\Conj{\varphi}{\psi}}$. 
      This implies that either $\pV\notin\evalE{\varphi}$ or $\pV\notin\evalE{\psi}$.
      \Wlog suppose the former.
      By the IH, we obtain $(\hstpV, \odF) \viol \varphi$.
      Our result, $(\hstpV, \odF) \viol \Conj{\varphi}{\psi}$, follows by rule \textsc{vAnd}.  
    \item Case \textsc{cOr}, \ie $\btrue \cons \Disj{\varphi}{\psi}$ because $\btrue \cons \varphi$ and $\btrue \cons \psi$.
      Assume $\pV\notin\evalE{\Disj{\varphi}{\psi}}$. 
      This implies that $\pV\notin\evalE{\varphi}$ and $\pV\notin\evalE{\psi}$.
      By the IH, we obtain $(\hstpV, \btrue) \viol \varphi$ and $(\hstpV, \btrue) \viol \psi$.
      Our result, $(\hstpV, \btrue) \viol \Disj{\varphi}{\psi}$, follows by rule \textsc{vOr}.  
    \item Case \textsc{cMax}, \ie $\odF \cons \Max{\XV}{\varphi}$ because $\odF \cons \varphi\subS{\Max{\XV}{\varphi}}{\XV}$.
      Assume $\pV\notin\evalE{\Max{\XV}{\varphi}}$.
      Since $\evalE{\Max{\XV}{\varphi}} = \evalE{\varphi\subS{\Max{\XV}{\varphi}}{\XV}}$, we also know $\pV\in\evalE{\varphi\subS{\Max{\XV}{\varphi}}{\XV}}$.
      By the IH, we obtain $(\hstpV, \odF) \viol \varphi\subS{\Max{\XV}{\varphi}}{\XV}$.
      Our result, $(\hstpV, \odF) \viol \Max{\XV}{\varphi}$, follows by rule \textsc{vMax}. 
    \qedhere
  \end{itemize}
\end{proof}

\section{Lower Bounds}
\label{sec:lowerbound-proofs}

We provide additional examples and results related to \Cref{sec:lowerbounds}. 
We start by \Cref{ex:rec-lower-bound} below, which complements \Cref{ex:lowerbound-invariant} by further illustrating the complexity of calculating lower bounds for formulae containing greatest fixed points.

\begin{example}\label{ex:rec-lower-bound}
  Recall $\varphi_4 \deftxt \Max{\XV}{\bigl(\Um{r}{\Um{s}{\XV}} \wedge (\Um{c}{\F} \vee \Um{a}{\F})\bigr)}$ from \Cref{ex:rechml}.
  The proposed syntactic analysis of this formula would determine that the history lower bound for $\varphi_4$ is 2.  
  Concretely, the conjunction sub-formula \Um{r}{\Um{s}{\XV}} can potentially contain an unbounded number of disjunctions due to recursion, whereas the right sub-formula contains 1 disjunction, meaning that 2 traces are required;  
  the lower bound across the conjunction is thus $2$.
  The unfolding of $\varphi_4$ is:  
  \begin{center}  
    $\varphi_4' \deftxtS \Bigl(\Um{r}{\Um{s}{ \bigl( \Max{\XV}{\bigl(\Um{r}{\Um{s}{\XV}} \wedge (\Um{c}{\F} \vee \Um{a}{\F})\bigr)} \bigr)}}\Bigr) \wedge (\Um{c}{\F} \vee \Um{a}{\F})$
  \end{center}    
  %
  where the history lower bound calculation is invariant at 2.
  \exqed
\end{example}

To prove \Cref{prop:shmlnf-disj-hst}, \Cref{thm:lowerbounds} and \Cref{cor:lb-inf} from \Cref{sec:lowerbounds}, we first give a few technical developments, starting with several properties for the function \lb{-} in \Cref{def:lb}, where the meta-function \fv{\varphi} returns the free recursion variables in $\varphi$.

\begin{lemma}\label{lemma:lb-ext}
  For all $\varphi,\psi,\chi\in\sHMLnf$:
  \begin{enumerate}
    \setlength\itemsep{0.5em} 
    \item $\lb{\varphi\subS{\psi}{\XV}} = \lb{\varphi\subS{\Max{\XVV}{\psi}}{\XV}}$ \label{lemma:lb-ext-max}
    \item $\lb{\psi}\leq\lb{\chi}$ implies $\lb{\varphi\subS{\psi}{\XV}} \leq \lb{\varphi\subS{\chi}{\XV}}$ \label{lemma:lb-sub-larger}
    \item $\lb{\varphi}\leq\lb{\psi}$ implies $\lb{\varphi} \leq \lb{\psi\subS{\varphi}{\XV}}$ \label{lemma:lb-ext-sub}
  \end{enumerate}
\end{lemma}

\begin{proof}
  We only give those for the main cases of (\ref{lemma:lb-ext-sub}); the others follow with a similar but more straightforward argument and can be proven independently.\\[0.5em]
  %
  \emph{The proof of (\ref{lemma:lb-ext-sub})}  
  proceeds by 
  induction on $\psi$.
  Assume $\lb{\varphi}\leq\lb{\psi}$.
  We show $\lb{\varphi} \leq \lb{\psi\subS{\varphi}{\XV}}$. 
  \begin{itemize}
    \setlength\itemsep{0.5em}
    \item When $\psi = \XVV$, we have $\lb{\XVV}\!=\!\infty$ and $\lb{\varphi}\leq\lb{\XVV}$.
    If $\XV\!=\!\XVV$, result follows immediately since 
    $\XV\subS{\varphi}{\XV} = \varphi$.
    If $\XV\!\neq\XVV$, result also follows immediately since $\XVV\subS{\varphi}{\XV} = \XVV$.
    \item When $\psi = \Um{\act}{\psi'}$, we have $\lb{\varphi} \leq \lb{\Um{\act}{\psi'}}\!=\!\lb{\psi'}$.
    By the IH, we deduce $\lb{\varphi} \leq \lb{\psi'\subS{\varphi}{\XV}}$, which implies 
    $\lb{\varphi} \leq \lb{(\Um{\act}{\psi'})\subS{\varphi}{\XV}}$.
    \item When $\psi = \Conj{\psi_1}{\psi_2}$, we have $\lb{\varphi} \leq \textsl{min}(\lb{\psi_1}, \lb{\psi_2})$, 
    which implies $\lb{\varphi} \leq \lb{\psi_1}$ and $\lb{\varphi} \leq \lb{\psi_2}$.
    By the IH, we obtain $\lb{\varphi} \leq \lb{\psi_1\subS{\varphi}{\XV}}$ and $\lb{\psi_2\subS{\varphi}{\XV}}$.
    Our result, $\lb{\varphi} \leq \lb{(\Conj{\psi_1}{\psi_2})\subS{\varphi}{\XV}}$, follows since $\lb{(\Conj{\psi_1}{\psi_2})\subS{\varphi}{\XV}} = \textsl{min}(\lb{\psi_1\subS{\varphi}{\XV}}, \lb{\psi_2\subS{\varphi}{\XV}})$.   
    \item When $\psi = \Disj{\psi_1}{\psi_2}$, then $\lb{\varphi} \!\leq\! \lb{\Disj{\psi_1}{\psi_2}} \!=\! \lb{\psi_1}\!+\!\lb{\psi_2}\!+\!1$.
    There are two subcases to consider:
    \begin{itemize}
      \item 
      When $\lb{\varphi} \!\leq\! \lb{\psi_1}$ or $\lb{\varphi} \!\leq\! \lb{\psi_2}$. 
      \Wlog suppose the former. 
      By the IH, we obtain $\lb{\varphi} \!\leq\! \lb{\psi_1\subS{\varphi}{\XV}}$, which implies that 
      $\lb{\varphi} \!\leq\! \lb{\psi_1\subS{\varphi}{\XV}} + \lb{\psi_2\subS{\varphi}{\XV}} + 1 = \lb{(\Disj{\psi_1}{\psi_2})\subS{\varphi}{\XV}}$.
      \item 
      When $\lb{\varphi} \!>\! \lb{\psi_1}$ and $\lb{\varphi} \!>\! \lb{\psi_2}$.
      By the IH, we obtain $\lb{\psi_1} \!\leq\! \lb{\psi_1\subS{\psi_1}{\XV}}$ and $\lb{\psi_2} \!\leq\! \lb{\psi_2\subS{\psi_2}{\XV}}$. 
      But by \Cref{lemma:lb-ext}(\ref{lemma:lb-sub-larger}), we also know $\lb{\psi_1\subS{\psi_1}{\XV}} \!\leq\! \lb{\psi_1\subS{\varphi}{\XV}}$ and $\lb{\psi_2\subS{\psi_2}{\XV}} \!\leq\! \lb{\psi_2\subS{\varphi}{\XV}}$.  
      This implies that 
      $\lb{\varphi} \leq \lb{\psi_1} + \lb{\psi_2} + 1  
      \!\leq\! \lb{\psi_1\subS{\varphi}{\XV}} + \lb{\psi_2\subS{\varphi}{\XV}} 
      = \lb{(\Disj{\psi_1}{\psi_2})\subS{\varphi}{\XV}}$, as required.
    \end{itemize}
    \item When $\psi \!=\! \Max{\XVV}{\psi'}$, then $\lb{\varphi} \!\leq\! \lb{\Max{\XVV}{\psi'}}\!=\!\lb{\psi'}$.
    If $\XV\!=\!\XVV$, our result is immediate since $(\Max{\XV}{\psi'})\subS{\varphi}{\XV} \!=\! \Max{\XV}{\psi'}$.
    If $\XV\!\neq\!\XVV$, then by the IH, we obtain $\lb{\varphi} \!\leq\! \lb{\psi'\subS{\varphi}{\XV}} \!=\! \lb{(\Max{\XVV}{\psi'})\subS{\varphi}{\XV}}$.
    \qedhere
  \end{itemize}
\end{proof}

\begin{corollary}\label{cor:lb-sub}
  For all $\varphi\!\in\!\sHMLnf$, $\lb{\varphi} \leq \lb{\varphi\subS{\varphi}{\XV}}$.  
\end{corollary}

\begin{proof}
  Follows from \Cref{lemma:lb-ext}(\ref{lemma:lb-ext-sub}) by letting $\varphi = \psi$.
\end{proof}

We give an alternative definition to the violation relation, \viol, in \Cref{def:viol-relation} that is specific to \sHMLnf formulae (in contrast to \viol\ which is defined over \sHMLw), namely \Cref{def:sep-viol-relation}. 
\Cref{thm:corr-viol-sviol} then shows that these two definitions correspond.

\begin{definition}\label{def:sep-viol-relation}
  The \emph{separation violation relation}, denoted as \sviol, is the least relation of the form $(\Hst \times \Bool \times \sHMLnf)$ satisfying the following rules:
  %
  \begin{mathpar}[\small]
    \inferrule[\rtitSS{svF}]
      { \hst \neq \emptyset }
      { (\hst, \odF) \sviol \F }
    \and
    \inferrule[\rtitSS{svMax}]
      { (\hst, \odF) \sviol \varphi\subS{\Max{\XV}{\varphi}}{\XV} }
      { (\hst, \odF) \sviol \Max{\XV}{\varphi} }
    \and
    \inferrule[\rtitSS{svUm}]
      { \hstt = \sub{\hst,\act} \quad \odF' = \odF \wedge \detAct{\act} \quad (\hstt, \odF') \sviol \varphi}
      { (\hst, \odF) \sviol \Um{\act}{\varphi} }
    \and
    \inferrule[\rtitSS{svUmPre}]
      { \hstt = \sub{\hst,\iact} \quad \odF' = \odF \wedge \detAct{\iact} \quad (\hstt, \odF') \sviol \Um{\act}{\varphi}}
      { (\hst, \odF) \sviol \Um{\act}{\varphi} }
    \and
    \inferrule[\rtitSS{svAndL}]
      { (\hst, \odF) \sviol \varphi }
      { (\hst, \odF) \sviol \Conj{\varphi}{\psi} }
    \and
    \inferrule[\rtitSS{svAndR}]
      { (\hst, \odF) \sviol \psi }
      { (\hst, \odF) \sviol \Conj{\varphi}{\psi} }
    \and
    \inferrule[\rtitSS{svOr}]
      { \hst=\hst_1\uplus\hst_2 \quad (\hst_1, \btrue)\sviol\varphi \quad (\hst_2, \btrue) \sviol \psi }
      { (\hst, \btrue) \sviol \Disj{\varphi}{\psi} }
  \end{mathpar}
  \\
  %
  We write 
  $\hst \sviol \varphi$ 
  to mean $(\hst, \btrue) \sviol \varphi$.
  \exqed
\end{definition}

\noindent
Width irrevocability also holds for the separation violation relation, \Cref{lemma:union-sviol-H}. 

\begin{lemma}\label{lemma:union-sviol-H}%
  For all $\varphi\in \sHMLnf$,
  if $(\hst,\odF) \sviol \varphi$ then $(\hst\cup\hstt, \odF) \sviol \varphi$. 
\end{lemma}

\begin{proof}
  The proof proceeds by rule induction.
  We only give the proof for the case when $(\hst,\odF)\sviol\varphi$ is derived via rule \textsc{svOr}; the other cases are straightforward.
  \begin{itemize}
    \item We know $(\hst,\odF) \sviol \Disj{\varphi_1}{\varphi_2}$ because $\hst=\hst_1\uplus\hst_2$ and $(\hst_1,\odF)\sviol\varphi_1$ and $(\hst_2,\odF)\sviol\varphi_2$.  
    We show that $(\hst\cup\hstt, \odF) \sviol \Disj{\varphi_1}{\varphi_2}$.
    Let $\hstt'=\hst\backslash\hstt$ where $\hst_1\cap\hstt=\emptyset$ and $\hst_2\cap\hstt=\emptyset$. 
    By the IH, we know $(\hst_1 \cup \hstt', \odF) \sviol \varphi_1$.
    Since $(\hst_1\cup\hstt') \cap \hst_2 = \emptyset$, we conclude $(\hst\cup\hstt, \odF) \sviol \Disj{\varphi_1}{\varphi_2}$ via rule \textsc{svOr}.
    %
    \qedhere 
  \end{itemize}
\end{proof}

\Cref{lemma:disjoint-hst} shows that all violating systems for $\bigvee_{i\in I} \Um{\act_i}{\varphi_i}$ from \sHMLnf\ can violate the sub-formulae \Um{\act_i}{\varphi_i} through disjoint histories.  
This result relies on the helper function $\start{\hst,\act} = \{ t \;|\; t=\act t' \in \hst \}$, returning the set of all traces in \hst\ that are prefixed with a sequence of internal actions $\iact\in\IAct^*$, followed by an \act\ action.
\Eg when $\hst = \{ \ut rsa, \uf rsc, ars\}$, then $\start{\hst,r} = \{ \ut rsa, \uf rsc\}$. 

%
\begin{lemma}\label{lemma:start-disjoint}
  For all $\act,\actt$ such that $\act\neq\actt$, $\start{\hst,\act} \cap \start{\hst,\actt} = \emptyset$. 
  \exqed
\end{lemma}

\begin{proof}
  Straightforward by definition.
\end{proof}

\begin{lemma}\label{lemma:disjoint-hst}
  For all formulae $\bigvee_{i\in I} \Um{\act_i}{\varphi_i} \in \sHMLnf$ and histories $\hst\in\Hst$:
  \begin{align*}
    \text{if } 
    (\hst, \odF) \sviol \bigvee_{i\in I} \Um{\act_i}{\varphi_i} 
    & \text{ then } 
    \bigl(\start{\hst,\act_i} , \odF \bigr)\sviol \Um{\act_i}{\varphi_i} \\[-1em]
    & \qquad \text{ for each } i\in I 
  \end{align*}
\end{lemma}

\begin{proof}
  %
  The proof proceeds by induction on the size of $I$.
  \begin{itemize}
    \setlength\itemsep{0.5em}
    %
    %
    \item 
      For the base case, $I=\{1\}$, \ie $(\hst, \odF) \sviol \Um{\act}{\varphi}$.
      Using \Cref{lemma:inversion}, we know $\exists t\in \IAct^*$ such that $\hstt=\sub{\hst,t\act}$ and $\odF'=\odF\wedge\detAct{t\act}$ and $(\hstt,\odF')\viol\varphi$.
      Let $\hstt' = \{ \act u \; | \; u \in \hstt \}$, \ie all traces in $\hstt$ prefixed with action $\act$.
      Applying rule \textsc{svUm}, we obtain $(\hstt', \odF \wedge \detAct{t}) \sviol \Um{\act}{\varphi}$.
      Let $\hstt'' = \{ t t' \; | \; t' \in \hstt \}$, \ie all traces in $\hstt'$ prefixed with trace $t$.
      Applying rule \textsc{svUmPre} $n$ times where $n$ is the length of trace $t$, we obtain $(\hst'',\odF) \viol \Um{\act}{\varphi}$.
      Since, by definition, $\hstt'' \subseteq \start{\hst, \act}$, we can use \Cref{lemma:union-sviol-H} to conclude that $(\start{\hst,\act} , \odF) \sviol \Um{\act}{\varphi}$. 
    \item 
      For the inductive case, $I{=}\{1,...,n{+}1\}$. 
      The judgment $(\hst,\odF) \!\sviol\! \bigvee_{i{\in} I} \Um{\act_i}{\varphi_i}$ can be expanded to 
      $$
        (\hst,\odF) \sviol \bigl(\Um{\act_1}{\varphi_1}\bigr) \vee \bigl( \bigvee_{j\in J} \Um{\act_j}{\varphi_j} \bigr) 
        \text{ where } 
        J=\{ 2,\ldots,n{+}1\} 
      $$
      By case analysis, this could have only been derived via rule \textsc{svOr}, which means that 
      there exist $\hst',\hst''$ such that $\hst=\hst'\uplus\hst''$ and   
      $(\hst', \odF) \sviol \Um{\act_1}{\varphi_1}$ and $(\hst'',\odF) \sviol \bigvee_{j\in J} \Um{\act_j}{\varphi_j}$. 
      Using \Cref{lemma:union-sviol-H}, we deduce 
      $(\pV,\hst) \sviol \Um{\act_1}{\varphi_1}$ and $(\pV, \hst) \sviol \bigvee_{j\in J} \Um{\act_j}{\varphi_j}$.
      By the IH, we obtain $(\start{\hst,\act_1}, \odF) \sviol \Um{\act_1}{\varphi_1}$ and $(\start{\hst,\act_j}, \odF) \sviol \Um{\act_j}{\varphi_j}$ for all $j\in J$.
      We can thus conclude $(\start{\hst,\act_i}, \odF) \sviol \Um{\act_i}{\varphi_i}$ for all $i\in I$, as required.
    \qedhere
  \end{itemize}
\end{proof}


\noindent
We show 
a similar result holds for the violation relation \viol. 

\begin{lemma}\label{lemma:disjoint-hst-viol}
  For all $\bigvee_{i\in I} \Um{\act_i}{\varphi_i} \in \sHMLnf$ and $\hst\subseteq\hstpV$:
  \begin{align*}
    \text{if }
    (\hst,\odF) \viol \bigvee_{i\in I} \Um{\act_i}{\varphi_i} 
    &\text{ then } 
    \bigl(\start{\hst,\act_i}, \odF \bigr) \viol \Um{\act_i}{\varphi_i} \\[-1em]
    & \qquad \text{ for each } i\in I 
  \end{align*}  
\end{lemma}

\begin{proof}
  Proof is similar, but more straightforward, to that for \Cref{lemma:disjoint-hst-viol}.
\end{proof}

\begin{theorem}[Correspondence]\label{thm:corr-viol-sviol}
  For all $\varphi{\in}\sHMLnf$ and $\hst{\in}\Hst$, $(\hst,\odF)\viol \varphi$ iff $(\hst, \odF) \sviol \varphi$.
\end{theorem}

\begin{proof}
  For the \textit{if} direction, we show $(\hst,\odF)\sviol \varphi$ implies $(\hst,\odF)\viol \varphi$.
  The proof is by rule induction. 
  We only give the case for when $(\hst,\odF)\viol \varphi$ is derived via rule \textsc{svOr}; all other cases are homogeneous.
  \begin{itemize}
    \item 
      We know $(\hst,\odF)\sviol \Disj{\varphi}{\psi}$ because $\exists\hst_1,\hst_2$ such that $\hst=\hst_1\uplus\hst_2$, 
      $(\hst_1, \odF) \sviol \varphi$ and $(\hst_2, \odF) \sviol \psi$.
      By the IH, we deduce $(\hst_1, \odF)\viol\varphi$ and $(\hst_2, \odF)\viol\psi$, 
      which imply $(\hst, \odF)\viol\varphi$ and $(\hst,\odF)\viol\psi$ by \Cref{prop:viol-w-irrevoc}. 
      Our result, $(\hst,\odF)\viol\Disj{\varphi}{\psi}$, follows via rule \textsc{vOr}. \\[-0.7em]
  \end{itemize} 
  For the \textit{only if} direction, we show $(\hst,\odF)\viol \varphi$ implies $(\hst,\odF)\sviol \varphi$.
  Again, the proof is by rule induction and we only give that for when $(\pV,\hst)\viol \varphi$ is derived via rule \textsc{vOr}.
  \begin{itemize}
    \item We know $(\hst,\odF) \!\viol\! \Disj{\varphi}{\psi}$ because $(\hst,\odF)\!\viol\!\varphi$ and $(\hst,\odF)\!\viol\!\psi$.
    By the IH, we deduce $(\hst,\odF)\sviol\varphi$ and $(\hst,\odF)\sviol\psi$. 
    Since $\Disj{\varphi}{\psi}\!\in\sHMLnf$, then $\varphi \!=\! \bigvee_{i\in I}\Um{\act_i}{\varphi_i}$ and $\psi \!=\! \bigvee_{j\in J}\Um{\act_j}{\psi_j}$ where $I\cap J = \emptyset$.
    By \Cref{lemma:disjoint-hst}, we obtain   
    $\bigl(\start{\hst,\act_i},\odF\bigr) \sviol \varphi_i$ and 
    $\bigl(\start{\hst,\act_j},\odF\bigr) \sviol \psi_j$ for each $i\in  I$ and $j\in J$.
    By \Cref{lemma:start-disjoint}, we also know $\bigcap_{k\in I\cup J} \start{\hst,\act_k} = \emptyset$.
    Repeatedly applying rule \textsc{svOr}, we deduce 
    $\bigl(\bigcup_{i\in I} \start{\hst,\act_i}, \odF \bigr) \sviol \bigvee_{i\in I}\Um{\act_i}{\varphi_i}$ and  
    $\bigl(\bigcup_{j\in J} \start{\hst,\act_j}, \odF \bigr) \sviol \bigvee_{j\in J}\Um{\act_i}{\psi_j}$.  
    By rule \textsc{svOr} again, we get $\bigl( \bigcup_{k\in I\cup J} \start{\hst,\act_k} , \odF \bigr) \sviol \Disj{\varphi}{\psi}$.
    Our result, $(\hst, \odF) \sviol \Disj{\varphi}{\psi}$, follows via \Cref{lemma:union-sviol-H} since $\bigcup_{k\in I\cup J} \start{\hst,\act_k} \subseteq \hst$.
    \qedhere
  \end{itemize}
\end{proof}

\noindent
Equipped with these technical results, we prove \Cref{prop:shmlnf-disj-hst} and \Cref{thm:lowerbounds}.
\Cref{thm:lowerbounds} is a direct consequence of \Cref{lemma:lowerbounds-sviol}.


\shmlnfdisjhst*

\begin{proof}
  Assume $(\hst, \odF)\viol\Disj{\varphi}{\psi}$. 
  Since $\Disj{\varphi}{\psi}\in\sHMLnf$, we know $\varphi = \bigvee_{i\in I} \Um{\act_i}{\varphi_i'}$ and $\psi = \bigvee_{j\in J} \Um{\act_j}{\psi_j'}$ where $I\cap J = \emptyset$.
  By \Cref{lemma:disjoint-hst-viol}, we deduce $(\start{\hst,\act_i}, \odF) \viol \Um{\act_i}{\varphi_i'}$ and $(\start{\hst,\act_j}, \odF)\viol\Um{\act_j}{\psi_j'}$ for each $i\in I$ and $j\in J$.
  Let $\hst_1=\bigcup_{i\in I}\start{\hst,\act_i}$ and $\hst_2=\bigcup_{j\in J}\start{\hst,\act_j}$.  
  Repeatedly applying rule \textsc{vOr}, we obtain $(\hst_1, \odF)\viol\varphi$ and $(\hst_2, \odF)\viol\psi$. 
  From \Cref{lemma:start-disjoint}, we know $\hst_1\cap \hst_2 = \emptyset$.
  Let $\hst_3 = \hst \backslash \hst_1$.
  By \Cref{prop:viol-w-irrevoc}, we get $(\hst_1, \odF) \viol \varphi$ and $(\hst_2\cup\hst_3, \odF) \viol \psi$ where $\hst=\hst_1\uplus(\hst_2\cup\hst_3)$ as required.
\end{proof}

\begin{lemma}\label{lemma:lowerbounds-sviol}
  For all 
  $\varphi\!\in\!\sHMLnf$ and $\hst\!\in\!\Hst$, if $(\hst, \odF) \!\sviol\! \varphi$ then 
  $|\hst|\geq \lb{\varphi}\!+\!1$.
\end{lemma}

\begin{proof}
  The proof is by rule induction.
  \begin{itemize}
    \setlength\itemsep{0.5em}
    \item Case \textsc{vF}, \ie $(\hst, \odF) \sviol \F$ where $\hst\neq\emptyset$. Thus $|\hst| \geq 1 = \lb{\F} + 1$.
    \item 
      Case \textsc{vUm}, \ie $(\hst, \odF) \sviol \Um{\act}{\varphi}$ because $(\hstt,\odF') \sviol\varphi$ where $\hstt=\sub{\hst,\act}$ and $\odF' = \odF\wedge \detAct{\act}$.
      By the IH, we deduce $|\hstt| \geq \lb{\varphi}\!+\!1$.
      Let $\hstt' = \{ \act t \; | \; t\in\hstt \}$.
      Since $\hstt' \subseteq \hst$, then $|\hst| \geq |\hstt'| = |\hstt| \geq \lb{\varphi}+1 = \lb{\Um{\act}{\varphi}}+1$. 
    \item 
      Case \textsc{vUmPre}, \ie $(\hst, \odF) \sviol \Um{\act}{\varphi}$ because $(\hstt,\odF') \sviol \Um{\act}{\varphi}$ where $\hstt=\sub{\hst,\iact}$ and $\odF' = \odF\wedge \detAct{\iact}$.
      By the IH, we deduce $|\hstt| \geq \lb{\Um{\act}{\varphi}}+1$.
      Let $\hstt' = \{ \iact t \; | \; t\in\hstt \}$.
      Since $\hstt' \subseteq \hst$, then $|\hst| \geq |\hstt'| = |\hstt| \geq \lb{\Um{\act}{\varphi}}+1$. 
    \item 
      Case \textsc{vAndL}, \ie $(\hst,\odF) \sviol \Conj{\varphi}{\psi}$ because $(\hst,\odF) \sviol \varphi$.  
      By the IH, $|\hst|\geq \lb{\varphi}+1 \geq \textsf{min}(\lb{\varphi}, \lb{\psi})+1 = \lb{\Conj{\varphi}{\psi}}+1$.
    \item Case \textsc{vAndR}. Analogous to previous case.
    \item 
      Case \textsc{vOr}, \ie $(\hst,\btrue) \sviol \Disj{\varphi}{\psi}$ because $\hst=\hst_1\uplus\hst_2$ and $(\hst_1,\btrue) \sviol \varphi$ and $(\hst_2,\btrue) \sviol \varphi$.
      By the IH, we obtain $|\hst_1|\geq \lb{\varphi}+1$ and $|\hst_2|\geq \lb{\psi}+1$, which means that   
      $|\hst| = |\hst_1| + |\hst_2| \geq \lb{\varphi}+\lb{\psi}+2 = \lb{\Disj{\varphi}{\psi}}+1$.
    \item 
      Case \textsc{vMax}, \ie $(\hst,\odF)\sviol\Max{\XV}{\varphi}$ because $(\hst,\odF)\sviol\varphi\subS{\Max{\XV}{\varphi}}{\XV}$.
      By the IH, \Cref{lemma:lb-ext}(\ref{lemma:lb-ext-max}) and \Cref{cor:lb-sub}, we conclude 
      $|\hst|\geq \lb{\varphi\subS{\Max{\XV}{\varphi}}{\XV}}+1 
      = \lb{\varphi\subS{\varphi}{\XV}}+1 \geq \lb{\varphi}+1 
      = \lb{\Max{\XV}{\varphi}}+1$, as required. 
    \qedhere
  \end{itemize}
\end{proof}

\lowerbounds*

\begin{proof}
  Follows from \Cref{lemma:lowerbounds-sviol} and \Cref{thm:corr-viol-sviol}.
\end{proof}

\lbinf*

\begin{proof}
  Suppose $\varphi\in\sHMLnf$ and $\hst\in\Hst$ such that $\lb{\varphi}=\infty$.
  Since histories are finite, $|\hst|\leq \lb{\varphi} < \lb{\varphi}+1 = \infty$.
  Our result, $\hst\nviol\varphi$, follows by the contrapositive of \Cref{thm:lowerbounds}.
\end{proof}

\end{document}